\documentclass[a4paper,11pt]{article}

\usepackage{amsmath,amsthm,amsbsy,amssymb,amsfonts,graphicx,mathrsfs,accents,cite}
\usepackage[all]{xy}

\makeatletter
\catcode`\@=11
\@addtoreset{equation}{section}

\makeatother

\renewcommand{\thefootnote}{\arabic{footnote}}

\newcommand{\Exp}[1]{\operatorname{e}^{#1}}
\newcommand{\rmd}{{\mathrm{d}}}
\newcommand{\rmT}{\mathrm{T}}
\newcommand{\nn}{\nonumber}
\newcommand{\Lie}{\pounds}
\newcommand{\gLie}{\hat{\pounds}}

\newcommand{\cH}{\mathcal H}
\newcommand{\cM}{\mathcal M}
\newcommand{\cT}{\mathcal T}

\newcommand{\sfa}{\mathsf a}
\newcommand{\sfb}{\mathsf b}
\newcommand{\sfc}{\mathsf c}
\newcommand{\sfd}{\mathsf d}
\newcommand{\sfe}{\mathsf e}
\newcommand{\sfm}{\mathsf m}
\newcommand{\sfn}{\mathsf n}
\newcommand{\sfp}{\mathsf p}
\newcommand{\sfq}{\mathsf q}
\newcommand{\sfr}{\mathsf r}
\newcommand{\sfx}{\mathsf x}
\newcommand{\sfy}{\mathsf y}
\newcommand{\sfA}{\mathsf A}
\newcommand{\sfB}{\mathsf B}
\newcommand{\sfM}{\mathsf M}
\newcommand{\sfN}{\mathsf N}
\newcommand{\sfP}{\mathsf P}
\newcommand{\sfQ}{\mathsf Q}

\newcommand{\GL}{\mathrm{GL}}
\newcommand{\SL}{\mathrm{SL}}
\newcommand{\SO}{\mathrm{SO}}
\newcommand{\OO}{\mathrm{O}}

\newcommand{\sMA}{{\mathtt{I}}}
\newcommand{\sMB}{{\mathtt{J}}}
\newcommand{\sMC}{{\mathtt{K}}}
\newcommand{\sBA}{{\mathtt{M}}}
\newcommand{\sBB}{{\mathtt{N}}}

\newcommand{\Ay}{{y}}
\newcommand{\Az}{{z}}
\newcommand{\By}{{\sfy}}
\newcommand{\bbeta}{{\boldsymbol\eta}}

\newcommand{\lambdaB}{{\mkern0.75mu\mathchar '26\mkern -9.75mu\lambda}}

\allowdisplaybreaks[3]

\begin{document}

\begin{titlepage}
\renewcommand{\thefootnote}{\fnsymbol{footnote}}

\vspace*{1.0cm}

\begin{center}
{\LARGE\textbf{$\eta$-symbols in exceptional field theory}}%
\end{center}
\vspace{1.0cm}

\centerline{
{Yuho Sakatani$^{a,b}$}%
\footnote{E-mail address: \texttt{yuho@koto.kpu-m.ac.jp}}
\ \ and \ \ 
{Shozo Uehara$^{a}$}%
\footnote{E-mail address: \texttt{uehara@koto.kpu-m.ac.jp}}
}

\vspace{0.2cm}

\begin{center}
${}^a${\it Department of Physics, Kyoto Prefectural University of Medicine,}\\
{\it Kyoto 606-0823, JAPAN}\\
and\\
${}^b${\it Fields, Gravity \& Strings, CTPU}\\
{\it Institute for Basic Sciences, Seoul 08826, KOREA}
\end{center}
\vspace*{1cm}
\begin{abstract}
We present the universal form of $\eta$-symbols that can be applied to an arbitrary $E_{d(d)}$ exceptional field theory (EFT) up to $d=7$. We then express the $Y$-tensor, which governs the gauge algebra of EFT, as a quadratic form of the $\eta$-symbols. The usual definition of the $Y$-tensor strongly depends on the dimension of the compactification torus while it is not the case for our $Y$-tensor. Furthermore, using the $\eta$-symbols, we propose a universal form of the linear section equation. In particular, in the $\SL(5)$ EFT, we explicitly show the equivalence to the known linear section equation.
\end{abstract}
\thispagestyle{empty}
\end{titlepage}

\setcounter{footnote}{0}

\section{Introduction}
In double field theory (DFT) \cite{Siegel:1993xq,Siegel:1993th,Siegel:1993bj,Hull:2009mi,Hull:2009zb,Hohm:2010jy,Hohm:2010pp}, for the purpose of the manifest $T$-duality covariance, we consider a $2d$-dimensional doubled space with the generalized coordinates $x^I$ ($I=1,\dotsc,2d$). 
In order to make contact with the conventional supergravity in $d$-dimensions, it is useful to decompose the generalized coordinates into the physical coordinates $x^i$ ($i=1,\dotsc,d$) and the dual coordinates $\tilde{x}_i$\,; $(x^I)=(x^i,\,\tilde{x}_i)$\,. 
By introducing the $\OO(d,d)$ $T$-duality-invariant metric,
\begin{align}
 (\eta_{IJ}) = \begin{pmatrix} 0 & \delta_i^j \\ \delta^i_j & 0 \end{pmatrix} , \qquad 
 (\eta^{IJ}) = \begin{pmatrix} 0 & \delta^i_j \\ \delta_i^j & 0 \end{pmatrix} , 
\end{align}
the consistency condition of DFT, the so-called the section condition, is expressed as
\begin{align}
 \eta^{IJ}\,\partial_I \otimes \partial_J = 0 \,. 
\end{align}
Here, $\otimes$ represents that
\begin{align}
 \eta^{IJ}\,\partial_I \partial_J A =0\,,\qquad 
 \eta^{IJ}\,\partial_I A\, \partial_J B = 0 \,,
\end{align}
are satisfied for arbitrary fields or gauge parameters $A$ and $B$. 
Under the section condition, the gauge algebra generated by the following generalized Lie derivative is closed:
\begin{align}
 \gLie_V W^I \equiv V^J\,\partial_J W^I - W^J\,\partial_J V^I + \eta^{IJ}\,\eta_{KL} \, \partial_J V^K \,W^L\,.
\end{align}

As a natural generalization of DFT, the $E_{d(d)}$ exceptional field theories (EFTs) \cite{Berman:2010is,Berman:2011cg,Berman:2011jh,Berman:2012vc,Hohm:2013pua,Hohm:2013vpa,Hohm:2013uia,Aldazabal:2013via,Hohm:2014fxa} have been formulated in a manifestly $E_{d(d)}$ $U$-duality covariant manner (see \cite{West:2001as,West:2003fc,West:2004st,Hillmann:2009ci} for the initial attempts). 
In EFT, the generalized coordinates $x^I$ ($I=1,\dotsc,D$) are defined to transform in a fundamental representation, called the $R_1$-representation (see Appendix \ref{app:Edd}). 
The generalized Lie derivative is defined by
\begin{align}
 \gLie_V W^I \equiv V^J\,\partial_J W^I - W^J\,\partial_J V^I + Y^{IJ}_{KL} \, \partial_J V^K \,W^L\,,
\end{align}
where the $Y$-tensor $Y^{IJ}_{KL}$ for each $d$ ($4\leq d\leq 7$) is given as follows \cite{Berman:2012vc} (see \cite{Pacheco:2008ps,Coimbra:2011ky,Coimbra:2012af} for the generalized Lie derivative in the context of exceptional generalized geometry):
\begin{align}
 \begin{tabular}{|c||c|c|c|c|c|}\hline
 $E_{d(d)}$ & $\SL(5)$ & $\SO(5,5)$ & $E_{6(6)}$ & $E_{7(7)}$ \\ \hline\hline
 $D=\dim R_1$& $10$ & $16$ & $27$ & $56$
\\\hline
 $Y^{IJ}_{KL}$& $\epsilon^{\sfe IJ}\,\epsilon_{\sfe KL}$ & $\frac{1}{2}\, \gamma_{\sfA}^{IJ}\,\bar{\gamma}^{\sfA}_{KL}$ & $10\,d^{IJM}\,d_{KLM}$ & $12\,c^{IJ}{}_{KL}+\frac{1}{2}\,\Omega^{IJ}\,\Omega_{KL}$
\\\hline
\end{tabular} \,.
\label{eq:Y-tensor-case-by-case}
\end{align}
Here, e.g., $\gamma_{\sfA}^{IJ}$ is the gamma matrix for the $\SO(5,5)$ group and $d^{IJK}$ is the totally symmetric tensor intrinsic to the $E_{6(6)}$ group (see Appendix \ref{app:eta-comparison} for the details of these $d$-dependent tensors). 
The gauge algebra of the generalized diffeomorphism is closed if the following section conditions are satisfied \cite{Coimbra:2011ky,Berman:2012vc}:
\begin{align}
\begin{split}
 d\leq 6\,: \qquad &Y^{IJ}_{KL}\, \partial_I\otimes \partial_J = 0 \,,
\\
 d=7\,: \qquad &Y^{IJ}_{KL}\, \partial_I\otimes \partial_J = 0\,,\qquad \Omega^{IJ}\,\partial_I\otimes \partial_J = 0\,,
\end{split}
\label{eq:EFT-SC}
\end{align}
where $\Omega^{IJ}$ is the antisymmetric tensor intrinsic to the $E_{7(7)}$ group. 
Under the section condition, all fields can depend on at most $d$ coordinates (see \cite{Bandos:2015rvs} for a proof in the $E_{7(7)}$ EFT). 

In the above conventional formulation, the $Y$-tensor and the section condition strongly depend on the dimension $d$, and when we consider applications of the $E_{d(d)}$ EFT, we need to specify the dimension $d$ explicitly. 
A hypothetical ``underlying EFT'' (or 11D EFT), which reproduces all $E_{d(d)}$ EFTs ($d\leq 8$) from simple truncations, has been proposed in \cite{Bandos:2016ppv}, but the program has not been completed yet. 
In this paper, we investigate such uniform formulations from a different approach. 
In our approach, the $Y$-tensor is expressed in terms of $\SL(d)$ [or $\SL(d-1)$] tensors and $E_{d(d)}$ tensors are not used. 
Accordingly, the truncation to lower $d$ can be easily performed. 

The present paper is organized as follows. 
In section \ref{sec:sketch}, we introduce $\eta$-symbols as a natural generalization of the $\OO(d,d)$-invariant metric $\eta_{IJ}$ in DFT and explain how the $\eta$-symbols are related to branes in M-theory/type IIB theory. 
The $Y$-tensor is expressed by using the $\eta$-symbols and the $\Omega$-tensor. 
In section \ref{sec:explicit-eta}, we find the explicit form of the $\eta$-symbols and the $\Omega$-tensor. 
In section \ref{sec:GeneralizedLie}, we show the explicit form of the section condition and the generalized Lie derivative. 
In section \ref{sec:Lienar-section}, we propose a new linear section equation, and show that it reproduces the known linear section equation \cite{Berman:2012vc} in the case of the $\SL(5)$ EFT. 
Section \ref{sec:conclusions} is devoted to conclusions and discussion.

\section{A sketch of the basic idea}
\label{sec:sketch}

The section condition in DFT has been proposed on the basis of the level-matching constraint in string sigma model \cite{Siegel:1993xq,Siegel:1993th,Siegel:1993bj},
\begin{align}
 S = -\frac{1}{4\pi\alpha'}\int_\Sigma\sqrt{-\gamma}\,\rmd^2\sigma\, \bigl(G_{ij}\,\gamma^{\bar{A}\bar{B}} + B_{ij}\,\epsilon^{\bar{A}\bar{B}}\bigr)\,\partial_{\bar{A}} X^i\,\partial_{\bar{B}} X^j \qquad (\bar{A},\bar{B}=\tau,\sigma)\,. 
\end{align}
In the canonical formulation, the level-matching constraint, or the momentum constraint $\cH_{\sigma}=0$, can be expressed as
\begin{align}
 \cH_{\sigma} = P_i\,\partial_\sigma X^i = \frac{1}{4\pi\alpha'}\,\eta^{IJ}\,Z_I\,Z_J = 0 \,,
\end{align}
where $P_i(\sigma)$ are the conjugate momenta to $X^i(\sigma)$ and $Z_I(\sigma)$ are the generalized momenta,
\begin{align}
 Z_I(\sigma) = \begin{pmatrix} 2\pi\alpha' P_i(\sigma)\\ \partial_{\sigma} X^i(\sigma) \end{pmatrix} . 
\end{align}
By supposing that the operator
\begin{align}
 \mathbb{L}_V \equiv \int\rmd \sigma\, V^I\bigl(X^J(\sigma)\bigr)\,Z_I(\sigma)\,,
\end{align}
acts as the generator of the diffeomorphism along $V^I\,\partial_I$, we can roughly identify $Z_I$ with $\partial_I$, and the momentum constraint corresponds to the section condition in DFT, $\eta^{IJ}\,\partial_I\otimes \partial_J = 0$\,.

A similar consideration has been given for M-theory branes in \cite{Hatsuda:2012vm,Hatsuda:2013dya}. 
In the case of an M2-brane wrapped on a 4-torus, the momentum constraint $\cH_A=0$ ($A=1,2$: index for spatial coordinates on the M2-brane) is rewritten as
\begin{align}
 \eta^{IJ;\,k} \, Z_I \, Z_J = 0 \,,
\end{align}
where 
\begin{align}
 \eta^k&\equiv (\eta^{IJ;\,k}) \equiv \begin{pmatrix}
 0 & \frac{2!\,\delta^{k i}_{j_1j_2}}{\sqrt{2!}} \\
 \frac{2!\,\delta^{k j}_{i_1i_2}}{\sqrt{2!}} & 0 
 \end{pmatrix} , \qquad 
 (Z_I) \equiv \begin{pmatrix} P_i \\ \frac{\frac{1}{2}\,\epsilon^{AB}\,\partial_A X^{[i_1}\,\partial_B X^{i_2]}}{\sqrt{2!}} \end{pmatrix} . 
\end{align}
Again by supposing the generalized momenta $Z_I$ to act as $\partial_I\equiv \partial/\partial x^I$ with $(x^I)=(x^i,\,\frac{y_{i_1i_2}}{\sqrt{2!}})$, the momentum constraint is expressed as the section condition,
\begin{align}
 \eta^{IJ;\,k} \, \partial_I \otimes \partial_J = 0 \,. 
\end{align}
Similarly, in the case of an M5-brane wrapped on a 5-torus, the momentum constraint, $\cH_A=0$ ($A=1,\dotsc,5$), has been expressed in a bilinear form (see \cite{Hatsuda:2013dya} for the details),
\begin{align}
 a_k\,\eta^{IJ;\,k} \, Z_I \, Z_J + b_{k_1\cdots k_4}\,\eta^{IJ;\,k_1\cdots k_4} \, Z_I \, Z_J = 0 \,,
\end{align}
where the matrices $\eta^k\equiv (\eta^{IJ;\,k})$ and $\eta^{k_1\cdots k_4}\equiv (\eta^{IJ;\,k_1\cdots k_4})$ have the form:
\begin{align}
 \eta^k \equiv \begin{pmatrix}
 0 & \frac{2!\,\delta^{k i}_{j_1j_2}}{\sqrt{2!}} & 0 \\
 \frac{2!\,\delta^{k j}_{i_1i_2}}{\sqrt{2!}} & 0 & 0 \\
 0 & 0 & 0 
 \end{pmatrix} , 
\qquad 
 \eta^{k_1\cdots k_4} 
 \equiv \begin{pmatrix}
 0 & 0 & \frac{5!\,\delta^{i k_1\cdots k_4}_{j_1\cdots j_5}}{\sqrt{5!}} \\
 0 & \frac{4!\,\delta^{k_1\cdots k_4}_{i_1i_2j_1j_2}}{\sqrt{2!\,2!}} & 0 \\
 \frac{5!\,\delta^{j k_1\cdots k_4}_{i_1\cdots i_5}}{\sqrt{5!}} & 0 & 0 
 \end{pmatrix} . 
\label{eq:E5-M2-M5-eta}
\end{align}
Here, $a_k$ and $b_{k_1\cdots k_4}\,(=b_{[k_1\cdots k_4]})$ are arbitrary constants and the section conditions can be decomposed into two parts,
\begin{align}
 \eta^{IJ;\,k} \, \partial_I \otimes \partial_J = 0 \,,\qquad 
 \eta^{IJ;\,k_1\cdots k_4} \, \partial_I \otimes \partial_J = 0 \,.
\end{align}
The former condition is the same as the section condition coming from the M2-brane and the latter is intrinsic to the M5-brane.

A similar consideration for a D$p$-brane in type II string theory was made in \cite{Hatsuda:2012uk} (though the $U$-duality covariance is not manifest there), and the general rule we observe is that each $p$-brane provides the corresponding $\eta$-symbol $\eta^{k_1\cdots k_{p-1}}$ and the associated section condition $\eta^{IJ;\,k_1\cdots k_{p-1}}\, \partial_I \otimes \partial_J = 0$\,. 
In fact, a set of multiple indices with one dimension fewer in the spatial dimension of branes is known to form the string multiplet of $E_{d(d)}$ group. 
The dimension of the string multiplet for each $U$-duality group is given as follows (see \cite{Obers:1998fb} for a concise review):
\begin{align}
 \begin{tabular}{|c|c|c|c|c|}\hline
 duality group & $\SL(5)$ & $\SO(5,5)$ & $E_{6(6)}$ & $E_{7(7)}$ \\ \hline
 dimension of string mult. & $5$ & $10$ & $27$ & $133$ \\ \hline
\end{tabular}\,. 
\end{align}
As is clear from the dimension, the string multiplet is the same as the $R_2$-representation that determines the section condition \cite{Obers:1998fb} (see also Appendix \ref{app:Edd}). 
Now, it is natural to expect that each brane in the string multiplet provides a particular $\eta$-symbol and the corresponding section condition, and the sum of all these section conditions is equivalent to the section condition in the $E_{d(d)}$ EFT. 
We thus introduce the following set of $\eta$-symbols associated with branes in the string multiplet in M-theory and type IIB theory:
\begin{align}
\footnotesize
 (\eta^{\sMA}) &\footnotesize
 = \biggl(\underbrace{\eta^k}_{\mathrm{M}2},\,\underbrace{\frac{\eta^{k_1\cdots k_4}}{\sqrt{4!}}}_{\mathrm{M}5},\,\underbrace{\frac{\eta^{k_1\cdots k_6,\,l}}{\sqrt{6!}}}_{\mathrm{KKM}/8},\,\underbrace{\frac{\eta^{k_1\cdots k_7,\,l_1l_2l_3}}{\sqrt{7!\,3!}}}_{5^3},\,\underbrace{\frac{\eta^{k_1\cdots k_7,\,l_1\cdots l_6}}{\sqrt{7!\,6!}}}_{2^6},\cdots\biggr) \,,
\nn\\
\footnotesize
 (\eta^{\sBA}) &\footnotesize
 = \biggl(\!\! \underbrace{\eta_\alpha}_{\mathrm{F}1/\mathrm{D}1}\!, \underbrace{\frac{\eta^{\sfm_1\sfm_2}}{\sqrt{2!}}}_{\mathrm{D}3}, \underbrace{\frac{\eta_\alpha^{\sfm_1\cdots \sfm_4}}{\sqrt{4!}}}_{\mathrm{NS}5/\mathrm{D}5}, \underbrace{\frac{\eta^{\sfm_1\cdots \sfm_5,\,\sfn}}{\sqrt{5!}}}_{\mathrm{KKM}/7_2}, \underbrace{\frac{\eta_{(\alpha\beta)}^{\sfm_1\cdots \sfm_6}}{\sqrt{6!}}}_{\mathrm{Q}7}, \underbrace{\frac{\eta_\alpha^{\sfm_1\cdots \sfm_6,\,\sfn_1\sfn_2}}{\sqrt{6!\,2!}}}_{5^2_2/5^2_3}, \underbrace{\frac{\eta^{\sfm_1\cdots \sfm_6,\,\sfn_1\cdots \sfn_4}}{\sqrt{6!\,4!}}}_{3^4_3}, \underbrace{\frac{\eta_\alpha^{\sfm_1\cdots \sfm_6,\,\sfn_1\cdots \sfn_6}}{\sqrt{6!\,6!}}}_{1^6_4/1^6_3},\cdots\!\biggr) \,,
\label{eq:eta-summary}
\end{align}
where the multiple indices are totally antisymmetrized and the ranges of the indices are $k,l=1,\dotsc,d$, $\sfm,\sfn=1,\dotsc,d-1$, and $\alpha,\beta=1,2$. 
Each $\eta$-symbol corresponds to a brane specified below the underbrace (see \cite{Obers:1998fb} and also \cite{Sakatani:2017nfr} for the notation of exotic branes $b^c_n$). 
The ellipses are relevant only for the $E_{d(d)}$ EFT with $d\geq 8$, which is not considered here. 

The above set of $\eta$-symbols would be essentially the same as the set of the $\eta$-symbols introduced in an ``F-theory'' \cite{Linch:2015lwa,Linch:2015fya,Linch:2015qva,Linch:2015fca,Linch:2016ipx}.
\footnote{The set of $\eta$-symbols was introduced in \cite{Coimbra:2011ky} as the projection $\times_N:\,E\times E\to N$, and in \cite{Malek:2017njj} as the wedge product $\wedge:\,R_1\otimes R_1\to R_{2}$. Note that the wedge product is defined for more general representations.}
There, the $\eta$-symbols were introduced as the Clebsch--Gordan--Wigner coefficients connecting $R_1\otimes R_1$ and the $R_2$-representation, and the Virasoro-like constraint was expressed as
\begin{align}
 \eta^{IJ;\,\sMA}\,\mathcal{P}_I\, \mathcal{P}_J = 0 \,. 
\end{align}
The generalized Lie derivative was obtained from the Virasoro-like constraint, and by comparing with the generalized Lie derivative, the $Y$-tensor in EFT was expressed as
\begin{align}
 Y^{IJ}_{KL} = \begin{cases} \eta^{IJ;\,\sMA} \, \eta_{KL;\,\sMA} & (d\leq 6)\\ \eta^{IJ;\,\sMA} \, \eta_{KL;\,\sMA} -\frac{1}{2}\,\Omega^{IJ}\,\Omega_{KL} & (d=7) \end{cases} \,,
\label{eq:Y-eta-Omega}
\end{align}
where the singlet constraint $\Omega^{IJ}\,\partial_I \otimes \partial_J=0$ was introduced for $d=7$ from consistency with the EFT. 
The explicit form of the $\eta$-symbol was found in \cite{Linch:2015fca} using a different convention from ours.

In this paper, instead of attempting to translate the $\eta$-symbols found in \cite{Linch:2015fca} into our convention, we utilize the linear map considered in \cite{Sakatani:2017nfr}. 
As has been well known \cite{Hohm:2013pua,Blair:2013gqa}, EFT can reproduce both M-theory and type IIB theory. 
Depending on which theory one has in mind, there are two natural parameterizations of the generalized coordinates: $x^I$ for M-theory and $x^{\sfM}$ for type IIB theory. 
The linear map in \cite{Sakatani:2017nfr} provides a relation between the two parameterizations:
\begin{align}
 x^I = S^I{}_{\sfN}\,x^{\sfN}\,,\qquad x^{\sfM} = (S^{-1})^{\sfM}{}_J\,x^J \,. 
\end{align}
When we consider the linear map, we decompose the physical coordinates $x^i$ ($i=1,\dotsc,d$) for M-theory and $\sfx^{\sfm}$ ($\sfm=1,\dotsc,d-1$) for type IIB theory as
\begin{align}
 (x^i)= (x^a,\,x^\alpha) \,,\qquad (\sfx^{\sfm}) = (\sfx^a,\,\sfx^\By) \qquad (a=1,\dotsc,d-2\,,\ \alpha =\Ay,\,\Az) \,. 
\end{align}
Here, $x^\Az$ in the M-theory side corresponds to the coordinate on the M-theory circle. 
If we adopt the type IIA picture (by compactifying the M-theory circle), the linear map corresponds to a single $T$-duality along the $x^\Ay$ or $\sfx^\By$ directions in type IIA/IIB theory. 
Indeed, in \cite{Sakatani:2017nfr}, it was shown that the linear map between two generalized metrics, $\cM_{IJ}$ (M-theory) and $\sfM_{\sfM\sfN}$ (type IIB theory),
\begin{align}
 \sfM_{\sfM\sfN} = S^I{}_{\sfM}\,S^J{}_{\sfN}\,\cM_{IJ} \,,
\end{align}
precisely reproduces the well-known $T$-duality transformation rules for supergravity fields. 
In this paper, we apply this linear map to $\eta$-symbols in M-theory/type IIB theory. 

To be more specific, following the convention used in \cite{Sakatani:2017nfr}, we parameterize the generalized coordinates as
\begin{align}
\begin{split}
 \text{M-theory:}\quad &(x^I) = \Biggl(\underbrace{x^i_{\vphantom{o}}}_{\mathrm{P}},\,\underbrace{\frac{y_{i_1i_2}}{\sqrt{2!}}}_{\mathrm{M}2},\,\underbrace{\frac{y_{i_1\cdots i_5}}{\sqrt{5!}}}_{\mathrm{M}5},\,\underbrace{\frac{y_{i_1\cdots i_7,\,j}}{\sqrt{7!}}}_{\mathrm{KKM}/8},\cdots\Biggr) \,,
\\
 \text{Type IIB:}\quad &(x^{\sfM}) = \Biggl(\underbrace{x^{\sfm}_{\vphantom{o}}}_{\mathrm{P}},\,\underbrace{\sfy^\alpha_{\sfm}}_{\mathrm{F}1/\mathrm{D}1},\,\underbrace{\frac{\sfy_{\sfm_1\sfm_2\sfm_3}}{\sqrt{3!}}}_{\mathrm{D}3},\,\underbrace{\frac{\sfy^\alpha_{\sfm_1\cdots \sfm_5}}{\sqrt{5!}}}_{\mathrm{NS}5/\mathrm{D}5},\,\underbrace{\frac{\sfy_{\sfm_1\cdots \sfm_6,\,\sfn}}{\sqrt{6!}}}_{\mathrm{KKM}/7_2},\cdots\Biggr) \,,
\end{split}
\label{eq:list-etas}
\end{align}
where the coordinates other than the physical coordinates are winding coordinates associated with some branes specified below the underbrace and ellipses again are relevant only for the $E_{d(d)}$ EFT with $d\geq 8$. 
In the above parameterized generalized coordinates $x^I$ for M-theory, we begin by considering two $\eta$-symbols,
\begin{align}
 \eta^k \equiv \begin{pmatrix}
 0 & \frac{2!\,\delta^{k i}_{j_1j_2}}{\sqrt{2!}} & 0 & 0 \\
 \frac{2!\,\delta^{k j}_{i_1i_2}}{\sqrt{2!}} & 0 & 0 & 0 \\
 0 & 0 & 0 & 0 \\
 0 & 0 & 0 & 0 
 \end{pmatrix} , \quad 
 \eta^{k_1\cdots k_4} \equiv \begin{pmatrix}
 0 & 0 & \frac{5!\,\delta^{i k_1\cdots k_4}_{j_1\cdots j_5}}{\sqrt{5!}} & 0 \\
 0 & \frac{4!\,\delta^{k_1\cdots k_4}_{i_1i_2j_1j_2}}{\sqrt{2!\,2!}} & 0 & 0 \\
 \frac{5!\,\delta^{j k_1\cdots k_4}_{i_1\cdots i_5}}{\sqrt{5!}} & 0 & 0 & 0 \\
 0 & 0 & 0 & 0
 \end{pmatrix} , 
\end{align}
which are trivial extensions of the $\eta$-symbols associated with M2-/M5-branes shown in \eqref{eq:E5-M2-M5-eta}. 
Under a compactification on the M-theory circle, an M2-brane becomes a D2-brane or an F-string in type IIA theory, and under a $T$-duality, it can become a D1/D3-brane or an F-string. 
Correspondingly, under the linear map, the $\eta$-symbol $(\eta^k)=(\eta^a,\eta^\alpha)$ can be mapped to an $\eta$-symbol, $\eta^{a\By}$ or $\eta_\alpha$, associated with a D3-brane or an F/D-string in type IIB theory. 
Similarly, the $\eta$-symbol $(\eta^{k_1\cdots k_4})=(\eta^{a_1\cdots a_4},\,\eta^{a_1a_2a_3\alpha},\,\eta^{a_1a_2\Ay\Az})$ can be mapped to an $\eta$-symbol, $\eta^{a_1\cdots a_4\By,\,\By}$, $\eta^{a_1\cdots a_3\By}_\alpha$, or $\eta^{a_1a_2}$, associated with a Kaluza--Klein monopole (KKM), an NS/D5-brane, or a D3-brane in type IIB theory. 
Repeating the linear map, we can find almost all of the $\eta$-symbols described in \eqref{eq:list-etas}. 
The only $\eta$-symbols that cannot straightforwardly be obtained from the linear map are $\eta^{[k_1\cdots k_6,\,l]}$ and $\eta^{[\sfm_1\cdots \sfm_5,\,\sfn]}$, which correspond to 8-branes in M-theory and $7_2$-branes in type IIB theory, respectively. 
These branes are not related to other branes described in \eqref{eq:list-etas} via $T$-duality transformations, but they are related to each other. 
In fact, by requiring the $\SL(d)$ or $\SL(d-1)$ covariance in the M-theory or type IIB theory sides, they also can be determined completely. 
Then, we find the explicit form of all $\eta$-symbols (and also the $\Omega$-tensor) and can construct the $Y$-tensor through \eqref{eq:Y-eta-Omega}. 

We expect that, in the same manner as \cite{Hatsuda:2012uk,Hatsuda:2012vm,Hatsuda:2013dya}, all of the $\eta$-symbols obtained in this paper will also be read off from the momentum constraint (i.e., Virasoro-like constraint) in worldvolume theories of branes appearing in \eqref{eq:eta-summary}, but we leave the task for future work, and here we will concentrate on the determination of the $\eta$-symbols for the $E_{d(d)}$ EFT ($d\leq 7$).

\section{Explicit form of $\eta$-symbols}
\label{sec:explicit-eta}

In this section, we begin by showing the explicit matrix form of $\eta$-symbols in two generalized coordinates, $x^I$ and $x^{\sfM}$, associated with M-theory and type IIB theory, respectively. 
Their derivations are explained in section \ref{sec:linear-map}. 
In Appendix \ref{app:eta-comparison}, we explain how to reproduce the known $Y$-tensors from our $\eta$-symbols. 

\subsection{M-theory parameterization}

When we adopt the M-theory description, the decomposition of $\eta^{\sMA}$ becomes
\begin{align}
 (\eta^{\sMA}) = \biggl(\eta^k,\, \frac{\eta^{k_1\cdots k_4}}{\sqrt{4!}},\, \frac{\eta^{k_1\cdots k_6,\,l}}{\sqrt{6!}},\, \frac{\eta^{k_1\cdots k_7,\,l_1l_2l_3}}{\sqrt{7!\,3!}},\, \frac{\eta^{k_1\cdots k_7,\,l_1\cdots l_6}}{\sqrt{7!\,6!}} \biggr)\,. 
\label{eq:etas-M}
\end{align}
The explicit forms of each matrix, $\eta^{\sMA}=(\eta^{IJ;\,\sMA})$, are
\begin{align}
 \eta^k 
 &\equiv \begin{pmatrix}
 0 & \frac{2!\,\delta^{k i}_{j_1j_2}}{\sqrt{2!}} & 0 & 0 \\
 \frac{2!\,\delta^{k j}_{i_1i_2}}{\sqrt{2!}} & 0 & 0 & 0 \\
 0 & 0 & 0 & 0 \\
 0 & 0 & 0 & 0 
 \end{pmatrix} , 
\\
 \eta^{k_1\cdots k_4} 
 &\equiv \begin{pmatrix}
 0 & 0 & \frac{5!\,\delta^{i k_1\cdots k_4}_{j_1\cdots j_5}}{\sqrt{5!}} & 0 \\
 0 & \frac{4!\,\delta^{k_1\cdots k_4}_{i_1i_2j_1j_2}}{\sqrt{2!\,2!}} & 0 & 0 \\
 \frac{5!\,\delta^{j k_1\cdots k_4}_{i_1\cdots i_5}}{\sqrt{5!}} & 0 & 0 & 0 \\
 0 & 0 & 0 & 0
 \end{pmatrix} , 
\\
 \eta^{k_1\cdots k_6,\,l} &\equiv \eta_{\text{KKM}}^{k_1\cdots k_6,\,l} + \eta^{k_1\cdots k_6l} \,, 
\\
 &{}{\tiny\hspace*{-20mm}{\eta_{\text{KKM}}^{k_1\cdots k_6,\,l}} 
 {\arraycolsep=-1.0mm \equiv \left(\begin{array}{cccc}
 0 & 0 & 0 & \frac{7! \Bigl(\delta^{k_1\cdots k_6 i}_{j_1\cdots j_7} \delta^l_j- \frac{\delta^{k_1\cdots k_6 l}_{j_1\cdots j_7} \delta^i_j}{7}\Bigr)}{\sqrt{7!}} \\
 0 & 0 & \frac{-6! 2! \Bigl(\delta^{k_1\cdots k_6}_{j_1\cdots j_5k}\delta^{kl}_{i_1i_2}- \delta^{k_1\cdots k_6 l}_{j_1\cdots j_5i_1i_2}\Bigr)}{\sqrt{2! 5!}} & 0 \\
 0 & \frac{-6! 2! \Bigl(\delta^{k_1\cdots k_6}_{i_1\cdots i_5k}\delta^{kl}_{j_1j_2}- \delta^{k_1\cdots k_6 l}_{i_1\cdots i_5j_1j_2}\Bigr)}{\sqrt{2! 5!}} & 0 & 0 \\
 \frac{7! \Bigl(\delta^{k_1\cdots k_6 j}_{i_1\cdots i_7} \delta^l_i- \frac{\delta^{k_1\cdots k_6 l}_{i_1\cdots i_7} \delta^j_i}{7}\Bigr)}{\sqrt{7!}} & 0 & 0 & 0
 \end{array}\right) ,}}
\\
 \eta^{k_1\cdots k_7} 
 &\equiv \frac{1}{7\sqrt{2}} \begin{pmatrix}
 0 & 0 & 0 & 3\,\frac{7!\,\delta^{k_1\cdots k_7}_{j_1\cdots j_7}\,\delta^i_j}{\sqrt{7!}} \\
 0 & 0 & \frac{7!\,\delta^{k_1\cdots k_7}_{j_1\cdots j_5i_1i_2}}{\sqrt{2!\,5!}} & 0 \\
 0 & \frac{7!\,\delta^{k_1\cdots k_7}_{i_1\cdots i_5j_1j_2}}{\sqrt{2!\,5!}} & 0 & 0 \\
 3\,\frac{7!\,\delta^{k_1\cdots k_7}_{i_1\cdots i_7}\,\delta^j_i}{\sqrt{7!}} & 0 & 0 & 0
\end{pmatrix} , 
\\
 \eta^{k_1\cdots k_7,\,l_1l_2l_3} 
 &\equiv \begin{pmatrix} 
 0 & 0 & 0 & 0 \\
 0 & 0 & 0 & \frac{-7!\,7!\,\delta^{l_1l_2l_3 m_1\cdots m_4}_{j_1\cdots j_7}\,\delta_{j i_1i_2 m_1\cdots m_4}^{k_1\cdots k_7}}{4!\sqrt{2!\,7!}} \\
 0 & 0 & \frac{7!\,5!\,\delta_{i_1\cdots i_5m_1m_2}^{k_1\cdots k_5k_6k_7}\,\delta^{m_1m_2 l_1l_2l_3}_{j_1\cdots j_5}}{2!\sqrt{5!\,5!}} & 0 \\
 0 & \frac{-7!\,7!\,\delta^{l_1l_2l_3 m_1\cdots m_4}_{i_1\cdots i_7}\,\delta_{i j_1j_2 m_1\cdots m_4}^{k_1\cdots k_7}}{4!\sqrt{2!\,7!}} & 0 & 0
 \end{pmatrix} , 
\\
 \eta^{k_1\cdots k_7,\,l_1\cdots l_6} 
 &\equiv \begin{pmatrix}
 0 & 0 & 0 & 0 \\
 0 & 0 & 0 & 0 \\
 0 & 0 & 0 & \frac{7!\,6!\,\delta^{k_1\cdots k_7}_{j_1\cdots j_7}\,\delta^{l_1\cdots l_6}_{j i_1\cdots i_5}}{\sqrt{5!\,7!}} \\
 0 & 0 & \frac{7!\,6!\,\delta^{k_1\cdots k_7}_{i_1\cdots i_7}\,\delta^{l_1\cdots l_6}_{i j_1\cdots i_5}}{\sqrt{5!\,7!}} & 0
 \end{pmatrix} .
\end{align}
Here, $\eta_{\text{KKM}}^{k_1\cdots k_6,\,l}$ is defined to satisfy $\eta_{\text{KKM}}^{[k_1\cdots k_6,\,l]}=0$. 
We also define the $\eta$-symbols $\eta_{\sMA} =(\eta_{IJ;\,\sMA})$ as 
\begin{align}
 \eta_{IJ;\,\sMA} = \eta^{IJ;\,\sMA} \,. 
\end{align}
For example, $\eta_k$ is defined as
\begin{align}
 \eta_k \equiv (\eta_{IJ;\,k})\equiv \begin{pmatrix}
 0 & \frac{2!\,\delta_{k i}^{j_1j_2}}{\sqrt{2!}} & 0 & 0 \\
 \frac{2!\,\delta_{k j}^{i_1i_2}}{\sqrt{2!}} & 0 & 0 & 0 \\
 0 & 0 & 0 & 0 \\
 0 & 0 & 0 & 0 
 \end{pmatrix} . 
\end{align}
The position of the indices is converted but $\eta_{\sMA}$ and $\eta^{\sMA}$ have the same components as a matrix. 

In the above expressions, we are supposing the case of the $E_{7(7)}$ EFT but the expressions for the $E_{d(d)}$ EFT with $d\leq 6$ can be obtained via a simple truncation of the above matrices. 
For example, in the $\SL(5)$ EFT (where $d=4$), the generalized coordinates are given by $x^I=(x^i,\,\frac{y_{i_1i_2}}{\sqrt{2!}})$ and the non-vanishing $\eta$-symbols become
\begin{align}
 \eta^k = \begin{pmatrix}
 0 & \frac{2!\,\delta^{k i}_{j_1j_2}}{\sqrt{2!}} \\
 \frac{2!\,\delta^{k j}_{i_1i_2}}{\sqrt{2!}} & 0 
 \end{pmatrix} , \qquad 
 \eta^{k_1\cdots k_4} = \begin{pmatrix}
 0 & 0 \\
 0 & \frac{4!\,\delta^{k_1\cdots k_4}_{i_1i_2j_1j_2}}{\sqrt{2!\,2!}} 
 \end{pmatrix} . 
\end{align}
The number of the generalized coordinates, the $\eta$-symbols, and the corresponding brane charges for $d\leq 7$ can be summarized as follows:
\begin{alignat}{3}
 d=2:\quad &(\underset{[3]\vphantom{\Big|}}{x^I}) = (\underbrace{x^i_{\vphantom{o}}}_{\mathrm{P}\,[2]},\,\underbrace{y_{i_1i_2}}_{\mathrm{M}2\,[1]}) \,,&&(\underset{[2]\vphantom{\Big|}}{\eta^{\sMA}}) = \bigl(\underbrace{\eta^k_{\vphantom{o}}}_{\mathrm{M2}\,[2]} \bigr)\,,
\\
 d=3:\quad &(\underset{[6]\vphantom{\Big|}}{x^I}) = (\underbrace{x^i_{\vphantom{o}}}_{\mathrm{P}\,[3]},\,\underbrace{y_{i_1i_2}}_{\mathrm{M}2\,[3]}) \,,&&(\underset{[3]\vphantom{\Big|}}{\eta^{\sMA}}) = \bigl(\underbrace{\eta^k_{\vphantom{o}}}_{\mathrm{M2}\,[3]} \bigr)\,,
\\
 d=4:\quad &(\underset{[10]\vphantom{\Big|}}{x^I}) = (\underbrace{x^i_{\vphantom{o}}}_{\mathrm{P}\,[4]},\,\underbrace{y_{i_1i_2}}_{\mathrm{M}2\,[6]}) \,,&&(\underset{[5]\vphantom{\Big|}}{\eta^{\sMA}}) = \bigl(\underbrace{\eta^k_{\vphantom{o}}}_{\mathrm{M2}\,[4]},\, \underbrace{\eta^{k_1\cdots k_4}}_{\mathrm{M5}\,[1]}\bigr)\,,
\\
 d=5:\quad &(\underset{[16]\vphantom{\Big|}}{x^I}) = (\underbrace{x^i_{\vphantom{o}}}_{\mathrm{P}\,[5]},\,\underbrace{y_{i_1i_2}}_{\mathrm{M}2\,[10]},\,\underbrace{y_{i_1\cdots i_5}}_{\mathrm{M}5\,[1]}) \,, &&(\underset{[10]\vphantom{\Big|}}{\eta^{\sMA}}) = \bigl(\underbrace{\eta^k_{\vphantom{o}}}_{\mathrm{M2}\,[5]},\, \underbrace{\eta^{k_1\cdots k_4}}_{\mathrm{M5}\,[5]} \bigr)\,,
\\
 d=6:\quad &(\underset{[27]\vphantom{\Big|}}{x^I}) = (\underbrace{x^i_{\vphantom{o}}}_{\mathrm{P}\,[6]},\,\underbrace{y_{i_1i_2}}_{\mathrm{M}2\,[15]},\,\underbrace{y_{i_1\cdots i_5}}_{\mathrm{M}5\,[6]}) \,,&&
 (\underset{[27]\vphantom{\Big|}}{\eta^{\sMA}}) = \bigl(\underbrace{\eta^k_{\vphantom{o}}}_{\mathrm{M2}\,[6]},\, \underbrace{\eta^{k_1\cdots k_4}}_{\mathrm{M5}\,[15]} ,\, \underbrace{\eta^{k_1\cdots k_6,\,l}}_{\mathrm{KKM}\,[6]} \bigr)\,,
\\
 d=7:\quad &(\underset{[56]\vphantom{\Big|}}{x^I}) = (\underbrace{x^i_{\vphantom{o}}}_{\mathrm{P}\,[7]},\,\underbrace{y_{i_1i_2}}_{\mathrm{M}2\,[21]},\,\underbrace{y_{i_1\cdots i_5}}_{\mathrm{M}5\,[21]},\, &&
 (\underset{[133]\vphantom{\Big|}}{\eta^{\sMA}}) = (\underbrace{\eta^k}_{\mathrm{M}2\, [7]},\,\underbrace{\eta^{k_1\cdots k_4}}_{\mathrm{M}5\, [35]},\,\underbrace{\eta^{k_1\cdots k_6,\,l}}_{\mathrm{KKM}/8\, [49]},\,
\nn\\
 & \qquad\qquad\qquad\qquad\quad \underbrace{y_{i_1\cdots i_7,\,j}}_{\mathrm{KKM}\,[7]})\,,\qquad && \qquad\qquad \underbrace{\eta^{k_1\cdots k_7,\,l_1l_2l_3}}_{5^3\, [35]},\,\underbrace{\eta^{k_1\cdots k_7,\,l_1\cdots l_6}}_{2^6\, [7]}) \,,
\end{alignat}
where the normalization coefficients like $(1/\sqrt{p!})$ are not displayed for simplicity.

In the case of $d=7$, in addition to the $\eta$-symbols, we also define antisymmetric matrices $\Omega^{IJ}$ and $\Omega_{IJ}$ appearing in \eqref{eq:Y-eta-Omega}. 
As we explain in section \ref{sec:linear-map}, their matrix forms, in our convention, are
\begin{align}
 (\Omega_{IJ}) &\equiv \begin{pmatrix}
 0 & 0 & 0 & \frac{\epsilon^{j_1\cdots j_7}\,\delta_i^j}{\sqrt{7!}} \\
 0 & 0 & \frac{\epsilon^{i_1i_2j_1\cdots j_5}}{\sqrt{2!\,5!}} & 0 \\
 0 & -\frac{\epsilon^{i_1\cdots i_5j_1j_2}}{\sqrt{2!\,5!}} & 0 & 0 \\
 -\frac{\epsilon^{i_1\cdots i_7}\,\delta_j^i}{\sqrt{7!}} & 0 & 0 & 0
 \end{pmatrix} ,
\\
 (\Omega^{IJ}) &\equiv \begin{pmatrix}
 0 & 0 & 0 & \frac{\epsilon_{j_1\cdots j_7}\,\delta^i_j}{\sqrt{7!}} \\
 0 & 0 & \frac{\epsilon_{i_1i_2j_1\cdots j_5}}{\sqrt{2!\,5!}} & 0 \\
 0 & -\frac{\epsilon_{i_1\cdots i_5j_1j_2}}{\sqrt{2!\,5!}} & 0 & 0 \\
 -\frac{\epsilon_{i_1\cdots i_7}\,\delta^j_i}{\sqrt{7!}} & 0 & 0 & 0
 \end{pmatrix} ,
\end{align}
where the totally antisymmetric symbols $\epsilon_{i_1\cdots i_7}$ and $\epsilon^{i_1\cdots i_7}$ are defined as $\epsilon^{1\cdots 7}=\epsilon_{1\cdots 7}=1$\,.

From the above $\eta$-symbols and the $\Omega$-tensor, we can obtain the $Y$-tensor as
\begin{align}
 Y^{IJ}_{KL} &= \eta^{IJ;\,\sMA}\,\eta_{KL;\,\sMA} - \frac{1}{2}\,\Omega^{IJ}\,\Omega_{KL}
\nn\\
 &= \eta^{IJ;\,k}\,\eta_{KL;\,k} 
 + \frac{\eta^{IJ;\,k_1\cdots k_4}\,\eta_{KL;\,k_1\cdots k_4}}{4!} 
 + \frac{\eta^{IJ;\,k_1\cdots k_6,\,l}\,\eta_{KL;\,k_1\cdots k_6,\,l}}{6!} 
\nn\\
 &+ \frac{\eta^{IJ;\,k_1\cdots k_7,\,l_1l_2l_3}\,\eta_{KL;\,k_1\cdots k_7,\,l_1l_2l_3}}{7!\,3!} 
 + \frac{\eta^{IJ;\,k_1\cdots k_7,\,l_1\cdots l_6}\,\eta_{KL;\,k_1\cdots k_7,\,l_1\cdots l_6}}{7!\,6!} - \frac{1}{2}\,\Omega^{IJ}\,\Omega_{KL} \,.
\end{align}

\subsection{Type IIB parameterization}

When we adopt the type IIB description, we consider the following decomposition of $\eta$-symbols:
\begin{align}
 (\eta^{\sBA}) = \biggr(\eta_\gamma,\,
 \frac{\eta^{\sfp_1\sfp_2}}{\sqrt{2!}},\,
 \frac{\eta^{\sfp_1\cdots\sfp_4}_\gamma}{\sqrt{4!}} ,\,
 \frac{\eta^{\sfp_1\cdots\sfp_5,\,\sfq}}{\sqrt{5!}},\,
 \frac{\eta^{\sfp_1\cdots\sfp_6}_{(\gamma_1\gamma_2)}}{\sqrt{6!}},\,
 \frac{\eta^{\sfp_1\cdots\sfp_6,\,\sfq_1\sfq_2}_{\gamma}}{\sqrt{6!\,2!}},\,
 \frac{\eta^{\sfp_1\cdots\sfp_6,\,\sfq_1\cdots\sfq_4}}{\sqrt{6!\,4!}},\,
 \frac{\eta^{\sfp_1\cdots\sfp_6,\,\sfq_1\cdots\sfq_6}_\gamma}{\sqrt{6!\,6!}} \biggr)\,,
\label{eq:etas-IIB}
\end{align}
where the matrices take the form
\begin{align}
 \eta_\gamma&\equiv \begin{pmatrix} 0 & \delta^\beta_\gamma\,\delta^{\sfm}_{\sfn} & 0 & 0 & 0 \\
 \delta^\alpha_\gamma\,\delta_{\sfm}^{\sfn} & 0 & 0 & 0 & 0 \\
 0 & 0 & 0 & 0 & 0 \\
 0 & 0 & 0 & 0 & 0 \\
 0 & 0 & 0 & 0 & 0
 \end{pmatrix},
\\
 \eta^{\sfp_1\sfp_2}&\equiv \begin{pmatrix}
 0 & 0 &\frac{3!\,\delta^{\sfm \sfp_1\sfp_2}_{\sfn_1\sfn_2\sfn_3}}{\sqrt{3!}} & 0 & 0 \\
 0 & -2!\,\epsilon^{\alpha\beta}\,\delta^{\sfp_1\sfp_2}_{\sfm\sfn} & 0 & 0 & 0 \\
 \frac{3!\,\delta^{\sfn\sfp_1\sfp_2}_{\sfm_1\sfm_2\sfm_3}}{\sqrt{3!}} & 0 & 0 & 0 & 0 \\
 0 & 0 & 0 & 0 & 0 \\
 0 & 0 & 0 & 0 & 0
 \end{pmatrix},
\\
 \eta^{\sfp_1\cdots\sfp_4}_\gamma 
 &\equiv \begin{pmatrix}
 0 & 0 & 0 & \frac{-5!\,\delta^{\beta}_{\gamma}\,\delta^{\sfp_1\cdots \sfp_4\sfm}_{\sfn_1\cdots \sfn_5}}{\sqrt{5!}} & 0 \\
 0 & 0 & \frac{4!\,\delta^{\alpha}_{\gamma}\,\delta^{\sfp_1\cdots \sfp_4}_{\sfn_1\sfn_2\sfn_3\sfm}}{\sqrt{3!}} & 0 & 0 \\
 0 &\frac{4!\,\delta^{\beta}_{\gamma}\,\delta^{\sfp_1\cdots \sfp_4}_{\sfm_1\sfm_2\sfm_3\sfn}}{\sqrt{3!}} & 0 & 0 & 0 \\
 \frac{-5!\,\delta^{\alpha}_{\gamma}\,\delta^{\sfp_1\cdots \sfp_4\sfn}_{\sfm_1\cdots \sfm_5}}{\sqrt{5!}} & 0 & 0 & 0 & 0 \\
 0 & 0 & 0 & 0 & 0
 \end{pmatrix} ,
\\
 \eta^{\sfp_1\cdots\sfp_5,\,\sfq} &\equiv \eta_{\text{KKM}}^{\sfp_1\cdots\sfp_5,\,\sfq} + \eta^{\sfp_1\cdots\sfp_5\sfq}\,,
\\
 &{}{\tiny\hspace*{-35mm}{\eta_{\text{KKM}}^{\sfp_1\cdots\sfp_5,\,\sfq}} 
 {\arraycolsep=-3mm \equiv 
 \left(\begin{array}{ccccc}
 0 & 0 & 0 & 0 &\frac{6!\bigl(\delta^{\sfm\sfp_1\cdots\sfp_5}_{\sfn_1\cdots\sfn_6} \delta^{\sfq}_{\sfn}
        + \frac{\delta^{\sfp_1\cdots\sfp_5\sfq}_{\sfn_1\cdots\sfn_6} \delta^{\sfm}_{\sfn}}{6}\bigr)}{\sqrt{6!}}~~\\
 0 & 0 & 0 &\frac{5! \epsilon^{\beta\alpha} \bigl(\delta^{\sfp_1\cdots\sfp_5}_{\sfn_1\cdots\sfn_5} \delta^{\sfq}_{\sfm}
     - \delta^{\sfp_1\cdots\sfp_5\sfq}_{\sfn_1\cdots\sfn_5\sfm}\bigr)}{\sqrt{5!}} & 0 \\
 0 & 0 &\frac{5! 3! \bigl(\delta^{\sfp_1\sfp_2\sfp_3\sfp_4\sfp_5}_{\sfm_1\sfm_2\sfm_3\sfr_1\sfr_2} \delta^{\sfr_1\sfr_2\sfq}_{\sfn_1\sfn_2\sfn_3}
    +\delta^{\sfp_1\sfp_2\sfp_3\sfp_4\sfp_5}_{\sfn_1\sfn_2\sfn_3\sfr_1\sfr_2} \delta^{\sfr_1\sfr_2\sfq}_{\sfm_1\sfm_2\sfm_3}\bigr)}{2\cdot 2!\sqrt{3! 3!}}& 0 & 0 \\
 0 &\frac{5!\epsilon^{\alpha\beta} \bigl(\delta^{\sfp_1\cdots\sfp_5}_{\sfm_1\cdots\sfm_5} \delta^{\sfq}_{\sfn}
     - \delta^{\sfp_1\cdots\sfp_5\sfq}_{\sfm_1\cdots\sfm_5\sfn} \bigr)}{\sqrt{5!}}& 0 & 0 & 0 \\
 ~~\frac{6! \bigl(\delta^{\sfn\sfp_1\cdots\sfp_5}_{\sfm_1\cdots\sfm_6} \delta^{\sfq}_{\sfm}
     + \frac{\delta^{\sfp_1\cdots\sfp_5\sfq}_{\sfm_1\cdots\sfm_6} \delta^{\sfn}_{\sfm}}{6}\bigr)}{\sqrt{6!}} & 0 & 0 & 0 & 0 
 \end{array}\right),}}
\\
 \eta^{\sfp_1\cdots\sfp_6}
 &\equiv \begin{pmatrix}
 0 & 0 & 0 & 0 &\frac{6!\,\delta^{\sfp_1\cdots \sfp_6}_{\sfn_1\cdots \sfn_6}\, \delta^{\sfm}_{\sfn}}{3\sqrt{6!}} \\
 0 & 0 & 0 &\frac{6!\,\epsilon^{\beta\alpha}\,\delta^{\sfp_1\cdots \sfp_6}_{\sfn_1\cdots \sfn_5\sfm}}{6\sqrt{5!}} & 0 \\
 0 & 0 & 0 & 0 & 0 \\
 0 & \frac{6!\,\epsilon^{\alpha\beta}\,\delta^{\sfp_1\cdots \sfp_6}_{\sfm_1\cdots \sfm_5\sfn}}{6\sqrt{5!}} & 0 & 0 & 0 \\
 \frac{6!\delta^{\sfp_1\cdots \sfp_6}_{\sfm_1\cdots \sfm_6}\,\delta^{\sfn}_{\sfm}}{3\sqrt{6!}} & 0 & 0 & 0 & 0
 \end{pmatrix},
\\
 \eta^{\sfp_1\cdots\sfp_6}_{(\gamma_1\gamma_2)}
 &\equiv \begin{pmatrix}
 0 & 0 & 0 & 0 & 0 \\
 0 & 0 & 0 &\frac{-6!\,\delta^{\alpha}_{(\gamma_1}\delta^{\beta}_{\gamma_2)}\,\delta^{\sfp_1\cdots \sfp_6}_{\sfn_1\cdots \sfn_5\sfm}}{\sqrt{5!}} & 0 \\
 0 & 0 & 0 & 0 & 0 \\
 0 &\frac{-6!\,\delta^{\alpha}_{(\gamma_1}\delta^{\beta}_{\gamma_2)}\,\delta^{\sfp_1\cdots\sfp_6}_{\sfm_1\cdots\sfm_5\sfn}}{\sqrt{5!}} & 0 & 0 & 0 \\
 0 & 0 & 0 & 0 & 0 
 \end{pmatrix},
\\
 \eta^{\sfp_1\cdots\sfp_6,\,\sfq_1\sfq_2}_{\gamma}
 &\equiv \begin{pmatrix}
 0 & 0 & 0 & 0 & 0 \\
 0 & 0 & 0 & 0 &\frac{-6!\,2!\,\delta^{\alpha}_{\gamma}\,\delta^{\sfp_1\cdots\sfp_6}_{\sfn_1\cdots\sfn_6}\,\delta^{\sfq_1\sfq_2}_{\sfm\sfn}}{\sqrt{6!}} \\
 0 & 0 & 0 &\frac{6!\,3!\,\delta^{\beta}_{\gamma}\,\delta^{\sfp_1\cdots\sfp_6}_{\sfn_1\cdots\sfn_5\sfr}\,\delta^{\sfr\sfq_1\sfq_2}_{\sfm_1\sfm_2\sfm_3}}{\sqrt{3!\,5!}}& 0 \\
 0 & 0 &\frac{6!\,3!\,\delta^{\alpha}_{\gamma}\,\delta^{\sfp_1\cdots\sfp_6}_{\sfm_1\cdots\sfm_5\sfr}\,\delta^{\sfr\sfq_1\sfq_2}_{\sfn_1\sfn_2\sfn_3}}{\sqrt{3!\,5!}}& 0 & 0 \\
 0 &\frac{-6!\,2!\,\delta^{\beta}_{\gamma}\,\delta^{\sfp_1\cdots\sfp_6}_{\sfm_1\cdots\sfm_6}\,\delta^{\sfq_1\sfq_2}_{\sfn\sfm}}{\sqrt{6!}}& 0 & 0 & 0
\end{pmatrix} ,
\\
 \eta^{\sfp_1\cdots\sfp_6,\,\sfq_1\cdots\sfq_4}
 &\equiv \begin{pmatrix}
 0 & 0 & 0 & 0 & 0 \\
 0 & 0 & 0 & 0 & 0 \\
 0 & 0 & 0 & 0 & \frac{6!\,4!\,\delta^{\sfp_1\cdots\sfp_6}_{\sfn_1\cdots\sfn_6}\,\delta^{\sfq_1\cdots\sfq_4}_{\sfn\sfm_1\sfm_2\sfm_3}}{\sqrt{6!\,3!}} \\
 0 & 0 & 0 &\frac{6!\,5!\,\epsilon^{\alpha\beta}\,\delta^{\sfp_1\cdots\sfp_6}_{\sfm_1\cdots\sfm_5\sfr}\,\delta^{\sfq_1\cdots\sfq_4\sfr}_{\sfn_1\cdots\sfn_5}}{\sqrt{5!\,5!}} & 0 \\
 0 & 0 &\frac{6!\,4!\,\delta^{\sfp_1\cdots\sfp_6}_{\sfm_1\cdots\sfm_6}\,\delta^{\sfq_1\cdots\sfq_4}_{\sfm\sfn_1\sfn_2\sfn_3}}{\sqrt{3!\,6!}}& 0 & 0 
\end{pmatrix} ,
\\
 \eta^{\sfp_1\cdots\sfp_6,\,\sfq_1\cdots\sfq_6}_\gamma
 &\equiv \begin{pmatrix}
 0 & 0 & 0 & 0 & 0 \\
 0 & 0 & 0 & 0 & 0 \\
 0 & 0 & 0 & 0 & 0 \\
 0 & 0 & 0 & 0 & \frac{6!\,6!\,\delta^{\alpha}_{\gamma}\,\delta^{\sfp_1\cdots\sfp_6}_{\sfn_1\cdots \sfn_6}\,\delta^{\sfq_1\cdots\sfq_6}_{\sfm_1\cdots\sfm_5\sfn}}{\sqrt{5!\,6!}}\\
 0 & 0 & 0 &\frac{6!\,6!\,\delta^{\beta}_{\gamma}\,\delta^{\sfp_1\cdots\sfp_6}_{\sfm_1\cdots\sfm_6}\,\delta^{\sfq_1\cdots\sfq_6}_{\sfn_1\cdots\sfn_5\sfm}}{\sqrt{5!\,6!}}& 0 
\end{pmatrix} .
\end{align}
We also define the $\eta$-symbols $\eta_{\sBA}=(\eta_{\sfM\sfN;\,\sBA})$ as
\begin{align}
 \eta_{\sfM\sfN;\,\sBA} = \eta^{\sfM\sfN;\,\sBA} \,.
\end{align}
A list of non-vanishing coordinates and $\eta$-symbols for each $d$ is
\begin{alignat}{3}
 d=2:\ &(\underset{[3]\vphantom{\Big|}}{x^{\sfM}}) = (\underbrace{\sfx^{\sfm}_{\vphantom{o}}}_{\mathrm{P}\,[1]},\,\underbrace{\sfy^\alpha_{\sfm}}_{\mathrm{F}1/\mathrm{D}1\,[2]}) &&
 (\underset{[2]\vphantom{\Big|}}{\eta^{\sBA}})
 = (\!\!\!\underbrace{\eta_\gamma}_{\mathrm{F}1/\mathrm{D}1\, [2]})\,,
\\
 d=3:\ &(\underset{[6]\vphantom{\Big|}}{x^{\sfM}}) = (\underbrace{\sfx^{\sfm}_{\vphantom{o}}}_{\mathrm{P}\,[2]},\,\underbrace{\sfy^\alpha_{\sfm}}_{\mathrm{F}1/\mathrm{D}1\,[4]}) &&
 (\underset{[3]\vphantom{\Big|}}{\eta^{\sBA}})
 = (\!\!\!\underbrace{\eta_\gamma}_{\mathrm{F}1/\mathrm{D}1\, [2]}\!\!,\,\underbrace{\eta^{\sfp_1\sfp_2}}_{\mathrm{D}3\, [1]})\,,
\\
 d=4:\ &(\underset{[10]\vphantom{\Big|}}{x^{\sfM}}) = (\underbrace{\sfx^{\sfm}_{\vphantom{o}}}_{\mathrm{P}\,[3]},\,\underbrace{\sfy^\alpha_{\sfm}}_{\mathrm{F}1/\mathrm{D}1\,[6]},\underbrace{\sfy_{\sfm_1\sfm_2\sfm_3}}_{\mathrm{D}3\,[1]}) &&
 (\underset{[5]\vphantom{\Big|}}{\eta^{\sBA}})
 = (\!\!\!\underbrace{\eta_\gamma}_{\mathrm{F}1/\mathrm{D}1\, [2]}\!\!,\,\underbrace{\eta^{\sfp_1\sfp_2}}_{\mathrm{D}3\, [3]})\,,
\\
 d=5:\ &(\underset{[16]\vphantom{\Big|}}{x^{\sfM}}) = (\underbrace{\sfx^{\sfm}_{\vphantom{o}}}_{\mathrm{P}\,[4]},\,\underbrace{\sfy^\alpha_{\sfm}}_{\mathrm{F}1/\mathrm{D}1\,[8]},\underbrace{\sfy_{\sfm_1\sfm_2\sfm_3}}_{\mathrm{D}3\,[4]}) &&
 (\underset{[10]\vphantom{\Big|}}{\eta^{\sBA}})
 = (\!\!\!\underbrace{\eta_\gamma}_{\mathrm{F}1/\mathrm{D}1\, [2]}\!\!,\,\underbrace{\eta^{\sfp_1\sfp_2}}_{\mathrm{D}3\, [6]},\,\underbrace{\eta_\gamma^{\sfp_1\cdots \sfp_4}}_{\mathrm{NS}5/\mathrm{D}5\, [2]})\,,
\\
 d=6:\ &(\underset{[27]\vphantom{\Big|}}{x^{\sfM}}) = (\underbrace{\sfx^{\sfm}_{\vphantom{o}}}_{\mathrm{P}\,[5]},\,\underbrace{\sfy^\alpha_{\sfm}}_{\mathrm{F}1/\mathrm{D}1\,[10]}, &&
 (\underset{[27]\vphantom{\Big|}}{\eta^{\sBA}})
 = (\!\!\!\underbrace{\eta_\gamma}_{\mathrm{F}1/\mathrm{D}1\, [2]}\!\!,\,\underbrace{\eta^{\sfp_1\sfp_2}}_{\mathrm{D}3\, [10]},\,\underbrace{\eta_\gamma^{\sfp_1\cdots \sfp_4}}_{\mathrm{NS}5/\mathrm{D}5\, [10]},\,\underbrace{\eta^{\sfp_1\cdots \sfp_5,\,\sfq}}_{\mathrm{KKM}\, [5]})\,,
\nn\\
 & \qquad\qquad \underbrace{\sfy_{\sfm_1\sfm_2\sfm_3}}_{\mathrm{D}3\,[10]},\,\underbrace{\sfy^\alpha_{\sfm_1\cdots \sfm_5}}_{\mathrm{NS}5/\mathrm{D}5\,[2]})\,,\quad&& 
\\
 d=7:\ &(\underset{[56]\vphantom{\Big|}}{x^{\sfM}}) = (\underbrace{\sfx^{\sfm}_{\vphantom{o}}}_{\mathrm{P}\,[6]},\,\underbrace{\sfy^\alpha_{\sfm}}_{\mathrm{F}1/\mathrm{D}1\,[12]},\,\underbrace{\sfy_{\sfm_1\sfm_2\sfm_3}}_{\mathrm{D}3\,[20]}, &&
 (\underset{[133]\vphantom{\Big|}}{\eta^{\sBA}})
 = (\!\!\!\underbrace{\eta_\gamma}_{\mathrm{F}1/\mathrm{D}1\, [2]}\!\!,\,\underbrace{\eta^{\sfp_1\sfp_2}}_{\mathrm{D}3\, [15]},\,\underbrace{\eta_\gamma^{\sfp_1\cdots \sfp_4}}_{\mathrm{NS}5/\mathrm{D}5\, [30]},\,\underbrace{\eta^{\sfp_1\cdots \sfp_5,\,\sfq}}_{\mathrm{KKM}/7_2\, [36]},\, \underbrace{\eta_{(\gamma_1\gamma_2)}^{\sfp_1\cdots \sfp_6}}_{\mathrm{Q}7\, [3]},
\nn\\
 & \qquad\qquad\quad \underbrace{\sfy^\alpha_{\sfm_1\cdots \sfm_5}}_{\mathrm{NS}5/\mathrm{D}5\,[12]} ,\,\underbrace{\sfy_{\sfm_1\cdots \sfm_6,\,\sfn}}_{\mathrm{KKM}\,[6]})\,,\quad&& \qquad\qquad
\underbrace{\eta_\gamma^{\sfp_1\cdots \sfp_6,\,\sfq_1\sfq_2}}_{5^2_2/5^2_3\, [30]},\, \underbrace{\eta^{\sfp_1\cdots \sfp_6,\,\sfq_1\cdots \sfq_4}}_{3^4_3\, [15]},\, \underbrace{\eta_\gamma^{\sfp_1\cdots \sfp_6,\,\sfq_1\cdots \sfq_6}}_{1^6_4/1^6_3\, [2]}) \,. 
\end{alignat}
Here again, the normalization coefficients like $(1/\sqrt{p!})$ are not displayed. 
On the other hand, the matrices $\Omega^{\sfM\sfN}$ and $\Omega_{\sfM\sfN}$ in the type IIB side have the form
\begin{align}
 (\Omega^{\sfM\sfN}) &\equiv \begin{pmatrix}
 0 & 0 & 0 & 0 & \frac{\epsilon_{\sfn_1\cdots \sfn_6}\,\delta^{\sfm}_{\sfn}}{\sqrt{6!}} \\
 0 & 0 & 0 & \frac{\epsilon^{\alpha\beta}\,\epsilon_{\sfm\sfn_1\cdots \sfn_5}}{\sqrt{5!}} & 0 \\
 0 & 0 & -\frac{\epsilon_{\sfm_1\sfm_2\sfm_3\sfn_1\sfn_2\sfn_3}}{\sqrt{3!\,3!}} & 0 & 0 \\
 0 & -\frac{\epsilon^{\alpha\beta}\,\epsilon_{\sfm_1\cdots \sfm_5\sfn}}{\sqrt{5!}} & 0 & 0 & 0 \\
 -\frac{\epsilon_{\sfm_1\cdots \sfm_6}\,\delta^{\sfn}_{\sfm}}{\sqrt{6!}} & 0 & 0 & 0 & 0 
\end{pmatrix} ,
\label{eq:Omega-IIB}
\\
 (\Omega_{\sfM\sfN}) &\equiv \begin{pmatrix}
 0 & 0 & 0 & 0 & \frac{\epsilon^{\sfn_1\cdots \sfn_6}\,\delta_{\sfm}^{\sfn}}{\sqrt{6!}} \\
 0 & 0 & 0 & \frac{\epsilon_{\alpha\beta}\,\epsilon^{\sfm\sfn_1\cdots \sfn_5}}{\sqrt{5!}} & 0 \\
 0 & 0 & -\frac{\epsilon^{\sfm_1\sfm_2\sfm_3\sfn_1\sfn_2\sfn_3}}{\sqrt{3!\,3!}} & 0 & 0 \\
 0 & -\frac{\epsilon_{\alpha\beta}\,\epsilon^{\sfm_1\cdots \sfm_5\sfn}}{\sqrt{5!}} & 0 & 0 & 0 \\
 -\frac{\epsilon^{\sfm_1\cdots \sfm_6}\,\delta_{\sfn}^{\sfm}}{\sqrt{6!}} & 0 & 0 & 0 & 0 
\end{pmatrix} ,
\end{align}
where the totally antisymmetric symbols $\epsilon_{\sfn_1\cdots \sfn_6}$ and $\epsilon^{\sfn_1\cdots \sfn_6}$ are defined as $\epsilon^{1\cdots 6}=\epsilon_{1\cdots 6}=1$\,. 

\subsection{The linear map}
\label{sec:linear-map}

We utilize the following linear map between generalized coordinates $x^I$ (M-theory) and $x^{\sfM}$ (type IIB) \cite{Sakatani:2017nfr}:
\begin{align}
 (x^{\sfM})&=
 \begin{pmatrix}
 \sfx^a \\[-2mm] \sfx^{\By} \\ \hline \sfy^\alpha_a \\[-2mm] \sfy^\alpha_{\By} \\ \hline \frac{\sfy_{a_1a_2a_3}}{\sqrt{3!}} \\[-2mm] \frac{\sfy_{a_1a_2 \By}}{\sqrt{2!}} \\ \hline \frac{\sfy^\alpha_{a_1\cdots a_5}}{\sqrt{5!}} \\[-2mm] \frac{\sfy^\alpha_{a_1\cdots a_4\By}}{\sqrt{4!}} \\ \hline \frac{\sfy_{a_1\cdots a_5\By,a}}{\sqrt{5!}} \\[-2mm] \frac{\sfy_{a_1\cdots a_5\By,\,\By}}{\sqrt{5!}} 
 \end{pmatrix} 
 = (S^{-1})^{\sfM}{}_J
 \begin{pmatrix}
 x^b \\[-2mm] x^\beta \\ \hline \frac{y_{b_1b_2}}{\sqrt{2!}} \\[-2mm] y_{b \beta} \\[-2mm] y_{\Ay\Az} \\ \hline \frac{y_{b_1\cdots b_5}}{\sqrt{5!}} \\[-2mm] \frac{y_{b_1\cdots b_4\beta}}{\sqrt{4!}} \\[-2mm] \frac{y_{b_1b_2b_3\Ay\Az}}{\sqrt{3!}} \\ \hline \frac{y_{b_1\cdots b_5\Ay\Az,\,b}}{\sqrt{5!}} \\[-2mm] \frac{y_{b_1\cdots b_5\Ay\Az,\,\beta}}{\sqrt{5!}} 
 \end{pmatrix} = (S^{-1})^{\sfM}{}_J\,x^J \,,
\\
 (S^{-1})^{\sfM}{}_J &\equiv {\footnotesize {\arraycolsep=0.7mm \left(\begin{array}{cc|ccc|ccc|cc}
 \delta^a_b & 0 & 0 & 0 & 0 & 0 & 0 & 0 & 0 & 0 \\
 0 & 0 & 0 & 0 & ~1~ & 0 & 0 & 0 & 0 & 0 \\ \hline
 0 & 0 & 0 & \epsilon^{\alpha\beta}\,\delta_a^b & 0 & 0 & 0 & 0 & 0 & 0 \\
 0 & \delta^\alpha_\beta & 0 & 0 & 0 & 0 & 0 & 0 & 0 & 0 \\ \hline
 0 & 0 & 0 & 0 & 0 & 0 & 0 & \delta_{a_1a_2a_3}^{b_1b_2b_3} & 0 & 0 \\
 0 & 0 & \delta_{a_1a_2}^{b_1b_2} & 0 & 0 & 0 & 0 & 0 & 0 & 0 \\ \hline
 0 & 0 & 0 & 0 & 0 & 0 & 0 & 0 & 0 & \epsilon^{\alpha\beta}\,\delta^{b_1\cdots b_5}_{a_1\cdots a_5} \\
 0 & 0 & 0 & 0 & 0 & 0 & \epsilon^{\alpha\beta}\,\delta_{a_1\cdots a_4}^{b_1\cdots b_4} & 0 & 0 & 0 \\ \hline
 0 & 0 & 0 & 0 & 0 & 0 & 0 & 0 & \delta_{a_1\cdots a_5}^{b_1\cdots b_5}\,\delta_a^b & 0 \\
 0 & 0 & 0 & 0 & 0 & \delta_{a_1\cdots a_5}^{b_1\cdots b_5} & 0 & 0 & 0 & 0 
 \end{array}\right)}} ,
\\
 S^I{}_{\sfN} &\equiv {\footnotesize {\arraycolsep=0.7mm \left(\begin{array}{cc|cc|cc|cc|cc}
 \delta^a_b & 0 & 0 & 0 & 0 & 0 & 0 & 0 & 0 & 0 \\
 0 & 0 & 0 & \delta^\alpha_\beta & 0 & 0 & 0 & 0 & 0 & 0 \\ \hline
 0 & 0 & 0 & 0 & 0 & \delta_{a_1a_2}^{b_1b_2} & 0 & 0 & 0 & 0 \\
 0 & 0 & \epsilon^\rmT_{\alpha\beta}\,\delta_a^b & 0 & 0 & 0 & 0 & 0 & 0 & 0 \\
 0 & ~1~ & 0 & 0 & 0 & 0 & 0 & 0 & 0 & 0 \\ \hline
 0 & 0 & 0 & 0 & 0 & 0 & 0 & 0 & 0 & \delta_{a_1\cdots a_5}^{b_1\cdots b_5} \\
 0 & 0 & 0 & 0 & 0 & 0 & 0 & \epsilon^\rmT_{\alpha\beta}\,\delta_{a_1\cdots a_4}^{b_1\cdots b_4} & 0 & 0 \\
 0 & 0 & 0 & 0 & \delta_{a_1a_2a_3}^{b_1b_2b_3} & 0 & 0 & 0 & 0 & 0 \\ \hline
 0 & 0 & 0 & 0 & 0 & 0 & 0 & 0 & \delta_{a_1\cdots a_5}^{b_1\cdots b_5}\,\delta_a^b & 0 \\
 0 & 0 & 0 & 0 & 0 & 0 & \epsilon^\rmT_{\alpha\beta} \,\delta_{a_1\cdots a_5}^{b_1\cdots b_5} & 0 & 0 & 0 
 \end{array}\right)}} .
\end{align}

Under the linear map, e.g., the matrix $\eta^a=(\eta^{IJ;\,a})$ associated with an M2-brane is mapped to a matrix $\eta^{a\By}=(\eta^{\sfM\sfN;\,a\By})$ associated with the D3-brane in type IIB theory,
\begin{align}
 \eta^{\sfM\sfN;\,a\By} = - (S^{-1})^{\sfM}{}_I\,\eta^{IJ;\,a}\,(S^{-\rmT})_J{}^{\sfN} \,,
\end{align}
where the minus sign is introduced by convention. 
Similarly, we can relate all of the $\eta$-symbols for M-theory and type IIB theory via the linear map $S$. 
By introducing a transformation matrix for the $R_2$-representation $T^{\sBA}{}_{\sMB}$, we can express the linear map for the $\eta$-symbols as
\begin{align}
 \eta^{\sfM\sfN;\,\sBA} = T^{\sBA}{}_{\sMB} \,(S^{-1})^{\sfM}{}_I\,\eta^{IJ;\,\sMB}\,(S^{-\rmT})_J{}^{\sfN} \,,\qquad 
 \eta^{IJ;\,\sMA} = (T^{-1})^{\sMA}{}_{\sBB} \,S^I{}_{\sfM}\,\eta^{\sfM\sfN;\,\sBB}\,(S^{\rmT})_{\sfN}{}^J \,,
\end{align}
Here, the matrix $T$ that maintains the $\SL(d-2)$ covariance can be summarized as follows:
\begin{align}
{\tiny
\begin{pmatrix}
 \eta_{\alpha}
\\\hline
 \frac{1}{\sqrt{2!}}\,\eta^{a_1a_2} \\
 \eta^{a\By}\\\hline
 \frac{1}{\sqrt{4!}}\,\eta^{a_1\cdots a_4}_\alpha \\
 \frac{1}{\sqrt{3!}}\,\eta^{a_1a_2a_3\By}_\alpha
\\[3pt]\hline
 \frac{1}{\sqrt{5!}}\,\eta^{a_1\cdots a_5,\,c} \\
 \frac{1}{\sqrt{5!}}\,\eta^{a_1\cdots a_5,\,\By} \\
 \frac{1}{\sqrt{4!}}\,\eta^{a_1\cdots a_4\By,\,c} \\
 \frac{1}{\sqrt{4!}}\,\eta^{a_1\cdots a_4\By,\,\By}
\\[3pt]\hline
 \frac{1}{\sqrt{5!}}\,\eta^{a_1\cdots a_5\By}_{(\alpha_1\alpha_2)}
\\[3pt]\hline
 \frac{1}{\sqrt{5!\,2!}}\,\eta^{a_1\cdots a_5\By,\,c_1c_2}_{\alpha} \\
 \frac{1}{\sqrt{5!}}\,\eta^{a_1\cdots a_5\By,\,c\By}_{\alpha}
\\\hline
 \frac{1}{\sqrt{5!\,4!}}\,\eta^{a_1\cdots a_5\By,\,c_1\cdots c_4} \\
 \frac{1}{\sqrt{5!\,3!}}\,\eta^{a_1\cdots a_5\By,\,c_1c_2c_3\By}
\\\hline
 \frac{1}{\sqrt{5!\,5!}}\eta^{a_1\cdots a_5\By,\,c_1\cdots c_5\By}_{\alpha}
\end{pmatrix}
 = {\normalsize T} \begin{pmatrix}
 \eta^b \\
 \eta^{\beta}
\\\hline
 \frac{1}{\sqrt{4!}}\,\eta^{b_1\cdots b_4}\\
 \frac{1}{\sqrt{3!}}\,\eta^{b_1b_2b_3\beta}\\
 \frac{1}{\sqrt{2!}}\,\eta^{b_1b_2yz}
\\[3pt]\hline
 \frac{1}{\sqrt{5!}}\,\eta^{b_1\cdots b_5\beta,\,d}\\
 \frac{1}{\sqrt{5!}}\,\eta^{b_1\cdots b_5(\beta_1,\,\beta_2)}\\
 \frac{1}{\sqrt{5!}}\,\eta^{b_1\cdots b_5[\Ay,\,\Az]}\\
 \frac{1}{\sqrt{4!}}\,\eta^{b_1\cdots b_4yz,\,d}\\
 \frac{1}{\sqrt{4!}}\,\eta^{b_1\cdots b_4yz,\,\beta}
\\[3pt]\hline
 \frac{1}{\sqrt{5!\,3!}}\,\eta^{b_1\cdots b_5yz,\,d_1d_2d_3}\\
 \frac{1}{\sqrt{5!\,2!}}\,\eta^{b_1\cdots b_5yz,\,d_1d_2\beta}\\
 \frac{1}{\sqrt{5!}}\,\eta^{b_1\cdots b_5yz,\,dyz}
\\[3pt]\hline
 \frac{1}{\sqrt{5!\,5!}}\,\eta^{b_1\cdots b_5yz,\,d_1\cdots d_5\beta}\\
 \frac{1}{\sqrt{5!\,4!}}\,\eta^{b_1\cdots b_5yz,\,d_1\cdots d_4yz}
\end{pmatrix},
}
\end{align}
\begin{align}
 &\hspace*{-20mm}{T\equiv \bigl(T^{\sBA}{}_{\sMB}\bigr)\equiv}
\nn\\
 &{}{\tiny\hspace*{-20mm}{ 
 {\arraycolsep=-1mm 
 \left(\begin{array}{cc|ccc|ccccc|ccc|cc}
  0&~\epsilon_{\beta\alpha}~ & 0&0&0 & 0&0&0&0&0 & 0&0&0 & 0&0 
\\\hline
  0&0 & 0&0&\delta^{a_1a_2}_{b_1b_2}~\, & 0&0&0&0&0 & 0&0&0 & 0&0 \\
  -\delta^a_b~&0 &0&0&0 & 0&0&0&0&0 & 0&0&0 & 0&0 
\\\hline
  0&0 & 0&0&0 &0&0&0&0&\!\!\!\!\epsilon_{\beta\alpha}\delta^{a_1\cdots a_4}_{b_1\cdots b_4}~ & 0&0&0 & 0&0 \\
  0&0 & 0&\!\!\!\epsilon_{\beta\alpha}\delta^{a_1a_2a_3}_{b_1b_2b_3}&0 & 0&0&0&0&0 & 0&0&0 & 0&0 
\\\hline
  0&0 & 0&0&0 & 0&0&0&0&0 & 0&0&\!\!\!\!\delta^{a_1\cdots a_5}_{b_1\cdots b_5}\delta^c_d~\, & 0&0 \\
  0&0 & 0&0&0 & 0&0&\!\!\!\!\!\!\!\!\!\!\!\!(1+2c_1)\delta^{a_1\cdots a_5}_{b_1\cdots b_5}&\sqrt{5}c_1\delta^{a_1\cdots a_5}_{b_1\cdots b_4d}&0 & 0&0&0 & 0&0 \\
  0&0 & 0&0&0 & 0&0&\!\!\!\!\!\!\!\!\!\!\sqrt{5}(1-2\sfc_2)\delta^{a_1\cdots a_4c}_{b_1\cdots b_5}&~~~\delta^{a_1\cdots a_4}_{b_1\cdots b_4}\delta^c_d-5\sfc_2\delta^{a_1\cdots a_4c}_{b_1\cdots b_4d}\!\!\!&0 & 0&0&0 &0&0 \\
  0&0 & ~\,\delta^{a_1\cdots a_4}_{b_1\cdots b_4}\!\!\!&0&0 & 0&0&0&0&0 & 0&0&0 & 0&0 
\\\hline
  0&0 & 0&0&0 & 0&\!\!\!\!\!\!\epsilon_{\alpha_1\beta_1}\epsilon_{\alpha_2\beta_2}\delta^{a_1\cdots a_5}_{b_1\cdots b_5}\!\!\!\!\!\!\!\!\!\!\!&0&0&0 & 0&0&0 & 0&0 
\\\hline
  0&0 & 0&0&0 & 0&0&0&0&0 & 0&\!\!\!\!\epsilon_{\beta\alpha}\delta^{a_1\cdots a_5}_{b_1\cdots b_5}\delta^{c_1c_2}_{d_1d_2}\!\!\!\!\!&0 & 0&0 \\
  0&0 & 0&0&0 & ~\,\epsilon_{\beta\alpha}\delta^{a_1\cdots a_5}_{b_1\cdots b_5}\delta^c_d\!\!\!\!\!\!&0&0&0&0 & 0&0&0 & 0&0 
\\\hline
  0&0 & 0&0&0 & 0&0&0&0&0 & 0&0&0 & 0&\!\!\delta^{a_1\cdots a_5}_{b_1\cdots b_5}\delta^{c_1\cdots c_4}_{d_1\cdots d_4}\!\\
  0&0 & 0&0&0 & 0&0&0&0&0 & ~-\delta^{a_1\cdots a_5}_{b_1\cdots b_5}\delta^{c_1c_2c_3}_{d_1d_2d_3}\!\!\!\!\!\!\!&0&0 & 0&0 
\\\hline
  0&0 & 0&0&0 & 0&0&0&0&0 & 0&0&0 & ~\,\epsilon_{\beta\alpha}\delta^{a_1\cdots a_5}_{b_1\cdots b_5}\delta^{c_1\cdots c_5}_{d_1\cdots d_5}\!\!\!\!\!\!\!\!\!\!&0 
 \end{array}\right)}}},
\label{eq:T-matrix}
\end{align}
where
\begin{align}
 \sfc_1\equiv \frac{3\sqrt{2}-2}{14} \,,\qquad \sfc_2\equiv \frac{\sqrt{2}+4}{14}\,.
\label{eq:c1-c2-def}
\end{align}

In the case of $d=7$, as we mentioned in section \ref{sec:sketch}, we cannot determine the matrix form of $\eta^{[k_1\cdots k_6,\,l]}= \eta^{k_1\cdots k_6l}$ and $\eta^{[\sfp_1\cdots\sfp_5,\,\sfq]}=\eta^{\sfp_1\cdots\sfp_5\sfq}$ only through the above mapping procedure. 
Assuming that these matrices are symmetric and constructed only from the Kronecker deltas, the possible form for $d\leq 7$ is
\begin{align}
 \eta^{k_1\cdots k_7} 
 &= \begin{pmatrix}
 0 & 0 & 0 & \chi_1\,\frac{7!\,\delta^{k_1\cdots k_7}_{j_1\cdots j_7}\,\delta^i_j}{\sqrt{7!}} \\
 0 & 0 & \chi_2\,\frac{7!\,\delta^{k_1\cdots k_7}_{j_1\cdots j_5i_1i_2}}{\sqrt{2!\,5!}} & 0 \\
 0 & \chi_2\,\frac{7!\,\delta^{k_1\cdots k_7}_{i_1\cdots i_5j_1j_2}}{\sqrt{2!\,5!}} & 0 & 0 \\
 \chi_1\,\frac{7!\,\delta^{k_1\cdots k_7}_{i_1\cdots i_7}\,\delta^j_i}{\sqrt{7!}} & 0 & 0 & 0
\end{pmatrix} , 
\\
 \eta^{\sfp_1\cdots\sfp_6}
 &= \begin{pmatrix}
 0 & 0 & 0 & 0 &\chi_3\,\frac{6!\,\delta^{\sfp_1\cdots \sfp_6}_{\sfn_1\cdots \sfn_6}\, \delta^{\sfm}_{\sfn}}{\sqrt{6!}} \\
 0 & 0 & 0 &\chi_4\,\frac{6!\,\epsilon^{\beta\alpha}\,\delta^{\sfp_1\cdots \sfp_6}_{\sfn_1\cdots \sfn_5\sfm}}{\sqrt{5!}} & 0 \\
 0 & 0 & 0 & 0 & 0 \\
 0 & \chi_4\,\frac{6!\,\epsilon^{\alpha\beta}\,\delta^{\sfp_1\cdots \sfp_6}_{\sfm_1\cdots \sfm_5\sfn}}{\sqrt{5!}} & 0 & 0 & 0 \\
 \chi_3\,\frac{6!\delta^{\sfp_1\cdots \sfp_6}_{\sfm_1\cdots \sfm_6}\,\delta^{\sfn}_{\sfm}}{\sqrt{6!}} & 0 & 0 & 0 & 0
 \end{pmatrix}.
\end{align}
From these ansatz, we define
\begin{align}
 \eta^{k_1\cdots k_6,\,l} \equiv \eta_{\text{KKM}}^{k_1\cdots k_6,\,l} + \eta^{k_1\cdots k_6l} \,, \qquad 
 \eta^{\sfp_1\cdots\sfp_5,\,\sfq} \equiv \eta_{\text{KKM}}^{\sfp_1\cdots\sfp_5,\,\sfq} + \eta^{\sfp_1\cdots\sfp_5\sfq}\,. 
\end{align}
Supposing that $\eta^{k_1\cdots k_6,\,l}$ and $\eta^{\sfp_1\cdots\sfp_5,\,\sfq}$ are related with each other by the linear map, under the decomposition $\{i\}\to\{a,\alpha\}$ and $\{\sfm\}\to\{a,\By\}$, the non-trivial components $\eta^{a_1\cdots a_5,\,\By}$ and $\eta^{a_1\cdots a_4\By,\,c}$ (which include the contribution from $\eta^{k_1\cdots k_6l}$) should be expanded with $\eta^{k_1\cdots k_6,\,l}$ in the following general forms
\begin{align}
\begin{split}
 \eta^{a_1\cdots a_5,\,\By} &= \lambda_1 \, \eta^{a_1\cdots a_5[\Ay,\,\Az]} + \lambda_2\,\eta^{[a_1\cdots a_4|\Ay\Az,\,|a_5]}\,,
\\
 \eta^{a_1\cdots a_4\By,\,c} &= \lambda_3 \, \eta^{a_1\cdots a_4yz,\,c} + \lambda_4\,\eta^{[a_1\cdots a_4|\Ay\Az,\,|c]} + \lambda_5\, \eta^{a_1\cdots a_4c[\Ay,\,\Az]}\,. 
\end{split}
\end{align}
This requires $\chi_1 = 3\,\chi_2$ and $\chi_3 = 2\,\chi_4$\,. 
We can determine the overall constant (up to sign) of $\eta^{k_1\cdots k_7}$ and $\eta^{\sfp_1\cdots\sfp_6}$ (i.e.~$\chi_2$ and $\chi_4$) by further requiring the conditions
\begin{align}
 \eta^{IJ;\,\sMA}\,\eta_{IJ;\,\sMB} \propto \delta^{\sMA}_{\sMB} \,,\qquad 
 \eta^{MN;\,\sBA}\,\eta_{MN;\,\sBB} \propto \delta^{\sBA}_{\sBB} \,. 
\end{align}
By choosing a sign convention, we obtain the $\eta$-symbols shown in the previous subsections. 
The coefficients $\lambda_1,\dotsc,\lambda_5$ also can be determined and the result is shown in \eqref{eq:T-matrix}.

Similarly, we can also determine the matrix form of the $\Omega$-tensors that appear in $d=7$. 
Supposing that they are also constructed from combinations of products of Kronecker deltas, the defining properties,
\begin{align}
 \Omega_{IJ} = \Omega^{IJ}\,,\qquad \Omega^{KI}\,\Omega_{KJ} = \delta^I_J\,,\qquad \Omega_{IJ} = \Omega_{[IJ]} \,,
\end{align}
require them to have the following form up to the overall sign convention:
\begin{align}
 (\Omega^{IJ}) &= \begin{pmatrix}
 0 & 0 & 0 & \frac{\epsilon_{j_1\cdots j_7}\,\delta^i_j}{\sqrt{7!}} \\
 0 & 0 & \pm \frac{\epsilon_{i_1i_2j_1\cdots j_5}}{\sqrt{2!\,5!}} & 0 \\
 0 & \mp \frac{\epsilon_{i_1\cdots i_5j_1j_2}}{\sqrt{2!\,5!}} & 0 & 0 \\
 -\frac{\epsilon_{i_1\cdots i_7}\,\delta^j_i}{\sqrt{7!}} & 0 & 0 & 0
 \end{pmatrix} .
\end{align}
In order for the $\Omega$-tensor in the type IIB side, namely $\Omega^{\sfM\sfN}\equiv (S^{-1})^{\sfM}{}_I\,\Omega^{IJ}\,(S^{-\rmT})_J{}^{\sfN}$, to be expressed covariantly by means of the Kronecker deltas, we shall choose the upper sign, and then the $\Omega$-tensor in the type IIB side becomes \eqref{eq:Omega-IIB}. 
In this manner, we have determined all of the $\eta$-symbols and the $\Omega$-tensor. 

\subsection{Properties of $\eta$-symbols}

We can easily check that the identities
\begin{align}
 \eta_{IJ;\,\sMA}\,\eta^{IJ;\,\sMB} = 2\,(d-1)\,\delta_{\sMA}^{\sMB} \,,\qquad
 \eta_{MN;\,\sBA}\,\eta^{MN;\,\sBB} = 2\,(d-1)\,\delta_{\sBA}^{\sBB} \,,
\label{eq:id-1}
\end{align}
are satisfied for $d=2,\dotsc,7$. 
We can also show the identities
\begin{align}
 \eta^{IK;\,\sMA}\, \eta_{KJ;\,\sMA} = D_{d-1} \, \delta^I_J \,, \qquad
 \eta^{\sfM\sfP;\,\sBA}\, \eta_{\sfP\sfN;\,\sBA} = D_{d-1} \, \delta^{\sfM}_{\sfN} \,, 
\label{eq:id-2}
\end{align}
where $D_{d-1}$ is given by $D_2 =2$, $D_3=3$, $D_4=5$, $D_5=10$, and $D_6= 57/2$. 
From identity \eqref{eq:id-1} or \eqref{eq:id-2}, the normalization of the $Y$-tensor in $E_{d(d)}$ EFT becomes
\begin{align}
 Y^{IJ}_{IJ} = n_d \qquad (n_3=12\,,\ n_4=30\,,\ n_5=80\,,\ n_6=270\,,\ n_7=1568) \,.
\end{align}

In the case of $E_{7(7)}$ EFT, we can check additional identities. 
If we define
\begin{align}
\begin{split}
 &(t^{\sMA})_I{}^J \equiv \Omega_{IK} \,\eta^{KJ;\,\sMA} \,,\qquad 
 K^{\sMA\sMB} \equiv \frac{1}{12}\, (t^{\sMA})_I{}^J\,(t^{\sMB})_J{}^I \,,
\\
 &(t_{\sMA})_I{}^J \equiv \Omega^{JK}\, \eta_{IK;\,\sMA} \,,\qquad 
 K_{\sMA\sMB} \equiv \frac{1}{12}\, (t_{\sMA})_I{}^J\,(t_{\sMB})_J{}^I \,,
\end{split}
\end{align}
we can show a relation that connects the two types of $\eta$-symbols, $\eta_{IJ;\,\sMA}$ and $\eta^{IJ;\,\sMA}$,
\begin{align}
 \eta_{IJ;\,\sMA} = - K_{\sMA\sMB}\,\Omega_{IK} \,\Omega_{JL} \,\eta^{KL;\,\sMB} \,,\qquad 
 t_{\sMA} = K_{\sMA\sMB}\, t^{\sMB} \,,\qquad 
 K^{\sMA\sMC}\,K_{\sMC\sMB} = \delta^{\sMA}_{\sMB} \,,
\end{align}
where $t_{\sMA}\equiv \bigl((t_{\sMA})_I{}^J\bigr)$ and $t^{\sMA}\equiv \bigl((t^{\sMA})_I{}^J\bigr)$\,. 
The matrix $K\equiv (K_{\sMA\sMB})$ becomes
\begin{align}
 K = \begin{pmatrix}
 0 & 0 & 0 & 0 & -\frac{\epsilon_{j_1\cdots j_7}\,\epsilon_{i l_1\cdots l_6}}{\sqrt{7!\,6!}} \\
 0 & 0 & 0 & -\frac{\epsilon_{j_1\cdots j_7}\,\epsilon_{i_1\cdots i_4 l_1l_2l_3}}{\sqrt{4!\,7!\,3!}} & 0 \\
 0 & 0 & \frac{\epsilon_{i_1\cdots i_6 l}\,\epsilon_{k j_1\cdots j_6}}{\sqrt{6!\,6!}} & 0 & 0 \\
 0 & -\frac{\epsilon_{i_1\cdots i_7}\,\epsilon_{j_1\cdots j_4 k_1k_2k_3}}{\sqrt{4!\,7!\,3!}} & 0 & 0 & 0 \\
 -\frac{\epsilon_{i_1\cdots i_7}\,\epsilon_{j k_1\cdots k_6}}{\sqrt{7!\,6!}} & 0 & 0 & 0 & 0 
 \end{pmatrix} ,
\end{align}
which has the eigenvalues $70$ ``$+1$'' and $63$ ``$-1$.'' 
The same relations are also satisfied in the type IIB side, and there the matrix $K=(K_{\sBA\sBB})$ becomes
\begin{align}
 {\footnotesize
 K = {\arraycolsep=-3mm \left(\begin{array}{cccccccc}
 0 & 0 & 0 & 0 & 0 & 0 & 0 & \frac{\epsilon^{\sfn_6\cdots\sfn_6}\,\epsilon^{\sfq_6\cdots\sfq_6}\,\epsilon^{\alpha\beta}}{\sqrt{6!\,6!}}~~ \\
 0 & 0 & 0 & 0 & 0 & 0 & \frac{\epsilon^{\sfm_1\sfm_2 \sfq_1\cdots\sfq_4}\,\epsilon^{\sfn_1\cdots\sfn_6}}{\sqrt{2!\,6!\,4!}} & 0 \\
 0 & 0 & 0 & 0 & 0 & \frac{\epsilon^{\beta\alpha}\,\epsilon^{\sfm_1\cdots\sfm_4\sfq_1\sfq_2}\,\epsilon^{\sfn_1\cdots\sfn_6}}{\sqrt{4!\,6!\,2!}} & 0 & 0 \\
 0 & 0 & 0 & \frac{\epsilon_{\sfm_1\cdots \sfm_5 \sfq}\,\epsilon_{\sfp \sfn_1\cdots \sfn_5}}{\sqrt{5!\,5!}} & 0 & 0 & 0 & 0 \\
 0 & 0 & 0 & 0 & -\epsilon_{\alpha_1\beta_1}\,\epsilon_{\alpha_2\beta_2} & 0 & 0 & 0 \\
 0 & 0 & \frac{\epsilon^{\alpha\beta}\,\epsilon^{\sfm_1\cdots\sfm_6}\,\epsilon^{\sfp_1\sfp_2\sfn_1\cdots\sfn_4}}{\sqrt{6!\,2!\,4!}} & 0 & 0 & 0 & 0 & 0 \\
 0 & \frac{\epsilon^{\sfm_1\cdots\sfm_6}\,\epsilon^{\sfp_1\cdots\sfp_4 \sfn_1\sfn_2}}{\sqrt{6!\,4!\,2!}} & 0 & 0 & 0 & 0 & 0 & 0 \\
 ~~\frac{\epsilon^{\sfm_6\cdots\sfm_6}\,\epsilon^{\sfp_6\cdots\sfp_6}\,\epsilon^{\beta\alpha}}{\sqrt{6!\,6!}} & 0 & 0 & 0 & 0 & 0 & 0 & 0 
 \end{array}\right)} .}
\end{align}

In fact, $t^{\sMA}$ corresponds to the generators of the $E_{7(7)}$ group. 
By using the generators of the $E_{7(7)}$ group shown in Appendix \ref{app:Edd}, $t^{\sMA}$ can be expressed as
\begin{align}
\begin{split}
 t^k &= \frac{1}{6!}\,\epsilon^{k l_1\cdots l_6}\, R_{l_1\cdots l_6} \,,
\\
 t^{k_1\cdots k_4} &= - \frac{1}{3!}\, \epsilon^{k_1\cdots k_4 l_1l_2l_3}\,R_{l_1l_2l_3} \,,
\\
 t^{k_1\cdots k_6,\,k} &= \Bigl(\epsilon^{k_1\cdots k_6 j} \, \delta_i^k- \frac{3-\sqrt{2}}{21}\,\epsilon^{k_1\cdots k_6 k} \, \delta_i^j\Bigr) \, K^i{}_j \,,
\\
 t^{k_1\cdots k_7,\,l_1l_2l_3} &= \epsilon^{k_1\cdots k_7}\,R^{l_1l_2l_3} \,,
\\
 t^{k_1\cdots k_7,\,l_1\cdots l_6} &= -\epsilon^{k_1\cdots k_7}\, R^{l_1\cdots l_6} \,,
\end{split}
\end{align}
Similarly, the $t^{\sBA}$ are related to the generators in the type IIB parameterization as
\begin{align}
\begin{split}
 t_\gamma &= \frac{1}{6!}\,\epsilon_{\gamma\delta}\,\epsilon^{\sfp_1 \cdots \sfp_6}\, R_{\sfp_1\cdots \sfp_6}^\delta \,,
\\
 t^{\sfp_1\sfp_2} &= -\frac{1}{4!}\, \epsilon^{\sfp_1\sfp_2\sfq_1\cdots \sfq_4}\,R_{\sfq_1\cdots \sfq_4} \,,
\\
 t^{\sfp_1\cdots \sfp_4}_\gamma 
 &= \epsilon_{\gamma\delta}\,\epsilon^{\sfp_1\cdots \sfp_4 \sfq_1\sfq_2} \, R_{\sfq_1\sfq_2}^\delta \,,
\\
 t^{\sfp_1\cdots \sfp_5,\,\sfp} 
 &= - \Bigl(\epsilon^{\sfp_1\cdots \sfp_5\sfq} \, \delta_{\sfr}^{\sfp} - \frac{1}{4}\,\epsilon^{\sfp_1\cdots \sfp_5\sfp} \, \delta_{\sfr}^{\sfq}\Bigr) \, K^{\sfr}{}_{\sfq} \,, 
\\
 t^{\sfp_1\cdots \sfp_6}_{(\gamma\delta)}
 &= \epsilon^{\sfp_1\cdots \sfp_6}\, R_{(\gamma\delta)} \,,
\\
 t^{\sfp_1\cdots \sfp_6,\,\sfq_1\sfq_2}_\gamma 
 &= -\epsilon^{\sfp_1\cdots \sfp_6}\,R^{\sfq_1\sfq_2}_\gamma \,,
\\
 t^{\sfp_1\cdots \sfp_6,\,\sfq_1\cdots \sfq_4}
 &= \epsilon^{\sfp_1\cdots \sfp_6}\,R^{\sfq_1\cdots \sfq_4} \,,
\\
 t^{\sfp_1\cdots \sfp_6,\,\sfq_1\cdots \sfq_6}_\gamma
 &= \epsilon^{\sfp_1\cdots \sfp_6}\,R^{\sfq_1\cdots \sfq_6}_\gamma \,. 
\end{split}
\end{align}
Then, $t^{[k_1\cdots k_6,\,k]}$ or $t^{[\sfp_1\cdots \sfp_5,\,\sfp]}$ and $t^{\sfp_1\cdots \sfp_6}_{(12)}$ are Cartan generators, and the matrix $K$ corresponds to the Cartan--Killing form.

We can also check the following identities \cite{Hohm:2013uia}:
\begin{align}
\begin{split}
 &(t_{\sMA})_K{}^I\,(t^{\sMA})_L{}^J 
  = \frac{1}{2}\, \delta_K^I \,\delta_L^J + \delta_K^J \,\delta_L^I 
 - \eta^{IJ;\,\sMA}\, \eta_{KL;\,\sMA} - \frac{1}{2}\,\Omega^{IJ}\,\Omega_{KL} \,,
\\
 &(t_{\sBA})_{\sfP}{}^{\sfM}\,(t^{\sBA})_{\sfQ}{}^{\sfN} 
  = \frac{1}{2}\, \delta_{\sfP}^{\sfM} \,\delta_{\sfQ}^{\sfN} + \delta_{\sfP}^{\sfN} \,\delta_{\sfQ}^{\sfM} 
 - \eta^{\sfM\sfN;\,\sBA}\, \eta_{\sfP\sfQ;\,\sBA} - \frac{1}{2}\,\Omega^{\sfM\sfN}\,\Omega_{\sfP\sfQ} \,, 
\end{split}
\label{eq:t-t-Y}
\\
 &t^{\sMA} \, t_{\sMB} \, t_{\sMA} = \frac{21}{2}\, t_{\sMB} \,,\qquad 
 t^{\sBA} \, t_{\sBB} \, t_{\sBA} = \frac{21}{2}\, t_{\sBB} \,. 
\end{align}

\section{Generalized Lie derivative}
\label{sec:GeneralizedLie}

By using the obtained $\eta$-symbols, the section condition $\eta^{IJ;\,\sMA}\,\partial_I\otimes\partial_J=0$ can be expressed as follows (see \cite{Coimbra:2011ky} for a quite similar section condition and also \cite{Bandos:2016ppv} for a section condition in the ``underlying EFT''):
\begin{align}
 &\partial_i \otimes \partial^{ki}+ \partial^{k i}\otimes \partial_i =0 \,,
\\
 &\partial_i \otimes \partial^{i k_1\cdots k_4} + 6\,\partial^{[k_1k_2}\otimes \partial^{k_3k_4]} + \partial^{i k_1\cdots k_4}\otimes \partial_i =0 \,,
\\
 &\partial_i\otimes \partial^{k_1\cdots k_6 i,\,l} + \sfc_1\, \partial_i\otimes \partial^{k_1\cdots k_6 l,\,i} 
 - 6\, \partial^{l[k_1}\otimes \partial^{k_2\cdots k_6]} 
 + 21\,\sfc_2\, \partial^{[k_1k_2}\otimes \partial^{k_3\cdots k_6l]} 
\nn\\
 &-6\,\partial^{[k_1\cdots k_5} \otimes \partial^{k_6]l}
 + 21\,\sfc_2\, \partial^{[k_1\cdots k_5}\otimes \partial^{k_6l]} 
 + \partial^{k_1\cdots k_6 j,\,l} \otimes \partial_j + \sfc_1\,\partial^{k_1\cdots k_6 l,\,i}\otimes \partial_i =0 \,,
\\
 &\epsilon_{j_1\cdots j_7}\, \bigl(21\, \partial_i \otimes \partial^{j_1\cdots j_7,\,i}
 + \partial^{[i_1i_2}\otimes \partial^{i_3\cdots i_7]}
 - \partial^{[i_1\cdots i_5}\otimes \partial^{i_6i_7]} 
 -21\, \partial^{i_1\cdots i_7,\,i}\otimes \partial_i\bigr) = 0\,,
\\
 &5\, \partial^{[k_1k_2|}\otimes \partial^{l_1l_2l_3|k_3\cdots k_6,\,k_7]}
 - \partial^{[k_1\cdots k_5}\otimes \partial^{k_6k_7] l_1l_2l_3}
 +5\, \partial^{l_1l_2l_3 [k_1\cdots k_4,\,k_5} \otimes\partial^{k_6k_7]} = 0 \,,
\\
 &\partial^{k_1\cdots k_7,\,[l_1}\otimes \partial^{l_2\cdots l_6]} + \partial^{[l_2\cdots l_6|} \otimes \partial^{k_1\cdots k_7,\,|l_1]} =0\,,
\end{align}
where $\sfc_1$ and $\sfc_2$ are defined in \eqref{eq:c1-c2-def}. 
In particular, when we consider, e.g., the $\SL(5)$ EFT, the above section conditions are truncated easily to get
\begin{align}
 \partial_i \otimes \partial^{ki}+ \partial^{k i}\otimes \partial_i =0 \,, \qquad
 \partial^{[k_1k_2}\otimes \partial^{k_3k_4]} =0 \,.
\end{align}
The section condition in the type IIB parameterization, $\eta^{\sfM\sfN;\,\sBA}\,\partial_{\sfM} \otimes\partial_{\sfN} =0$, can also be rewritten in a similar manner, though we will not show this explicitly. 

There are two well-known solutions to the section condition. 
One is the solution, called the M-theory section, where 
\begin{align}
 \partial^{i_1i_2} = 0 \,,\qquad \partial^{i_1\cdots i_5} = 0\,,\qquad \partial^{i_1\cdots i_7,\,j} = 0\,, 
\end{align}
are satisfied; namely, on the M-theory section, all fields depend only on the $d$ coordinates $x^i$. 
The other solution is called the type IIB section, where
\begin{align}
 \partial_\alpha^{\sfm} = 0\,,\qquad 
 \partial^{\sfm_1\sfm_2\sfm_3} = 0 \,,\qquad 
 \partial_\alpha^{\sfm_1\cdots \sfm_5} = 0 \,,\qquad 
 \partial^{\sfm_1\cdots \sfm_6,\,\sfn} = 0\,,
\end{align}
are satisfied. 
In the type IIB section, fields depend only on the $d-1$ coordinates $\sfx^{\sfm}$ (see \cite{Blair:2013gqa} for the type IIB section in the $\SL(5)$ EFT and also \cite{Hohm:2013pua,Hohm:2013vpa,Hohm:2013uia,Hohm:2014fxa} for a higher $E_{d(d)}$ EFT). 

On the M-theory section, the generalized Lie derivative reduces to the exceptional Dorfman bracket \cite{Pacheco:2008ps,Coimbra:2011ky,Coimbra:2012af,Rosabal:2014rga}. 
Indeed, by using the $Y$-tensor,
\begin{align}
 Y^{IJ}_{KL} = \eta^{IJ;\,\sMA}\,\eta_{KL;\,\sMA} - \frac{1}{2}\,\Omega^{IJ}\,\Omega_{KL} \,,
\end{align}
and the explicit form of the $\eta$-symbols and the $\Omega$-tensor, we obtain
\begin{align}
 \begin{pmatrix}
 \delta_V w^i\\
 \frac{\delta_V w_{i_1i_2}}{\sqrt{2!}}\\
 \frac{\delta_V w_{i_1\cdots i_5}}{\sqrt{5!}}\\
 \frac{\delta_V w_{i_1\cdots i_7,\,i}}{\sqrt{7!}}
 \end{pmatrix}
 = 
 \begin{pmatrix}
 \Lie_v w^i \\
 \frac{\Lie_v w_{i_1i_2} - w^k\,(\rmd v_2)_{ki_1i_2}}{\sqrt{2!}} \\
 \frac{\Lie_v w_{i_1\cdots i_5} +10\,(\rmd v_2)_{[i_1i_2i_3}\,w_{i_4i_5]}- w^k\,(\rmd v_5)_{ki_1\cdots i_5}}{\sqrt{5!}} \\
 \frac{\Lie_v w_{i_1\cdots i_7,\,i}+21\,(\rmd v_2)_{i[i_1i_2}\,w_{i_3\cdots i_7]}+7\,w_{i[i_1}\,(\rmd v_5)_{i_2\cdots i_7]}}{\sqrt{7!}}
 \end{pmatrix},
\end{align}
where $(\rmd v_2)_{i_1i_2i_3} \equiv 3\,\partial_{[i_1}v_{i_2i_3]}$ and $(\rmd v_5)_{i_1\cdots i_6} \equiv 6\,\partial_{[i_1}v_{i_2\cdots i_6]}$\,.
In the last line, we have repeatedly used the Schouten-like identities such as
\begin{align}
 \partial_{[i_1} v^k\, w_{i_2\cdots i_8],\,k} = 0 \,,
\end{align}
which are satisfied for $d\leq 7$\,. 
This result precisely matches with the known result \cite{Pacheco:2008ps,Coimbra:2011ky,Coimbra:2012af,Rosabal:2014rga}. 

For a gauge parameter of the form $V^I=\eta^{IJ;\,\sMA}\,\partial_J f_{\sMA} = \partial_J f^{IJ}$, where $f^{IJ}\equiv \eta^{IJ;\,\sMA}\, f_{\sMA}$\,, the generalized Lie derivative becomes
\begin{align}
 \gLie_V W^I = \bigl(Y^{IJ}_{KL} \, Y^{KP}_{RS} - Y^{IJ}_{RS}\,\delta_L^P\bigr)\,\partial_{(J} \partial_{P)} f^{RS} \,W^L\,. 
\end{align}
In fact, a condition,
\begin{align}
 \bigl(Y^{IJ}_{KL} \, Y^{KP}_{RS} - Y^{IJ}_{RS}\,\delta_L^P\bigr)\,\partial_{(J} \otimes \partial_{P)} = 0\,,
\end{align}
is necessary for the closure of the gauge algebra \cite{Berman:2012vc}, and for $d\leq 7$, it is indeed satisfied under the section condition \eqref{eq:EFT-SC} \cite{Berman:2012vc}. 
Therefore, a gauge parameter of the form $V^I=\eta^{IJ;\,\sMA}\,\partial_J f_{\sMA}$ is a generalized Killing vector for an arbitrary $f_{\sMA}$\,. 
Moreover, $V^I=\Omega^{IJ}\, \chi_J$ with $\chi_J$ satisfying
\begin{align}
 \eta^{IJ;\,\sMA}\,\chi_I\otimes \partial_J = 0 \,,\qquad 
 \Omega^{IJ}\,\chi_I\otimes \partial_J = 0 \,,
\label{eq:chi-cond}
\end{align}
is also a trivial generalized Killing vector \cite{Hohm:2013uia},
\begin{align}
 \gLie_V W^I 
 &= \bigl(Y^{IJ}_{PL} \,\Omega^{PK} - \Omega^{IK}\,\delta_L^J\bigr)\, \partial_J \chi_K \,W^L 
\nn\\
 &= -\Bigl[\eta^{KJ;\,\sMA}\,\bigl(t_{\sMA}\bigr)_L{}^I - \frac{1}{2}\, \delta_L^I\,\Omega^{JK} \Bigr]\, \partial_J \chi_K \,W^L = 0 \,,
\end{align}
where the identity \eqref{eq:t-t-Y} is used in the second equality and \eqref{eq:chi-cond} is used in the last equality. 

On the other hand, if we choose the type IIB section, the generalized Lie derivative takes the form
\begin{align}
 \begin{pmatrix}
 \delta_V w^{\sfm} \\
 \delta_V w_{\sfm}^\alpha \\
 \frac{\delta_V w_{\sfm_1\sfm_2\sfm_3}}{\sqrt{3!}}\\
 \frac{\delta_V w^\alpha_{\sfm_1\cdots\sfm_5}}{\sqrt{5!}}\\
 \frac{\delta_V w_{\sfm_1\cdots \sfm_6,\,\sfm}}{\sqrt{6!}}
 \end{pmatrix}
 = 
 \begin{pmatrix}
 \Lie_v w^{\sfm} \\
 \Lie_v w_{\sfm}^\alpha -w^{\sfn}\,(\rmd v_1^\alpha)_{\sfn\sfm} \\
 \frac{\Lie_v w_{\sfm_1\sfm_2\sfm_3} - 3\,\epsilon_{\gamma\delta}\,(\rmd v_1^\gamma)_{[\sfm_1\sfm_2}\,w^\delta_{\sfm_3]} - w^{\sfn}\,(\rmd v_3)_{\sfn\sfm_1\sfm_2\sfm_3}}{\sqrt{3!}}\\
 \frac{\Lie_v w^\alpha_{\sfm_1\cdots\sfm_5} + 10\,(\rmd v_1^\alpha)_{[\sfm_1\sfm_2}\,w_{\sfm_3\sfm_4\sfm_5]} - 5\,(\rmd v_3)_{[\sfm_1\cdots\sfm_4}\,w^\alpha_{\sfm_5]} - w^{\sfn}\,(\rmd v_5^\alpha)_{\sfn\sfm_1\cdots\sfm_5}}{\sqrt{5!}}\\
 \frac{\Lie_v w_{\sfm_1\cdots \sfm_6,\,\sfm}+6\,\epsilon_{\gamma\delta}\,(\rmd v^\gamma_1)_{\sfm[\sfm_1}\,w^\delta_{\sfm_2\cdots \sfm_6]} -20\,(\rmd v_3)_{\sfm[\sfm_1\sfm_2\sfm_3}\,w_{\sfm_4\sfm_5\sfm_6]}+\epsilon_{\gamma\delta}\,(\rmd v^\gamma_5)_{\sfm_1\cdots\sfm_6}\,w^\delta_{\sfm}}{\sqrt{6!}}
 \end{pmatrix}.
\end{align}
Again, $V^{\sfM} = \eta^{\sfM\sfN;\,\sBA}\,\partial_{\sfN} f_{\sBA}$ and $V^{\sfM} =\Omega^{\sfM\sfN}\, \chi_{\sfN}$ are trivial gauge parameters. 

\section{Linear section equation}
\label{sec:Lienar-section}

In the $\OO(d,d)$ DFT, the section condition is expressed as $\eta^{IJ}\,\partial_I\otimes \partial_J = 0$. 
This condition states that $\partial_I$ is restricted to a $d$-dimensional maximal null subspace in the generalized tangent bundle. 
We can specify the maximal null subspace by introducing a set of independent $d$ generalized vectors $\lambda^a=(\lambda^a_I)=(\lambda^a_i,\,\lambda^{i;\,a})$ ($a=1,\dotsc,d$) satisfying
\begin{align}
 \lambda^a_I\,\eta^{IJ}\,\lambda^b_J = 0\,, \qquad \lambda^a_I\, \eta^{IJ}\,\partial_J = 0 \,. 
\label{eq:DFT-linear-section}
\end{align}
If we consider a particular solution to the first equation, $\lambda^a=\hat{\lambda}^a$, that takes the form,
\begin{align}
 \hat{\lambda}^a = (\hat{\lambda}^a_I) = \begin{pmatrix} \delta_i^a \\ 0 \end{pmatrix} ,
\end{align}
the second equation in \eqref{eq:DFT-linear-section} gives $\tilde{\partial}^i = 0$, which is the commonly used section to reproduce the usual supergravity from DFT. 
More generally, if the $d\times d$ matrix $\lambda^a_i$ is invertible, we can always realize $\lambda^a_i=\delta^a_i$ by a redefinition of $\lambda^a$; $\lambda^a\to \Lambda^a{}_b\,\lambda^b$. 
Then, by introducing an antisymmetric tensor $\beta^{ij}=\beta^{[ij]}$, the general solution to the first equation in \eqref{eq:DFT-linear-section} becomes
\begin{align}
 \lambda^a = \begin{pmatrix} \delta_i^a \\ \beta^{ia} \end{pmatrix} = \begin{pmatrix}
 \delta_i^j & 0 \\ \beta^{ij} & \delta^i_j 
 \end{pmatrix} \begin{pmatrix}
 \delta^a_j \\ 0
 \end{pmatrix} ,
\end{align}
which is just an $\OO(d,d)$ rotation of the generalized vector $\hat{\lambda}^a$. 
For this general $\lambda^a$, the second equation in \eqref{eq:DFT-linear-section} becomes
\begin{align}
 \tilde{\partial}^i = \beta^{ij}\,\partial_j \,.
\end{align}
We can easily show that this leads to the section condition,
\begin{align}
 \eta^{IJ}\,\partial_I\otimes \partial_J 
 = \partial_i \otimes \tilde{\partial}^i + \tilde{\partial}^i\otimes \partial_i = \bigl(\beta^{ij}+\beta^{ji}\bigr)\,\partial_i \otimes \partial_j = 0 \,. 
\end{align}
In fact, in the context of generalized geometry, essentially the same set of generalized vectors has been considered in \cite{Courant} (see also \cite{Rey:2015mba}). 
There, the maximal null subspace has been called the Dirac manifold or the Dirac structure, and the set of generalized vectors $\lambda^a$ has been called the basis representation of the Dirac structure. 
In addition, it has been shown that the Dirac structure can be characterized by an antisymmetric tensor, which is denoted by $\beta^{ij}$ here. 
Alternatively, we can also characterize the Dirac structure by using a pure spinor \cite{Gualtieri:2003dx}. 

A linear differential equation similar to \eqref{eq:DFT-linear-section}, which reproduces the section condition, is called the linear section equation in \cite{Berman:2012vc}. 
There, a linear section equation in $E_{d(d)}$ EFT for $d\leq 7$ was proposed, but the equation strongly depends on the dimension $d$ and it becomes complicated for higher $d$. 
Here, using the $\eta$-symbols, we propose a simple linear section equation, and show that it is equivalent to the proposal of \cite{Berman:2012vc} for the $\SL(5)$ EFT.

Our linear section equations take the form
\begin{align}
\setlength{\fboxsep}{2.5\fboxsep}
 \boxed{
\lambda^a\,\bbeta\, \partial = 0\,, \qquad \lambda^a\,\Omega\,\partial = 0
 }\,,
\end{align}
where $\lambda^a$ ($a=1,\dotsc,N$) is a set of generalized vectors satisfying the null conditions
\begin{align}
\setlength{\fboxsep}{2.5\fboxsep}
 \boxed{
\lambda^a\,\bbeta\, \lambda^b = 0\,, \qquad \lambda^a\,\Omega\,\lambda^b = 0
 }\,. 
\end{align}
If we show all of the indices explicitly, the linear section equations in the M-theory/type IIB parameterization become
\begin{align}
\begin{split}
 \text{M-theory}:\qquad&\lambda^a_I\,\eta^{IJ;\,\sMA}\, \partial_J = 0\,, \qquad \lambda^a_I\,\Omega^{IJ}\,\partial_J = 0\,,
\\
 \text{Type IIB}:\qquad&\lambda^a_{\sfM}\,\eta^{\sfM\sfN;\,\sBA}\, \partial_{\sfN} = 0\,, \qquad \lambda^a_{\sfM}\,\Omega^{\sfM\sfN}\,\partial_{\sfN} = 0\,.
\end{split}
\end{align}
The number of independent null generalized vectors $N$ depends on the choice of the section. 
As was shown in \cite{Bandos:2015rvs}, $N$ cannot be greater than $d$, but we can always choose $N=d$ or $N=d-1$, which correspond to the M-theory section and the type IIB section, respectively. 
In fact, the M-theory section and the type IIB section can be described by the following set of null vectors, $\hat{\lambda}^a$ ($N=d$) and $\hat{\lambdaB}^a$ ($N=d-1$):
\begin{align}
 \hat{\lambda}^a = \begin{pmatrix} \hat{\lambda}^a_i \\ \frac{(\hat{\lambda}^a)^{i_1i_2}}{\sqrt{2!}} \\ \frac{(\hat{\lambda}^a)^{i_1\cdots i_5}}{\sqrt{5!}} \\ \frac{(\hat{\lambda}^a)^{i_1\cdots i_7,\,i}}{\sqrt{7!}} \end{pmatrix} = \begin{pmatrix} \delta^a_i \\ 0 \\ 0 \\ 0 \end{pmatrix} ,
\qquad
 \hat{\lambdaB}^a = \begin{pmatrix} \hat{\lambdaB}^a_{\sfm} \\ (\hat{\lambdaB}^a)_\alpha^{\sfm} \\ \frac{(\hat{\lambdaB}^a)^{\sfm_1\sfm_2\sfm_3}}{\sqrt{3!}} \\ \frac{(\hat{\lambdaB}^a)^{\sfm_1\cdots \sfm_5}}{\sqrt{5!}} \\ \frac{(\hat{\lambdaB}^a)^{\sfm_1\cdots \sfm_6,\,\sfm}}{\sqrt{6!}} \end{pmatrix} = \begin{pmatrix} \delta^a_{\sfm} \\ 0 \\ 0 \\ 0 \\ 0 \end{pmatrix} . 
\end{align}
In the former case, $\hat{\lambda}^a\,\eta^k\, \partial =0$ and $\hat{\lambda}^a\,\eta^{k_1\cdots k_4}\, \partial =0$ require $\partial^{i_1i_2}=\partial^{i_1\cdots i_5}=0$\,. 
On the other hand, $\hat{\lambda}^a\,\eta^{k_1\cdots k_7,\,l_1l_2l_3}\, \partial =0$ and $\hat{\lambda}^a\,\eta^{k_1\cdots k_7,\,l_1\cdots l_6}\, \partial=0$ are trivially satisfied. 
The remaining conditions, $\hat{\lambda}^a\,\eta^{k_1\cdots k_6,\,l}\, \partial =0$ and $\hat{\lambda}^a\,\Omega\, \partial =0$, require $\partial^{i_1\cdots i_7,\,i}=0$. 
Therefore, $\hat{\lambda}^a$ describes the M-theory section where all fields depend only on $x^i$. 
The quadratic section condition $\eta^{IJ;\,\sMA}\,\partial_I\otimes \partial_J=0$ is trivially satisfied on this section. 
Similarly, in the latter case, we can easily show that all fields depend only on $\sfx^{\sfm}$, and $\hat{\lambdaB}^a$ describes the type IIB section.

In order to describe a more general section, we can rotate the above canonical sections, $\hat{\lambda}^a$ and $\hat{\lambdaB}^a$, by $U$-duality transformations:
\begin{align}
 \hat{\lambda}^a_I \to \lambda^a_I \equiv a_I{}^J\, \hat{\lambda}^a_J\,,\qquad 
 \hat{\lambdaB}^a_{\sfM} \to \lambdaB^a_{\sfM} \equiv b_{\sfM}{}^{\sfN}\, \hat{\lambdaB}^a_{\sfN} \,. 
\end{align}
Since $\eta^{IJ;\,\sMA}$ and $\eta^{\sfM\sfN;\,\sBA}$ behave as the Clebsch--Gordan--Wigner coefficients, they satisfy
\begin{align}
 a_K{}^I\,a_L{}^J\,\eta^{KL;\,\sMA} = \hat{a}^{\sMA}{}_{\sMB} \, \eta^{IJ;\,\sMB} \,,\qquad 
 b_{\sfP}{}^{\sfM}\,b_{\sfQ}{}^{\sfN}\,\eta^{\sfP\sfQ;\,\sBA} = \hat{b}^{\sBA}{}_{\sBB} \, \eta^{\sfM\sfN;\,\sBB} \,,
\end{align}
where $\hat{a}^{\sMA}{}_{\sMB}$ and $\hat{b}^{\sBA}{}_{\sBB}$ are certain $U$-duality-transformation matrices in the $R_2$-representation. 
Moreover, the $\Omega$-tensor is invariant under the $U$-duality transformations. 
Then, the transformed generalized vector, $\lambda^a_I$ or $\lambdaB^a_{\sfM}$, also satisfies the null conditions
\begin{align}
\begin{alignedat}{2}
 \lambda^a \,\eta^{\sMA}\, \lambda^b &= \hat{a}^{\sMA}{}_{\sMB}\,\bigl(\hat{\lambda}^a \,\eta^{\sMB}\, \hat{\lambda}^b\bigr) = 0\,, \qquad& 
 \lambda^a \,\Omega \,\lambda^b &= \hat{\lambda}^a \,\Omega \,\hat{\lambda}^b = 0\,,
\\
 \lambdaB^a \,\eta^{\sBA}\, \lambdaB^b &= \hat{b}^{\sBA}{}_{\sBB}\,\bigl(\hat{\lambdaB}^a \,\eta^{\sBB}\, \hat{\lambdaB}^b\bigr) = 0\,, \qquad &
 \lambdaB^a \,\Omega \, \lambdaB^b &= \hat{\lambdaB}^a \,\Omega \,\hat{\lambdaB}^b = 0\,,
\end{alignedat}
\end{align}
which are the required properties for writing the linear section equations. 
In fact, the linear section equations specified by the transformed $\lambda^a$ or $\lambdaB^a$ lead to the quadratic section condition $\eta^{IJ;\,\sMA}\,\partial_I\otimes \partial_J =0$ or $\eta^{\sfM\sfN;\,\sBA}\,\partial_{\sfM}\otimes \partial_{\sfN} =0$\,. 
Indeed, the linear section equations require that the dual components of the transformed derivatives $\partial'_I \equiv (a^{-1})_I{}^J\,\partial_J$ or $\partial'_{\sfM} \equiv (b^{-1})_{\sfM}{}^{\sfN}\,\partial_{\sfN}$ vanish, and from this, we can easily show $\eta^{IJ;\,\sMA}\,\partial'_I\otimes \partial'_J =0$ and $\Omega^{IJ}\,\partial'_I\otimes \partial'_J =0$ or $\eta^{\sfM\sfN;\,\sBA}\,\partial'_{\sfM}\otimes \partial'_{\sfN} =0$ and $\Omega^{\sfM\sfN}\,\partial'_{\sfM}\otimes \partial'_{\sfN} =0$, which are equivalent to the quadratic section condition.

In an example of the $E_{6(6)}$ EFT, the null conditions for the generalized vectors, $(\lambda^a_I)=\bigl(\lambda^a_i,\,\frac{\lambda^{i_1i_2;\,a}}{\sqrt{2!}},\,\frac{\lambda^{i_1\cdots i_5;\,a}}{\sqrt{5!}}\bigr)$, are
\begin{align}
\begin{split}
 &\lambda^a_i\, \lambda^{ki;\,b}+ \lambda^{k i;\,a}\,\lambda^b_i =0 \,, \qquad
 \lambda^a_i\,\lambda^{i k_1\cdots k_4;\,b} + 6\,\lambda^{[k_1k_2|;\,a}\,\lambda^{|k_3k_4];\,b} + \lambda^{i k_1\cdots k_4;\,a}\,\lambda^b_i =0 \,,
\\
 &\lambda^{l[k_1|;\,a}\,\lambda^{|k_2\cdots k_6];\,b} - \lambda^{[k_1\cdots k_5|;\,a}\, \lambda^{|k_6]l;\,b} = 0 \,. 
\end{split}
\label{eq:null-E6}
\end{align}
If $\lambda^a_i$ is invertible, we can choose $\lambda^a_i=\delta^a_i$ and then the first equation requires $\lambda^{ka;\,b}=-\lambda^{kb;\,a}$\,. 
Then, we can express $\lambda^{ij;\,k}$ by using a 3-vector (i.e.~antisymmetric third-rank tensor) $\omega^{ijk}$, $\lambda^{ij;\,k}=-\omega^{ijk}$\,, and the second equation becomes
\begin{align}
 \lambda^{k_1\cdots k_4k_5;\,k_6} + 6\,\omega^{k_5[k_1k_2}\,\omega^{k_3k_4]k_6} + \lambda^{k_6 k_1\cdots k_4;\,k_5} =0 \,. 
\end{align}
This leads to
\begin{align}
 \lambda^{k_1\cdots k_5;\,k_6} = \omega^{k_1\cdots k_6} - 5\, \omega^{[k_1k_2k_3}\,\omega^{k_4k_5] k_6} \,,
\end{align}
where $\omega^{k_1\cdots k_6}$ is an arbitrary 6-vector. 
The last equation in \eqref{eq:null-E6} is trivially satisfied. 
Therefore, the most general parameterization is given by
\begin{align}
 \bigl(\lambda^a_I\bigr) &= \begin{pmatrix}
 \lambda^a_i\\ \frac{\lambda^{i_1i_2;\,a}}{\sqrt{2!}}\\ \frac{\lambda^{i_1\cdots i_5;\,a}}{\sqrt{5!}}
 \end{pmatrix}
 = \begin{pmatrix}
 \delta^a_i\\ -\frac{\omega^{i_1i_2a}}{\sqrt{2!}}\\ \frac{\omega^{i_1\cdots i_5 a} - 5\, \omega^{[i_1i_2i_3}\,\omega^{i_4i_5] a}}{\sqrt{5!}} \end{pmatrix}
\nn\\
 &= \Exp{\frac{1}{6!}\,\omega^{i_1\cdots i_6}\,R_{i_1\cdots i_6}}\,\Exp{\frac{1}{3!}\,\omega^{ijk}\,R_{ijk}} \, \hat{\lambda}^a \,.
\end{align}
In this way, when $\lambda^a_i$ is invertible, the most general parameterization of $\lambda^a$ is obtained from $\hat{\lambda}^a$ via a $U$-duality transformation generated only by negative-root generators $R_{i_1i_2i_3}$ and $R_{i_1\cdots i_6}$ ($\GL(d)$ generators $K^i{}_j$ are not necessary).\footnote{The positive-root generators $R^{i_1i_2i_3}$ and $R^{i_1\cdots i_6}$ do not rotate $\hat{\lambda}^a$.} 
It is also the case for lower exceptional groups $d\leq 5$\,. 
The same will be the case for $E_{7(7)}$ also, and in that case, $\lambda^a$ will be specified by $42\,(=35+7)$ parameters $\omega^{i_1i_2i_3}$ and $\omega^{i_1\cdots i_6}$\,. 
Similarly, in the case of the type IIB section, if $\lambdaB^a_\sfm$ is invertible, the most general parameterization of $\lambdaB^a$ will be given by
\begin{align}
 \lambdaB^a = \Exp{\frac{1}{2!}\,\omega^{\sfm_1\sfm_2}_\alpha\,R^\alpha_{\sfm_1\sfm_2}} \Exp{\frac{1}{4!}\,\omega^{\sfm_1\cdots \sfm_4}\,R_{\sfm_1\cdots \sfm_4}} \Exp{\frac{1}{6!}\,\omega^{\sfm_1\cdots\sfm_6}_\alpha\, R_{\sfm_1\cdots\sfm_6}^\alpha} \,\hat{\lambdaB}^a\,.
\end{align}
In the following, we show that our linear section reproduces the known linear section equation in the $\SL(5)$ EFT, both for the M-theory and the type IIB sections.

\subsection{M-theory section in $\SL(5)$ EFT}

In the $\SL(5)$ EFT, the generalized coordinates $x^I$ ($I=1,\dotsc,10$) are frequently parameterized as $x^I= x^{\sfa\sfb}\,(=x^{[\sfa\sfb]})$ $(\sfa,\sfb=1,\dotsc,5)$. 
In this parameterization, the section condition takes the form \cite{Berman:2010is}
\begin{align}
 \epsilon^{\sfa\sfb\sfc\sfd\sfe}\,\partial_{\sfb\sfc} \otimes \partial_{\sfd\sfe} = 0 \,. 
\label{eq:section-SL5}
\end{align}
On the other hand, the linear section equation is expressed as \cite{Berman:2012vc}
\begin{align}
 \Lambda_{[\sfa}\,\partial_{\sfb\sfc]} =0 \quad (\sfa,\sfb,\sfc=1,\dotsc,5) \,,
\label{eq:linear-section-SL5}
\end{align}
where $\Lambda_{\sfa}$ are arbitrary parameters that specify the section (which is considered to be a generalized notion of the pure spinor that specifies a generalized notion of the Dirac structure \cite{Berman:2012vc}). 
Before comparing this equation with our linear section equation, let us consider the number of independent equations. 
In order for the linear section equation to be meaningful, $\Lambda_{\sfa}$ should not be a zero-vector, and let us suppose $\Lambda_5\neq 0$\,. 
Then, we can decompose the linear section equation as
\begin{align}
 \Lambda_{[i}\,\partial_{jk]} =0\,,\qquad 
 \partial_{ij} = - \frac{2}{\Lambda_5}\, \Lambda_{[i}\,\partial_{j]5} \,. 
\label{eq:SL5-linear-section}
\end{align}
Since the first equation in \eqref{eq:SL5-linear-section} is satisfied when the second equation is satisfied, the second equation is equivalent to the linear section equation (although the $\SL(5)$ covariance is lost). 
Moreover, the linear section \eqref{eq:linear-section-SL5} is sufficient for the section condition \eqref{eq:section-SL5} since \eqref{eq:section-SL5} is automatically satisfied from the second equation.

On the other hand, our linear section equations are given by
\begin{align}
 \lambda^a_I\, \eta^{IJ;\,\sMA}\,\partial_J = 0 \,,
\end{align}
where $\lambda^a_I=(\lambda^a_i,\, \frac{\lambda^{i_1i_2;\,a}}{\sqrt{2}})$ satisfies
\begin{align}
 \lambda^a_I\,\eta^{IJ;\,\sMA}\,\lambda^b_J = 0 \,,
\end{align}
namely
\begin{align}
\begin{split}
 &\lambda^a_I\,\eta^{IJ;\,k}\, \lambda^b_J = \lambda^a_k\,\lambda^{ik;\,b} + \lambda^{ik;\,a}\,\lambda^b_k =0 \,,
\\
 &\lambda^a_I\,\eta^{IJ;\,k_1\cdots k_4}\, \lambda^b_I = \epsilon_{i_1i_2j_1j_2}\,\lambda^{i_1i_2;\,a}\,\lambda^{j_1j_2;\,b} = 0 \,. 
\end{split}
\end{align}
If we consider a case where $\lambda^a_k$ is invertible, we can choose $\lambda^a_k=\delta^a_k$ and the first equation shows $\lambda^{ij;\,a}=-\omega^{ija}$ with $\omega^{ijk}=\omega^{[ijk]}$. 
The second equation is then automatically satisfied since the following identity is satisfied in $d=4$:
\begin{align}
 \epsilon_{k_1\cdots k_4}\,\omega^{k_1k_2 i}\,\omega^{k_3k_4 j} = 0\,.
\end{align}
Therefore, the set of the null vectors becomes
\begin{align}
 \lambda^a =\begin{pmatrix}
 \lambda^a_i \\ \frac{\lambda^{i_1i_2;\,a}}{\sqrt{2}} 
 \end{pmatrix} 
 = \begin{pmatrix}
 \delta^a_i \\ - \frac{\omega^{i_1i_2 a}}{\sqrt{2}} 
 \end{pmatrix}
 = \begin{pmatrix}
 \delta_i^j & 0 \\
 - \frac{\omega^{i_1i_2 j}}{\sqrt{2}} & \delta^{i_1i_2}_{j_1j_2}
 \end{pmatrix} \begin{pmatrix}
 \delta^a_j \\ 0
 \end{pmatrix} . 
\end{align}
The linear section equations $\lambda^a_I\,\eta^{IJ;\,k}\, \partial_J = 0$ and $\lambda^a_I\, \eta^{IJ;\,k_1\cdots k_4}\, \partial_J = 0$ then become
\begin{align}
\begin{split}
 &\lambda^a_k\,\partial^{ik} - \lambda^{ik;\,a}\,\partial_k = \partial^{ia} + \omega^{iak} \,\partial_k = 0 \,,
\\
 &\epsilon_{i_1i_2j_1j_2}\,\lambda^{i_1i_2;\,a}\,\partial^{j_1j_2} = -\epsilon_{i_1i_2j_1j_2}\,\omega^{i_1i_2 a} \,\partial^{j_1j_2} = 0 \,.
\end{split}
\end{align}
The first condition is precisely the second equation in \eqref{eq:SL5-linear-section} if we make the identifications
\begin{align}
 \frac{\Lambda_i}{\Lambda_5} \equiv \frac{1}{3!}\,\epsilon_{i j_1j_2j_3}\,\omega^{j_1j_2j_3}\quad (\Lambda_5\neq 0)\,,\quad 
 \partial_{i5}\equiv -\partial_{5i}\equiv \partial_i\,,\quad 
 \partial_{ij}\equiv \frac{1}{2!}\,\epsilon_{ijkl}\,\partial^{kl}\,. 
\end{align}
The second condition follows from the first. 
In this sense, when $\lambda^a_i$ is invertible, our linear section equations are equivalent to \eqref{eq:linear-section-SL5}. 

For completeness, let us see the number of independent parameters that specify a section. 
The linear section equation \eqref{eq:linear-section-SL5} includes 5 parameters $\Lambda_{\sfa}$, but as we can see from \eqref{eq:SL5-linear-section}, only the 4 ratios $\Lambda_i/\Lambda_5$ specify the section. 
This matches with the number of independent parameters $\omega^{ijk}$ entering in our section equations. 

If we consider a case where $\lambda^a_i$ is not invertible, we may find an inequivalent section. 
For example, when $(\lambda^a_i)=\mathrm{diag}(1,1,0,0)$, $(\lambda^{34;\,3})=1$, and other components vanish ($\lambda^4$ is a zero-vector in this case), the null condition is trivially satisfied, and the linear section equation shows fields can depend only on $x^1$, $x^2$, and $y_{34}$\,. 
This is the well-known type IIB section considered in the next section with a different parameterization of the generalized coordinates. 
Note that if the number of non-vanishing components of $\lambda^a$ is too small, the linear section equation is not sufficient to reproduce the section condition. 

\subsection{Type IIB section in $\SL(5)$ EFT}

The known linear section equation for the IIB section is \cite{Cederwall:IIB,Blair:2017gwn}
\begin{align}
 \Lambda^{\sfa\sfb}\,\partial_{\sfb\sfc}=0 \,,
\label{eq:LSE-IIB-SL5}
\end{align}
where $\Lambda^{\sfa\sfb}$ is defined to satisfy $\epsilon_{\sfe\sfa\sfb\sfc\sfd}\,\Lambda^{\sfa\sfb}\,\Lambda^{\sfc\sfd}=0$\,. 
If $\Lambda^{34}\neq 0$, the condition $\epsilon_{\sfe\sfa\sfb\sfc\sfd}\,\Lambda^{\sfa\sfb}\,\Lambda^{\sfc\sfd}=0$ determines components $\Lambda^{st}$ ($s,t=1,2,5$) as
\begin{align}
 \Lambda^{st} = \frac{2\,\Lambda^{[s|3}\,\Lambda^{|t]4}}{\Lambda^{34}} \,.
\end{align}
Then, the linear section equations become
\begin{align}
 \partial_{s3} = \frac{\Lambda^{t4}}{\Lambda^{34}}\,\partial_{ts} \,,\qquad 
 \partial_{s4} = -\frac{\Lambda^{t3}}{\Lambda^{34}}\,\partial_{ts} \,,\qquad 
 \partial_{34} = \frac{\Lambda^{s3}\,\Lambda^{t4}}{(\Lambda^{34})^2}\, \partial_{st} \,,
\label{sec:LS-E4-IIB}
\end{align}
and from these, we can show the section condition, $\epsilon^{\sfe\sfa\sfb\sfc\sfd}\,\partial_{\sfa\sfb}\,\partial_{\sfc\sfd}=0$. 
In this approach, a section is specified by 6 parameters $\Lambda^{s3}/\Lambda^{34}$ and $\Lambda^{s4}/\Lambda^{34}$. 
In particular, $\Lambda^{s3}/\Lambda^{34}=\Lambda^{s4}/\Lambda^{34}=0$ corresponds to the type IIB section where fields depend on 3 coordinates $\{x^{15},\, x^{25},\, x^{12}\}$. 
The generalized coordinates $x^{\sfa\sfb}$ in the literature are related to our generalized coordinates $(x^I)=\bigl(x^i,\,\frac{y_{i_1i_2}}{\sqrt{2!}}\bigr)$ or $(\sfx^{\sfM})=\bigl(\sfx^{\sfm},\,\sfy_{\sfm}^\alpha,\,\frac{\sfy_{\sfm_1\sfm_2\sfm_3}}{\sqrt{3!}}\bigr)$ as follows, and the coordinates $\{x^{15},\, x^{25},\, x^{12}\}$ correspond to the physical coordinates $\{\sfx^1,\,\sfx^2,\,\sfx^3\}$ in the type IIB parameterization:
\begin{align}
\begin{array}{|c||c|c|c|c|c|c|c|c|c|c|}\hline
 x^{\sfa\sfb} & x^{15} & x^{25} & x^{12}& x^{13} & x^{23} & x^{53} & x^{14} & x^{24} & x^{34} & x^{45} \\ \hline
 \text{M-theory} & x^1 & x^2 & y_{34}& -y_{24} & y_{14} & -x^3 & y_{23} & -y_{13} & y_{12} & x^4 \\ \hline
 \text{type IIB theory} & \sfx^1 & \sfx^2 & \sfx^3 & -\sfy^1_2 & \sfy^1_1 & -\sfy^1_3 & -\sfy^2_2 & \sfy^2_1 & \sfy_{123} & \sfy^2_3 \\ \hline
\end{array}
\end{align}

Our linear section equations are specified by
\begin{align}
\footnotesize
 \lambdaB^a =\begin{pmatrix}
 \lambdaB^a_{\sfm} \\ \lambdaB_\alpha^{\sfm;\,a} \\ \frac{\lambdaB_\alpha^{\sfm_1\sfm_2\sfm_3;\,a}}{\sqrt{3!}} 
 \end{pmatrix} 
 = \begin{pmatrix}
 \delta_{\sfm}^{\sfn} & 0 & 0 \\
 \omega^{\sfm\sfn}_\alpha & \delta_\alpha^\beta\,\delta^{\sfm}_{\sfn} & 0 \\
 -\frac{\frac{3}{2}\,\epsilon^{\gamma\delta}\,\omega^{[\sfm_1\sfm_2}_\gamma\,\omega^{\sfm_3]\sfn}_\delta}{\sqrt{3!}} & \frac{3\,\epsilon^{\beta\gamma}\,\delta^{[\sfm_1}_{\sfn}\,\omega^{\sfm_2\sfm_3]}_\gamma}{\sqrt{3!}} & \delta^{\sfm_1\sfm_2\sfm_3}_{\sfn_1\sfn_2\sfn_3}
 \end{pmatrix} \begin{pmatrix}
 \delta^a_{\sfn} \\ 0 \\ 0 
 \end{pmatrix} = \begin{pmatrix}
 \delta_{\sfm}^a \\ \omega^{\sfm a}_\alpha \\ -\frac{\frac{3}{2}\,\epsilon^{\gamma\delta}\,\omega^{[\sfm_1\sfm_2}_\gamma\,\omega^{\sfm_3] a}_\delta}{\sqrt{3!}}
 \end{pmatrix} ,
\end{align}
which satisfies the null conditions
\begin{align}
 \lambdaB^a_{\sfm} \, \lambdaB_\gamma^{\sfm;\,b} + \lambdaB_\gamma^{\sfm;\,a}\,\lambdaB^b_{\sfm} =0 \,,\qquad 
 \lambdaB^a_{\sfm}\, \lambdaB^{\sfm \sfp_1\sfp_2;\,b} 
 -2!\,\epsilon^{\alpha\beta}\,\lambdaB^{[\sfp_1|;\,a}_\alpha\,\lambdaB^{|\sfp_2];\,b}_\beta
 +\lambdaB^{\sfn \sfp_1\sfp_2;\,a} \,\lambdaB^b_{\sfn} = 0 \,.
\end{align}
By using the explicit form of $\lambdaB^a$, the linear section equations become
\begin{align}
 \partial^{\sfm}_\alpha = -\omega^{\sfm\sfn}_\alpha\,\partial_{\sfn} \,,\qquad 
 \partial^{123} 
 = \epsilon^{\gamma\delta}\,\bigl(\omega^{12}_\gamma\,\omega^{13}_\delta \,\partial_1+\omega^{12}_\gamma\,\omega^{23}_\delta \,\partial_2+\omega^{13}_\gamma\,\omega^{23}_\delta \,\partial_3\bigr) \,.
\end{align}
These are precisely equations \eqref{sec:LS-E4-IIB} if we make the following identifications:
\begin{align}
 \omega_1^{s3} = \frac{\Lambda^{s4}}{\Lambda^{34}}\,,\quad
 \omega_2^{s3} = -\frac{\Lambda^{s3}}{\Lambda^{34}}\,,\quad
 \omega_1^{12} = -\frac{\Lambda^{54}}{\Lambda^{34}}\,,\quad
 \omega_2^{12} = -\frac{\Lambda^{35}}{\Lambda^{34}}\qquad 
 (s=1,2)\,. 
\end{align}
In this sense, our linear section equations are equivalent to the linear section equation \eqref{eq:LSE-IIB-SL5} for the type IIB section in the $\SL(5)$ EFT. 

\section{Conclusions and discussion}
\label{sec:conclusions}

In this paper, we obtained a set of $\eta$-symbols associated with branes in the string multiplet, and reproduced the known $Y$-tensor in $E_{d(d)}$ EFT with $d\leq 7$. 
Our expression does not depend on the $E_{d(d)}$ tensors for a particular $d$, and a reduction to lower $d$ can be easily performed. 
Using the $\eta$-symbols (and the $\Omega$-tensor), we proposed linear section equations that reproduce the usual quadratic section condition. 
Equivalence to the known linear section for the M-theory and the type IIB sections in the $\SL(5)$ EFT are shown. 

\begin{table}[t]
 \begin{tabular}{|c|c|c||c|c|c|}\hline
 brane & tension $\cT$ & ${\text{number of} \atop \text{degeneracy}}$ & brane & tension $\cT$ & ${\text{number of} \atop \text{degeneracy}}$
\\ \hline \hline
 M2 & $\frac{R_i}{\ell_{11}^3}$ & 8 & 
 $1^{(7,1,0)}$ & $\frac{R_{i_1}^4\cdots R_{i_7}^4\,R_{j}^3}{\ell_{11}^{33}}$ & 8 
\\ \hline
 M5 & $\frac{R_{i_1}\cdots R_{i_4}}{\ell_{11}^6}$ & 70 & 
 $1^{(4,4,0)}$ & $\frac{R^4_{i_1}\cdots R^4_{i_4}\,R^3_{j_1}\cdots R_{j_4}^3}{\ell_{11}^{30}}$ & 70 
\\ \hline
 KKM & $\frac{R_{i}^2\,R_{j_1}\cdots R_{j_5}}{\ell_{11}^9}$ & 168 & 
 $1^{(2,5,1)}$ & $\frac{R_{i_1}^4\,R_{i_2}^4\,R_{j_1}^3\cdots R_{j_5}^3\,R_{k}^2}{\ell_{11}^{27}}$ & 168 
\\ \hline
 $7\times$M8 & $\frac{R_{i_1} \cdots R_{i_7}}{\ell_{11}^9}$ & $7\times 8$ & 
 $7\times 1^{(1,7,0)}$ & $\frac{R_i^4\,R_{j_1}^3 \cdots R_{j_7}^3}{\ell_{11}^{27}}$ & $7\times 8$ 
\\ \hline
 $5^3$ & $\frac{R^2_{i_1}\,R^2_{i_2}\,R^2_{i_3}\,R_{j_1}\cdots R_{j_4}}{\ell_{11}^{12}}$ & 280 & 
 $1^{(1,4,3)}$ & $\frac{R_i^4\,R_{j_1}^3\cdots R_{j_4}^3\,R^2_{k_1}\,R^2_{k_2}\,R^2_{k_3}}{\ell_{11}^{24}}$ & 280 
\\ \hline
 $8^{(1,0)}$ & $\frac{R^3_{i}\,R_{j_1} \cdots R_{j_7}}{\ell_{11}^{12}}$ & 8 & 
 $2^{(7,0)}$ & $\frac{R^3_{i_1}\cdots R_{i_7}^3\,R_j}{\ell_{11}^{24}}$ & 8 
\\ \hline
 $7\times 7^2$ & $\frac{R^2_{i_1}\,R^2_{i_2}\,R_{j_1} \cdots R_{j_6}}{\ell_{11}^{12}}$ & $7\times 28$ & 
 $7\times 1^{(6,2)}$ & $\frac{R_{i_1}^3 \cdots R_{i_6}^3\,R^2_{j_1}\,R^2_{j_2}}{\ell_{11}^{24}}$ & $7\times 28$ 
\\ \hline
 $2^6$ & $\frac{R^2_{i_1}\cdots R^2_{i_6}\,R_{j}}{\ell_{11}^{15}}$ & 56 & 
 $1^{(1,1,6)}$ & $\frac{R_{i}^4\,R_{j}^3\,R^2_{k_1}\cdots R^2_{k_6}}{\ell_{11}^{21}}$ & 56 
\\ \hline
 $5^{(1,3)}$ & $\frac{R^3_{i}\,R^2_{j_1}\,R^2_{j_2}\,R^2_{j_3}\,R_{k_1}\cdots R_{k_4}}{\ell_{11}^{15}}$ & 280 & 
 $2^{(4,3)}$ & $\frac{R_{i_1}^3\cdots R_{i_4}^3\,R^2_{j_1}\,R^2_{j_2}\,R^2_{j_3}\,R_{k}}{\ell_{11}^{21}}$ & 280 
\\ \hline
 $7\times 4^5$ & $\frac{R^2_{i_1}\cdots R^2_{i_5}\,R_{j_1}\,R_{j_2}\,R_{j_3}}{\ell_{11}^{15}}$ & $7\times 56$ & 
 $7\times 1^{(3,5)}$ & $\frac{R_{i_1}^3\,R_{i_2}^3\,R_{i_3}^3\,R^2_{j_1}\cdots R^2_{j_5}}{\ell_{11}^{21}}$ & $7\times 56$ 
\\ \hline
 $3^{(2,4)}$ & $\frac{R^3_{i_1}\,R^3_{i_2}\, R^2_{j_1}\cdots R^2_{j_4}\,R_{k_1}\,R_{k_2}}{\ell_{11}^{18}}$ & 420 &&&
\\ \hline
 $7\times 2^{(1,6)}$ & $\frac{R^3_{i}\, R^2_{j_1}\cdots R^2_{j_6}\,R_{k}}{\ell_{11}^{18}}$ & $7\times 56$ &&&
\\ \hline
 $35\times 1^8$ & $\frac{R^2_{i_1}\cdots R^2_{i_8}}{\ell_{11}^{18}}$ & $35\times 1$ &&& \\ \hline
 \end{tabular}
\caption{M-theory branes in the string multiplet for the $E_{8(8)}$ EFT. In each column, all of the indices $\{i_1,\dotsc, i_p, j_1,\dotsc,j_q,k_1,\dotsc,k_r\}$ must be different. Branes on the left and right column are dual to each other \cite{Obers:1998fb}.}
\label{table:E8-string}
\end{table}

Our considerations are limited to $d\leq 7$, but we can also consider the $E_{8(8)}$ EFT, where the number of $\eta$-symbols will be the same as the dimension of the $R_2$-representation of $E_{8(8)}$, namely $3875$. 
According to \cite{Obers:1998fb}, the branes in the string multiplet can be summarized as in Table \ref{table:E8-string}. 
There, each brane in the table is wrapping a certain cycle in the 8-torus $T^8$ and behaves as a string with a tension $\cT$ in the uncompactified spacetime. 
We call the brane a ``$b^{(c,d,e)}$-brane'' if the tension of the string takes the form
\begin{align}
 \cT = \frac{(R_{i_1}\cdots R_{i_c})^4\,(R_{j_1}\cdots R_{j_d})^3\,(R_{k_1}\cdots R_{k_e})^2\,R_{l_1}\cdots R_{l_{b-1}}}{\ell_{11}^{b+4c+3d+2e+1}}\,,
\end{align}
where $R_i$ denotes the radius along the $x^i$-direction ($i=1,\dotsc,8$), and $\ell_{11}$ is the 11-dimensional Planck length. 
We also define $b^{(d,e)}\equiv b^{(0,d,e)}$ and $b^{e}\equiv b^{(0,e)}$. 
It will be interesting to determine all of the $\eta$-symbols associated with the $3875$ branes. 

In this paper, we have not discussed the role of the $\eta$-symbols in worldvolume theories in detail, but in fact, they play an important role. 
In the $T$-duality manifest formulation of the string, the equations of motion can be expressed as the self-duality relation \cite{Duff:1989tf},
\begin{align}
 \cM_{IJ}\, *_\gamma \mathcal{P}^J = \eta_{IJ}\, \mathcal{P}^J \,,
\end{align}
where $*_\gamma$ is the Hodge star operator on the worldsheet associated with the metric $\gamma$, and
\begin{align}
 (\cM_{IJ})\equiv \begin{pmatrix}
 G_{ij}-B_{ik}\,G^{kl}\,B_{lj} & B_{ik}\,G^{kj} \cr
 -G^{ik}\,B_{kj} & G^{ij}
 \end{pmatrix}, \qquad 
 (\mathcal{P}^I) \equiv \begin{pmatrix}
 \rmd X^i \cr
 \rmd \tilde{X}_i
 \end{pmatrix} . 
\end{align}
As a generalization of this relation, if we consider a membrane theory in the approach of \cite{Sakatani:2016sko}, the equations of motion can be expressed as
\begin{align}
 \cM_{IJ}\,*_\gamma \mathcal{P}^J = \eta^{\text{\tiny(M2)}}_{IJ} \wedge \mathcal{P}^J \,,\qquad 
 \eta^{\text{\tiny(M2)}}_{IJ}\equiv \frac{1}{2}\,\eta_{IJ;\,k}\,\rmd X^k\,,
\end{align}
by using the $\eta$-symbol $\eta_k$ associated with an M2-brane. 
It will be interesting to see whether this kind of self-duality relation is satisfied for all of the branes in the string multiplet. 
It is also interesting to see how the $\Omega$-tensor appears in the brane worldvolume theories. 

\appendix

\section{Conventions and formulas}
\label{app:conventions}

\subsection{Combinatoric factors}

We shall use the following convention for multiple indices. 
When we consider M-theory, the generalized vector is parameterized as
\begin{align}
 (V^I) = \Bigl(v^i,\, \frac{v_{i_1i_2}}{\sqrt{2!}} ,\, \frac{v_{i_1\cdots i_5}}{\sqrt{5!}} ,\, \frac{v_{i_1\cdots i_7,\,j}}{\sqrt{7!}} \Bigr) \,,
\qquad
 (W_I) = \Bigl(w_i,\, \frac{w^{i_1i_2}}{\sqrt{2!}} ,\, \frac{w^{i_1\cdots i_5}}{\sqrt{5!}} ,\, \frac{w^{i_1\cdots i_7,\,j}}{\sqrt{7!}} \Bigr) \,.
\end{align}
The combinatoric factors are introduced such that the indices are summed with weight 1 when we consider the ordered multiple indices $\overline{i_1\cdots i_p}$, which satisfy $i_1<\cdots <i_p$. 
For example, the inner product between $V^I$ and $W_I$ becomes
\begin{align}
 V^I\,W_I &= v^i\,w_i + \frac{1}{2!}\, v_{i_1i_2}\,w^{i_1i_2} + \frac{1}{5!}\,v_{i_1\cdots i_5}\,w^{i_1\cdots i_5}
           + \frac{1}{7!}\,v_{i_1\cdots i_7,\,j}\,w^{i_1\cdots i_7,\,j} 
\nn\\
 &= v^i\,w_i + v_{\overline{i_1i_2}}\,w^{\overline{i_1i_2}} + v_{\overline{i_1\cdots i_5}}\,w^{\overline{i_1\cdots i_5}}
            + v_{\overline{i_1\cdots i_7},\,j}\,w^{\overline{i_1\cdots i_7},\,j} \,,
\end{align}
and in the second line, all components are summed with weight 1. 
Similarly, the generalized coordinates and derivatives are defined as
\begin{alignat}{2}
 (x^I) &= \Bigl( x^i ,\, \frac{y_{i_1i_2}}{\sqrt{2!}} ,\, \frac{y_{i_1\cdots i_5}}{\sqrt{5!}} ,\, \frac{y_{i_1\cdots i_7,\,j}}{\sqrt{7!}} \Bigr) \,, 
\quad
 &(\partial_I) &= \Bigl( \partial_i ,\, \frac{\partial^{i_1i_2}}{\sqrt{2!}} ,\, \frac{\partial^{i_1\cdots i_5}}{\sqrt{5!}} ,\, \frac{\partial^{i_1\cdots i_7,\,j}}{\sqrt{7!}} \Bigr) \,,
\\
 (x^{\bar{I}}) &= \bigl( x^i ,\, y_{\overline{i_1i_2}} ,\, y_{\overline{i_1\cdots i_5}} ,\, y_{\overline{i_1\cdots i_7},\,j} \bigr) \,,\quad 
 &(\partial_{\bar{I}}) &= \bigl( \partial_i ,\, \partial^{\overline{i_1i_2}} ,\, \partial^{\overline{i_1\cdots i_5}} ,\, \partial^{\overline{i_1\cdots i_7},\,j} \bigr) \,. 
\end{alignat}
We define the derivative as
\begin{align}
 \partial^{i_1\cdots i_p}y_{j_1\cdots j_p}=\delta^{i_1\cdots i_p}_{j_1\cdots j_p}\equiv \delta^{[i_1}_{j_1}\cdots \delta^{i_p]}_{j_p}\,,\qquad 
 \partial^{\overline{i_1\cdots i_p}}y_{\overline{j_1\cdots j_p}}=\delta^{\overline{i_1\cdots i_p}}_{\overline{j_1\cdots j_p}}\equiv p!\,\delta^{i_1\cdots i_p}_{j_1\cdots j_p}\,,
\end{align}
which gives, e.g., $\partial^{12}y_{12}=1/2$ and $\partial^{\overline{12}}y_{\overline{12}}=1$\,. 
If we define the Kronecker delta as
\begin{align}
 (\delta_I^J) = \begin{pmatrix} \delta_i^j & 0 & 0 & 0 \\ 0 & \delta^{i_1i_2}_{j_1j_2} & 0 & 0 \\ 0 & 0 & \delta^{i_1\cdots i_5}_{j_1\cdots j_5} & 0 \\ 0 & 0 & 0 & \delta^{i_1\cdots i_7}_{j_1\cdots j_7}\,\delta^i_j
 \end{pmatrix} ,\quad
 (\delta_{\bar{I}}^{\bar{J}}) = \begin{pmatrix} \delta_i^j & 0 & 0 & 0 \\ 0 & \delta^{\overline{i_1i_2}}_{\overline{j_1j_2}} & 0 & 0 \\ 0 & 0 & \delta^{\overline{i_1\cdots i_5}}_{\overline{j_1\cdots j_5}} & 0 \\ 0 & 0 & 0 & \delta^{\overline{i_1\cdots i_7}}_{\overline{j_1\cdots j_7}}\,\delta^i_j
 \end{pmatrix} ,
\end{align}
they satisfy 
\begin{align}
 \partial_I x^J=\delta_I^J\,,\qquad \partial_{\bar{I}} x^{\bar{J}}=\delta_{\bar{I}}^{\bar{J}}\,,\qquad 
 \delta_I^I = \delta_{\bar{I}}^{\bar{I}} = D \,.
\end{align}
If we use the ordered indices, e.g., the matrix $\eta^{k_1\cdots k_4}$ has a simpler form. 
Indeed, the complicated numerical factors disappear:
\begin{align}
 \eta^{\bar{I}\bar{J};\,\overline{k_1\cdots k_4}} 
 &\equiv \begin{pmatrix}
 0 & 0 & \delta^{\overline{i k_1\cdots k_4}}_{\overline{j_1\cdots j_5}} & 0 \\
 0 & \delta^{\overline{k_1\cdots k_4}}_{\overline{i_1i_2j_1j_2}} & 0 & 0 \\
 \delta^{\overline{j k_1\cdots k_4}}_{\overline{i_1\cdots i_5}} & 0 & 0 & 0 \\
 0 & 0 & 0 & 0
 \end{pmatrix} .
\end{align}
In fact, all of the $\eta$-symbols except those associated with KKM and 8-branes (or KKM and $7_2$-branes in the type IIB side) have a simple form without complicated numerical factors. 
If we stick to the unordered multiple indices, as in the main text, the rule for the numerical factor is as follows: 
For a $\{i_1\cdots i_p,\,k_1\cdots k_q\}$--$\{j_1\cdots j_r,\,l_1\cdots l_s\}$ component of $\eta^{\sMA}$, we introduce $\frac{1}{\sqrt{p!\,q!\,r!\,s!}}$ (where $q$ or $s$ may be 0). 
For each $\delta^{i_1\cdots i_p}_{j_1\cdots j_p}$ inside $\eta^{\sMA}$, we introduce $p!$. 
If there are contractions of multiple indices in the Kronecker deltas like $\delta^{..i_1\cdots i_p}_{.......}\,\delta^{.......}_{..i_1\cdots i_p}$, we additionally introduce $1/p!$. 
This rule reproduces (almost) all of the numerical factors in $\eta^{\sMA}$ and $\Omega$.

\subsection{$E_{d(d)}$ group}
\label{app:Edd}

The simple roots of the $E_{d(d)}$ group are denoted by $\alpha_n$ ($n=1,\dotsc,d$) and their relation is shown in the following Dynkin diagram:
$$
 \xygraph{
    *{R_1}*\cir<8pt>{} ([]!{+(0,-.4)} {\alpha_1}) - [r]
    *\cir<8pt>{} ([]!{+(0,-.4)} {\alpha_2}) - [r]
    \cdots ([]!{+(0,-.4)} {}) - [r]
    *\cir<8pt>{} ([]!{+(0,-.4)} {\alpha_{d-4}}) - [r]
    *\cir<8pt>{} ([]!{+(0,-.4)} {\alpha_{d-3}})
(
        - [u] *\cir<8pt>{} ([]!{+(.5,0)} {\alpha_{d}}),
        - [r] *\cir<8pt>{} ([]!{+(0,-.4)} {\alpha_{d-2}})
        - [r] *{R_2}*\cir<8pt>{} ([]!{+(0,-.4)} {\alpha_{d-1}})
)}
$$
In this convention, the $R_1$-/$R_2$-representations are defined by the following Dynkin labels:
\begin{align}
\begin{split}
 &\text{$R_1$-representation (particle multiplet):} \quad (1,0,\dotsc,0) \,,
\\
 &\text{$R_2$-representation (string multiplet):} \quad (0,\dotsc,0,1,0) \,.
\end{split}
\end{align}

The generators of the $E_{d(d)}$ group $(d\leq 7)$ can be parameterized in two different ways, depending on whether we are considering M-theory or type IIB theory \cite{West:2003fc,Berman:2011jh,Tumanov:2014pfa,Lee:2016qwn}:
\begin{align}
\begin{split}
 \text{M-theory:}\quad &\{K^i{}_j ,\, R^{i_1i_2i_3},\,R_{i_1i_2i_3},\, R^{i_1\cdots i_6},\,R_{i_1\cdots i_6}\}\,, 
\\
 \text{Type IIB:}\quad &\{K^{\sfm}{}_{\sfn} ,\, R_{\alpha\beta},\,R_\alpha^{\sfm_1\sfm_2},\,R^\alpha_{\sfm_1\sfm_2},\, R^{\sfm_1\cdots \sfm_4},\, R_{\sfm_1\cdots \sfm_4},\,R^{\sfm_1\cdots\sfm_6}_\alpha,\,R_{\sfm_1\cdots\sfm_6}^\alpha\}\,, 
\end{split}
\end{align}
where $i,j=1,\dotsc,d$, $\sfm,\sfn=1,\dotsc,d-1$, and $\alpha,\beta=1,2$\,. 
By considering $R_{\alpha\beta}=R_{(\alpha\beta)}$, the number of the above generators is the same as the dimension of the $E_{d(d)}$ group $(d\leq 7)$. 

In the M-theory parameterization, the explicit forms of the generators are given by
\begin{align}
 &(K^{k_1}{}_{k_2})_I{}^J \equiv {\footnotesize
 \begin{pmatrix}
 \delta^{k_1}_i\,\delta^j_{k_2} & 0 & 0 & 0 \\
 0 & -\frac{2!\,2!\,\delta^{i_1i_2}_{k_2l}\,\delta^{k_1l}_{j_1j_2}}{\sqrt{2!\,2!}} & 0 & 0 \\
 0 & 0 & -\frac{5!\,5!\,\delta^{i_1\cdots i_5}_{k_2l_1\cdots l_4}\,\delta^{k_1l_1\cdots l_4}_{j_1\cdots j_5}}{4!\sqrt{5!\,5!}} & 0 \\
 0 & 0 & 0 & -\frac{2\cdot 7!\,\delta^{i_1\cdots i_7}_{j_1\cdots j_7}\,\delta^{(i}_j\,\delta^{k_1)}_{k_2}}{\sqrt{7!\,7!}}
 \end{pmatrix} + \frac{\delta^{k_1}_{k_2}}{9-d}}\,\delta_I^J \,, 
\\
 &(R^{k_1k_2k_3})_I{}^J \equiv {\footnotesize
 \begin{pmatrix}
 0 & -\frac{3!\,\delta_{i j_1j_2}^{k_1k_2k_3}}{\sqrt{2!}} & 0 & 0 \\
 0 & 0 & \frac{5!\, \delta^{i_1i_2 k_1k_2k_3}_{j_1\cdots j_5}}{\sqrt{2!\,5!}} & 0 \\
 0 & 0 & 0 & \frac{7!\,3!\,\delta^{i_1\cdots i_5 l_1l_2}_{j_1\cdots j_7}\,\delta_{l_1l_2j}^{k_1k_2k_3}}{2!\sqrt{5!\,7!}} \\
 0 & 0 & 0 & 0
 \end{pmatrix} }\,, 
\\
 &(R_{k_1k_2k_3})_I{}^J \equiv {\footnotesize
 \begin{pmatrix}
 0 & 0 & 0 & 0 \\
 -\frac{3!\,\delta^{i_1i_2 j}_{k_1k_2k_3}}{\sqrt{2!}} & 0 & 0 & 0 \\
 0 & \frac{5!\, \delta_{j_1j_2 k_1k_2k_3}^{i_1\cdots i_5}}{\sqrt{2!\,5!}} & 0 & 0 \\
 0 & 0 & \frac{7!\,3!\,\delta_{j_1\cdots j_5 l_1l_2}^{i_1\cdots i_7}\,\delta^{l_1l_2i}_{k_1k_2k_3}}{2!\sqrt{5!\,7!}} & 0
 \end{pmatrix} }\,, 
\\
 &(R^{k_1\cdots k_6})_I{}^J \equiv {\footnotesize
 \begin{pmatrix}
 0 & 0 & -\frac{6!\,\delta_{i j_1\cdots j_5}^{k_1\cdots k_6}}{\sqrt{5!}} & 0 \\
 0 & 0 & 0 & \frac{7!\,6!\,\delta^{i_1i_2 l_1\cdots l_5}_{j_1\cdots j_7}\,\delta_{l_1\cdots l_5j}^{k_1\cdots k_6}}{5!\sqrt{2!\,7!}} \\
 0 & 0 & 0 & 0 \\
 0 & 0 & 0 & 0
 \end{pmatrix} }\,,
\\
 &(R_{k_1\cdots k_6})_I{}^J \equiv {\footnotesize
 \begin{pmatrix}
 0 & 0 & 0 & 0 \\
 0 & 0 & 0 & 0 \\
 \frac{6!\,\delta^{i_1\cdots i_5 j}_{k_1\cdots k_6}}{\sqrt{5!}} & 0 & 0 & 0\\
 0 & \frac{7!\,6!\,\delta_{j_1j_2 l_1\cdots l_5}^{i_1\cdots i_7}\,\delta^{l_1\cdots l_5i}_{k_1\cdots k_6}}{5!\sqrt{2!\,7!}} & 0 & 0
 \end{pmatrix} }\,.
\end{align}

On the other hand, in the type IIB parameterization, the explicit forms of the generators are given by
\begin{align}
 &(K^{\sfp_1}{}_{\sfp_2})_{\sfM}{}^{\sfN} \equiv {\arraycolsep=-1mm {\footnotesize\left(\begin{array}{ccccc} 
 \delta^{\sfp_1}_{\sfm}\,\delta^{\sfn}_{\sfp_2} & 0 & 0 & 0 & 0 \\
 0 & -\delta_\alpha^\beta\,\delta^{\sfm}_{\sfp_2} \,\delta^{\sfp_1}_{\sfn} & 0 & 0 & 0 \\
 0 & 0 & -\frac{3!\,3!\,\delta^{\sfm_1\sfm_2\sfm_3}_{\sfp_2 \sfq_1\sfq_2} \,\delta^{\sfp_1\sfq_1\sfq_2}_{\sfn_1\sfn_2\sfn_3}}{2!\sqrt{3!\,3!}} & 0 & 0 \\
 0 & 0 & 0 & -\frac{5!\,5!\,\delta_\alpha^\beta\,\delta^{\sfm_1\cdots \sfm_5}_{\sfp_2 \sfq_1\cdots \sfq_4} \,\delta^{\sfp_1\sfq_1\cdots\sfq_4}_{\sfn_1\cdots\sfn_5}}{4!\sqrt{5!\,5!}} & 0 \\
 0 & 0 & 0 & 0 & - \frac{2\cdot 6!\,\delta^{\sfm_1\cdots\sfm_6}_{\sfn_1\cdots\sfn_6}\,\delta^{(\sfm}_{\sfp_2}\,\delta^{\sfp_1)}_{\sfn}}{\sqrt{6!\,6!}}
 \end{array}\right)}} + \frac{\delta^{\sfp_1}_{\sfp_2}}{9-d}\,\delta_{\sfM}^{\sfN} \,,
\\
 &(R_{\gamma\delta})_{\sfM}{}^{\sfN} \equiv {\arraycolsep=0.5mm {\footnotesize\left(\begin{array}{ccccc} 
 ~0~ & 0 & ~0~ & 0 & ~0~ \\
 0 & \epsilon_{\alpha(\gamma}\,\delta^\beta_{\delta)}\,\delta^{\sfm}_{\sfn} & 0 & 0 & 0 \\
 0 & 0 & 0 & 0 & 0 \\
 0 & 0 & 0 & \epsilon_{\alpha(\gamma}\,\delta^\beta_{\delta)}\,\delta^{\sfm_1\cdots \sfm_5}_{\sfn_1\cdots \sfn_5} & 0 \\
 0 & 0 & 0 & 0 & 0 
 \end{array}\right)}} ,
\\
 &(R^{\sfp_1\sfp_2}_\gamma)_{\sfM}{}^{\sfN} \equiv {\arraycolsep=0.5mm {\footnotesize\left(\begin{array}{ccccc} 
 ~0~ & -2!\,\delta^\alpha_\gamma\,\delta_{\sfm\sfn}^{\sfp_1\sfp_2} & 0 & 0 & 0 \\
 0 & 0 & \frac{3!\,\epsilon_{\alpha\gamma} \,\delta_{\sfn_1\sfn_2\sfn_3}^{\sfm \sfp_1\sfp_2}}{\sqrt{3!}} & 0 & 0 \\
 0 & 0 & 0 & \frac{5!\,\delta^\beta_\gamma\,\delta_{\sfn_1\cdots \sfn_5}^{\sfm_1\sfm_2\sfm_3\sfp_1\sfp_2}}{\sqrt{3!\,5!}} & 0 \\
 0 & 0 & 0 & 0 & -\frac{6!\,2!\,\epsilon_{\alpha\gamma} \,\delta_{\sfn_1\cdots \sfn_6}^{\sfm_1\cdots \sfm_5\sfq}\,\delta_{\sfq\sfn}^{\sfp_1\sfp_2}}{\sqrt{5!\,6!}} \\
 0 & 0 & 0 & 0 & 0 
 \end{array}\right)}} ,
\\
 &(R_{\sfp_1\sfp_2}^\gamma)_{\sfM}{}^{\sfN} \equiv {\arraycolsep=0.5mm {\footnotesize\left(\begin{array}{ccccc} 
 0 & 0 & 0 & 0 & ~0~ \\
 2!\,\delta_\alpha^\gamma\,\delta^{\sfm\sfn}_{\sfp_1\sfp_2} & 0 & 0 & 0 & 0 \\
 0 & \frac{3!\,\epsilon^{\beta\gamma} \,\delta^{\sfm_1\sfm_2\sfm_3}_{\sfn \sfp_1\sfp_2}}{\sqrt{3!}} & 0 & 0 & 0 \\
 0 & 0 & \frac{5!\,\delta_\alpha^\gamma\,\delta^{\sfm_1\cdots \sfm_5}_{\sfn_1\sfn_2\sfn_3\sfp_1\sfp_2}}{\sqrt{3!\,5!}} & 0 & 0 \\
 0 & 0 & 0 & -\frac{6!\,2!\,\epsilon^{\beta\gamma} \,\delta^{\sfm_1\cdots \sfm_6}_{\sfn_1\cdots \sfn_5\sfq}\,\delta^{\sfq\sfm}_{\sfp_1\sfp_2}}{\sqrt{5!\,6!}} & 0 
 \end{array}\right)}} ,
\\
 &(R^{\sfp_1\cdots\sfp_4})_{\sfM}{}^{\sfN} \equiv {\arraycolsep=0.5mm {\footnotesize\left(\begin{array}{ccccc} 
 ~0~ & ~0~ & -\frac{4!\,\delta_{\sfm \sfn_1\sfn_2\sfn_3}^{\sfp_1\cdots\sfp_4}}{\sqrt{3!}} & 0 & 0 \\
 0 & 0 & 0 & -\frac{5!\,\delta^\beta_\alpha \,\delta_{\sfn_1\cdots\sfn_5}^{\sfm\sfp_1\cdots\sfp_4}}{\sqrt{5!}} & 0 \\
 0 & 0 & 0 & 0 & -\frac{6!\,4!\,\delta_{\sfn_1\cdots \sfn_6}^{\sfm_1\sfm_2\sfm_3\sfq_1\sfq_2\sfq_3}\,\delta_{\sfq_1\sfq_2\sfq_3\sfn}^{\sfp_1\cdots \sfp_4}}{3!\sqrt{3!\,6!}} \\
 0 & 0 & 0 & 0 & 0 \\
 0 & 0 & 0 & 0 & 0 
 \end{array}\right)}} , 
\\
 &(R_{\sfp_1\cdots\sfp_4})_{\sfM}{}^{\sfN} \equiv {\arraycolsep=0.5mm {\footnotesize\left(\begin{array}{ccccc} 
 0 & 0 & 0 & ~0~ & ~0~ \\
 0 & 0 & 0 & 0 & 0 \\
 \frac{4!\,\delta^{\sfm_1\sfm_2\sfm_3 \sfn}_{\sfp_1\cdots\sfp_4}}{\sqrt{3!}} & 0 & 0 & 0 & 0 \\
 0 & -\frac{5!\,\delta_\alpha^\beta \,\delta^{\sfm_1\cdots\sfm_5}_{\sfn\sfp_1\cdots\sfp_4}}{\sqrt{5!}} & 0 & 0 & 0 \\
 0 & 0 & -\frac{6!\,4!\,\delta^{\sfm_1\cdots \sfm_6}_{\sfn_1\sfn_2\sfn_3\sfq_1\sfq_2\sfq_3}\,\delta^{\sfq_1\sfq_2\sfq_3\sfm}_{\sfp_1\cdots \sfp_4}}{3!\sqrt{3!\,6!}} & 0 & 0 
 \end{array}\right)}} , 
\\
 &(R^{\sfp_1\cdots\sfp_6}_\gamma)_{\sfM}{}^{\sfN} \equiv {\arraycolsep=0.5mm {\footnotesize\left(\begin{array}{ccccc} 
 ~0~ & ~0~ & ~0~ & -\frac{6!\,\delta^\beta_\gamma\,\delta_{\sfm\sfn_1\cdots \sfn_5}^{\sfp_1\cdots\sfp_6}}{\sqrt{5!}} & 0 \\
 0 & 0 & 0 & 0 & \frac{6!\,6!\,\epsilon_{\alpha\gamma}\,\delta_{\sfn_1\cdots \sfn_6}^{\sfm \sfq_1\cdots \sfq_5}\, \delta_{\sfq_1\cdots \sfq_5\sfn}^{\sfp_1\cdots \sfp_6}}{5!\sqrt{6!}} \\
 0 & 0 & 0 & 0 & 0 \\
 0 & 0 & 0 & 0 & 0 \\
 0 & 0 & 0 & 0 & 0
 \end{array}\right)}} ,
\\
 &(R_{\sfp_1\cdots\sfp_6}^\gamma)_{\sfM}{}^{\sfN} \equiv {\arraycolsep=0.5mm {\footnotesize\left(\begin{array}{ccccc} 
 0 & 0 & ~0~ & ~0~ & ~0~ \\
 0 & 0 & 0 & 0 & 0 \\
 0 & 0 & 0 & 0 & 0 \\
 \frac{6!\,\delta_\alpha^\gamma\,\delta^{\sfm_1\cdots \sfm_5 \sfn}_{\sfp_1\cdots\sfp_6}}{\sqrt{5!}} & 0 & 0 & 0 & 0 \\
 0 & \frac{6!\,6!\,\epsilon^{\beta\gamma}\,\delta^{\sfm_1\cdots \sfm_6}_{\sfn \sfq_1\cdots \sfq_5}\, \delta^{\sfq_1\cdots \sfq_5\sfm}_{\sfp_1\cdots \sfp_6}}{5!\sqrt{6!}} & 0 & 0 & 0
 \end{array}\right)}} .
\end{align}

\section{Comparison with known $Y$-tensors}
\label{app:eta-comparison}

In this appendix, we reproduce known $Y$-tensors from our result. 

\subsection{$Y$-tensor in $\SL(5)$ EFT}

In the $\SL(5)$ EFT, we have 5 non-vanishing $\eta$-symbols, which can be redefined as
\begin{align}
 \epsilon^k \equiv \eta^k = \begin{pmatrix}
 0 & \frac{2!\,\delta^{k i}_{j_1j_2}}{\sqrt{2!}} \\
 \frac{2!\,\delta^{k j}_{i_1i_2}}{\sqrt{2!}} & 0
 \end{pmatrix} , \qquad 
 \epsilon^5\equiv \frac{1}{4!}\,\epsilon_{k_1\cdots k_4}\,\eta^{k_1\cdots k_4} = \begin{pmatrix}
 0 & 0 \\
 0 & \frac{\epsilon_{i_1i_2j_1j_2}}{\sqrt{2!\,2!}} 
 \end{pmatrix} . 
\end{align}
If we redefine the coordinates as $(x^I)=\bigl(x^i,\,\frac{x^{i_1i_2}}{\sqrt{2!}}\bigr)$ with $x^{i_1i_2}\equiv \frac{1}{2!}\,\epsilon^{i_1i_2j_1j_2}\,y_{j_1j_2}$\,, they become
\begin{align}
 \epsilon^k = \begin{pmatrix}
 0 & \frac{\epsilon^{ki j_1j_2}}{\sqrt{2!}} \\
 \frac{\epsilon^{kj i_1i_2}}{\sqrt{2!}} & 0
 \end{pmatrix} , \qquad 
 \epsilon^5 = \begin{pmatrix}
 0 & 0 \\
 0 & \frac{\epsilon^{i_1i_2j_1j_2}}{\sqrt{2!\,2!}}
 \end{pmatrix} . 
\end{align}
These can be neatly summarized as follows by introducing indices $\sfa,\sfb,\sfc=1,\dotsc,5$ and a totally antisymmetric tensor $\epsilon^{\sfa_1\cdots \sfa_5}$ satisfying $\epsilon^{i_1\cdots i_45}= \epsilon^{i_1\cdots i_4}$\,:
\begin{align}
 (\epsilon^{\sfc}) = (\epsilon^k,\,\epsilon^5)\,,\qquad 
 \epsilon^{\sfc} = \begin{pmatrix}
 0 & \frac{\epsilon^{\sfc i5 j_1j_2}}{\sqrt{2!}} \\
 \frac{\epsilon^{\sfc j5 i_1i_2}}{\sqrt{2!}} & \frac{\epsilon^{\sfc i_1i_2j_1j_2}}{\sqrt{2!\,2!}}
 \end{pmatrix} . 
\end{align}
By further using the conventional parameterization $x^I\equiv x^{\sfa_1\sfa_2}$ ($x^{i5}\equiv x^i$), they become
\begin{align}
 \epsilon^{\sfc} = \bigl(\epsilon^{\sfc\,IJ} \bigr) = \biggl(\frac{\epsilon^{\sfc\sfa_1\sfa_2\sfb_1\sfb_2}}{\sqrt{2!\,2!}} \biggr) \,. 
\end{align}
We also define $\epsilon_{\sfc} = \bigl(\epsilon_{\sfc\,IJ} \bigr) = \Bigl(\frac{\epsilon_{\sfc\sfa_1\sfa_2\sfb_1\sfb_2}}{\sqrt{2!\,2!}} \Bigr)$\,, and then $Y^{IJ}_{KL}$ becomes
\begin{align}
 Y^{IJ}_{KL} = \frac{\epsilon^{\sfe\sfa_1\sfa_2\sfb_1\sfb_2}}{\sqrt{2!\,2!}}\,\frac{\epsilon_{\sfe\sfc_1\sfc_2\sfd_1\sfd_2}}{\sqrt{2!\,2!}} \,,
\end{align}
which is summarized as $Y^{IJ}_{KL}=\epsilon^{\sfe IJ}\,\epsilon_{\sfe KL}$ in \eqref{eq:Y-tensor-case-by-case}. 

\subsection{$Y$-tensor in $\SO(5,5)$ EFT}
\label{app:Y-E5}

In the $\SO(5,5)$ EFT, we have 10 $\eta$-symbols, which can be redefined as
\begin{align}
 \gamma^k&\equiv \sqrt{2}\,\eta^k = \sqrt{2}\,\begin{pmatrix}
 0 & \frac{2!\,\delta^{k i}_{j_1j_2}}{\sqrt{2!}} & 0 \\
 \frac{2!\,\delta^{k j}_{i_1i_2}}{\sqrt{2!}} & 0 & 0 \\
 0 & 0 & 0 
 \end{pmatrix} , 
\\
 \gamma_k&\equiv \sqrt{2}\,\frac{\epsilon_{kk_1\cdots k_4}\,\eta^{k_1\cdots k_4}}{4!} = \sqrt{2}\,\begin{pmatrix}
 0 & 0 & \frac{5\,\delta^i_{[j_1}\,\epsilon_{j_2 \cdots j_5]k}}{\sqrt{5!}} \\
 0 & \frac{\epsilon_{k i_1i_2j_1j_2}}{\sqrt{2!\,2!}} & 0 \\
 \frac{5\,\delta^j_{[i_1}\,\epsilon_{i_2 \cdots i_5]k}}{\sqrt{5!}} & 0 & 0 
 \end{pmatrix} . 
\end{align}
In the coordinates $(x^I)=\bigl(x^i,\, \frac{y_{i_1i_2}}{\sqrt{2}},\,z \bigr)$ with $z \equiv \frac{1}{5!}\,\epsilon^{j_1\cdots j_5}\,y_{j_1\cdots j_5}$, they become
\begin{align}
 \gamma^k = \sqrt{2}\,\begin{pmatrix}
 0 & \frac{2!\,\delta^{k i}_{j_1j_2}}{\sqrt{2!}} & 0 \\
 \frac{2!\,\delta^{k j}_{i_1i_2}}{\sqrt{2!}} & 0 & 0 \\
 0 & 0 & 0 
 \end{pmatrix} , \quad 
 \gamma_k = \sqrt{2}\,\begin{pmatrix}
 0 & 0 & \delta^i_k \\
 0 & \frac{\epsilon_{ki_1i_2j_1j_2}}{\sqrt{2!\,2!}} & 0 \\
 \delta^j_k & 0 & 0
 \end{pmatrix} .
\end{align}

We also define matrices
\begin{align}
 \bar{\gamma}^k \equiv \bigl(\bar{\gamma}^k_{IJ}\bigr) \equiv \sqrt{2}\,\frac{\epsilon^{kk_1\cdots k_4}\, \eta_{k_1\cdots k_4}}{4!} \,,\qquad 
 \bar{\gamma}_k \equiv \bigl(\bar{\gamma}_{k\,IJ}\bigr) \equiv \sqrt{2}\,\eta_k \,,
\end{align}
or more explicitly,
\begin{align}
 \bar{\gamma}^k = \sqrt{2}\,\begin{pmatrix}
 0 & 0 & \delta^k_i \\
 0 & \frac{\epsilon^{ki_1i_2j_1j_2}}{\sqrt{2!\,2!}} & 0 \\
 \delta^k_j & 0 & 0
 \end{pmatrix} , \qquad 
 \bar{\gamma}_k = \sqrt{2}\,\begin{pmatrix}
 0 & \frac{2!\,\delta_{k i}^{j_1j_2}}{\sqrt{2!}} & 0 \\
 \frac{2!\,\delta_{k j}^{i_1i_2}}{\sqrt{2!}} & 0 & 0 \\
 0 & 0 & 0
 \end{pmatrix} . 
\end{align}
Then, the $Y$-tensor can be expressed as
\begin{align}
 Y^{IJ}_{KL} = \eta^{kIJ}\,\eta_{kKL} + \frac{1}{4!}\,\eta^{k_1\cdots k_4IJ}\,\eta_{k_1\cdots k_4KL}
 = \frac{1}{2}\,\bigl(\gamma^{kIJ}\,\bar{\gamma}_{kKL} + \gamma_k^{IJ}\,\bar{\gamma}^k_{IJ}\bigr) \,.
\end{align}
If we further define
\begin{align}
 \bigl(\gamma^{\sfA}\bigr) \equiv \bigl(\gamma^i,\,\gamma_i\bigr)\,,\qquad 
 \bigl(\bar{\gamma}^{\sfA}\bigr) \equiv \bigl(\bar{\gamma}^i,\,\bar{\gamma}_i\bigr)\,,\qquad 
 \bigl(\eta^{\sfA\sfB}\bigr) \equiv \begin{pmatrix} 0 & \delta^i_j \\ \delta_i^j & 0 \end{pmatrix} ,
\end{align}
which satisfy the relation
\begin{align}
 \bigl(\gamma^{\sfA}\,\bar{\gamma}^{\sfB} + \gamma^{\sfB} \,\bar{\gamma}^{\sfA}\bigr){}^I{}_J = 2\,\eta^{\sfA\sfB}\,\delta^I_J \,, 
\end{align}
the $Y$-tensor can be expressed in the conventional form \eqref{eq:Y-tensor-case-by-case},
\begin{align}
 Y^{IJ}_{KL} = \frac{1}{2}\, \gamma_{\sfA}^{IJ}\,\bar{\gamma}^{\sfA}_{KL} \qquad \bigl(\gamma_{\sfA}\equiv \eta_{\sfA\sfB}\,\gamma^{\sfB}\bigr)\,.
\end{align}

\subsection{$Y$-tensor in $E_{6(6)}$ EFT}
\label{app:Y-E6}

In the $E_{6(6)}$ EFT, we have 27 $\eta$-symbols, which can be redefined as
\begin{align}
 d^k &\equiv -\frac{\epsilon_{k_1\cdots k_6}\,\eta^{k_1\cdots k_6,\,k}}{6!\sqrt{10}} 
 = \frac{1}{\sqrt{10}}\,\begin{pmatrix}
 0 & 0 & 0 \\
 0 & 0 & \frac{2!\,\epsilon_{j_1\cdots j_5 k}\delta^{kl}_{i_1i_2}}{\sqrt{2!\,5!}} \\
 0 & \frac{2!\,\epsilon_{i_1\cdots i_5 k}\delta^{kl}_{j_1j_2}}{\sqrt{2!\,5!}} & 0 
\end{pmatrix} ,
\\
 d_{k_1k_2} &\equiv \frac{\epsilon_{k_1k_2l_1\cdots l_4}\, \eta^{l_1\cdots l_4}}{4!\sqrt{10}} 
 = \frac{1}{\sqrt{10}}\,\begin{pmatrix}
 0 & 0 & \frac{5\,\delta^{i}_{[j_1}\,\epsilon_{j_2\cdots j_5]k_1k_2}}{\sqrt{5!}} \\
 0 & \frac{\epsilon_{k_1k_2i_1i_2j_1j_2}}{\sqrt{2!\,2!}} & 0 \\
 \frac{5\,\delta^{j}_{[i_1}\,\epsilon_{i_2\cdots i_5]k_1k_2}}{\sqrt{5!}} & 0 & 0 
 \end{pmatrix} , 
\\
 d^{\bar{k}} &\equiv \frac{1}{\sqrt{10}}\,\eta^k = \frac{1}{\sqrt{10}}\,\begin{pmatrix}
 0 & \frac{2!\,\delta^{k i}_{j_1j_2}}{\sqrt{2!}} & 0 \\
 \frac{2!\,\delta^{k j}_{i_1i_2}}{\sqrt{2!}} & 0 & 0 \\
 0 & 0 & 0 
 \end{pmatrix} . 
\end{align}
In the coordinates $\bigl(x^i,\,\frac{y_{i_1i_2}}{\sqrt{2}},\,z^i\bigr)$ with $z^i\equiv -\frac{1}{5!}\,\epsilon^{i j_1\cdots j_5}\,y_{j_1\cdots j_5}$, the matrices become
\begin{align}
 &d^k = \frac{1}{\sqrt{10}}\,\begin{pmatrix}
 0 & 0 & 0 \\
 0 & 0 & \frac{2!\,\delta^{kj}_{i_1i_2}}{\sqrt{2!}} \\
 0 & \frac{2!\,\delta^{ki}_{j_1j_2}}{\sqrt{2!}} & 0 
 \end{pmatrix} , \qquad
 d^{\bar{k}} = \frac{1}{\sqrt{10}}\,\begin{pmatrix}
 0 & \frac{2!\,\delta^{k i}_{j_1j_2}}{\sqrt{2!}} & 0 \\
 \frac{2!\,\delta^{k j}_{i_1i_2}}{\sqrt{2!}} & 0 & 0 \\
 0 & 0 & 0
 \end{pmatrix} ,
\\
 &d_{k_1k_2} = \frac{1}{\sqrt{10}}\,\begin{pmatrix}
 0 & 0 & 2!\,\delta^{ij}_{k_1k_2} \\
 0 & \frac{\epsilon_{k_1k_2 i_1i_2j_1j_2}}{\sqrt{2!\,2!}} & 0 \\
 2!\,\delta^{ji}_{k_1k_2} & 0 & 0 
 \end{pmatrix} .
\end{align}
They are components of the conventional totally symmetric tensor $d^{IJK}\equiv \bigl(d^{IJ;\,k},\,\frac{d^{IJ}_{k_1k_2}}{\sqrt{2!}},\,d^{IJ;\,\bar{k}}\bigr)$ (see, e.g., Eq.~(4.42) in \cite{Hohm:2013vpa}). 
By defining $d_{IJK}$ in a similar manner, they satisfy
\begin{align}
 d^{IKL}\,d_{JKL} = \delta^I_J \,,\qquad d^{IJK}\,d_{IJK}=27 \,. 
\end{align}
They also satisfy the relation
\begin{align}
 10\,d^{M(I|P} \,d_{QPL}\,d^{Q|JK)} - d^{M(IJ}\,\delta^{K)}_L = \frac{1}{3}\,d^{IJK}\,\delta^M_L \,,
\end{align}
which ensures the following relation for the $Y$-tensor:
\begin{align}
 Y_{RS}^{(I|P} \, Y_{PL}^{|JK)} - Y_{RS}^{(IJ}\,\delta^{K)}_L = \frac{10}{3}\,d_{LRS}\,d^{IJK} \,.
\end{align}

\subsection{$Y$-tensor in $E_{7(7)}$ EFT}
\label{app:Y-E7}

In the $E_{7(7)}$ case, we have 133 $\eta$-symbols, which can be redefined as
\begin{align}
 t_{k_1\cdots k_4}
 &\equiv \frac{\epsilon_{i_1\cdots i_7}\,\epsilon_{k_1\cdots k_4 j_1j_2j_3}\,\eta^{i_1\cdots i_7,\,j_1j_2j_3}}{3!\,7!} 
\nn\\
 &= \begin{pmatrix}
 0 & 0 & 0 & 0 \\
 0 & 0 & 0 & -\frac{7!\,\epsilon_{i_1i_2 j [j_1\cdots j_4}\,\epsilon_{j_5j_6j_7] k_1\cdots k_4}}{3!\,4!\sqrt{2!\,7!}} \\
 0 & 0 & \frac{5!\,\epsilon_{i_1\cdots i_5[j_1j_2}\,\epsilon_{j_3j_4j_5]k_1\cdots k_4}}{2!\,3!\sqrt{5!\,5!}} & 0 \\
 0 & -\frac{7!\,\epsilon_{j_1j_2 i [i_1\cdots i_4}\,\epsilon_{i_5i_6i_7] k_1\cdots k_4}}{3!\,4!\sqrt{2!\,7!}} & 0 & 0
 \end{pmatrix} , 
\\
 t_{k_1k_2k_38}
 &\equiv -\frac{1}{4!}\,\epsilon_{k_1k_2k_3l_1\cdots l_4}\, \eta^{l_1\cdots l_4} 
\nn\\
 &= \begin{pmatrix}
 0 & 0 & -\frac{5}{\sqrt{5!}}\,\delta^i_{[j_1}\,\epsilon_{j_2\cdots j_5] k_1k_2k_3} & 0 \\
 0 & -\frac{\epsilon_{k_1k_2k_3 i_1i_2j_1j_2}}{\sqrt{2!\,2!}} & 0 & 0 \\
 -\frac{5}{\sqrt{5!}}\,\delta^j_{[i_1}\,\epsilon_{i_2\cdots i_5] k_1k_2k_3} & 0 & 0 & 0 \\
 0 & 0 & 0 & 0
 \end{pmatrix} , 
\\
 t_8{}^k &\equiv - \eta^k
 = \begin{pmatrix}
 0 & -\frac{2!}{\sqrt{2!}}\,\delta^{k i}_{j_1j_2} & 0 & 0 \\
 -\frac{2!}{\sqrt{2!}}\,\delta^{k j}_{i_1i_2} & 0 & 0 & 0 \\
 0 & 0 & 0 & 0 \\
 0 & 0 & 0 & 0 
 \end{pmatrix} , 
\\
 t_k{}^l &\equiv \frac{1}{6!}\, \epsilon_{kk_1\cdots k_6}\,\Bigl(\eta^{k_1\cdots k_6,\,l}-\frac{\sqrt{2}}{4}\,\eta^{[k_1\cdots k_6,\,l]}\Bigr)
\nn\\
 &\equiv {\tiny
 {\arraycolsep=-4.0mm \left(\begin{array}{cccc}
 0 & 0 & 0 & \frac{\delta_k^i\epsilon_{j_1\cdots j_7}\delta_j^l - \frac{1}{4}\delta_k^l\epsilon_{j_1\cdots j_7}\delta_j^i}{\sqrt{7!}}~~~~ \\
 0 & 0 & -\frac{2!\delta^{lm}_{i_1i_2}\epsilon_{j_1\cdots j_5km}-\frac{1}{4}\delta_k^l\epsilon_{i_1i_2j_1\cdots j_5}}{\sqrt{2!5!}} & 0 \\
 0 & -\frac{2!\delta^{lm}_{j_1j_2}\epsilon_{i_1\cdots i_5km}-\frac{1}{4}\delta_k^l\epsilon_{j_1j_2i_1\cdots i_5}}{\sqrt{2!5!}} & 0 & 0 \\
 ~~~~\frac{\delta_k^j\epsilon_{i_1\cdots i_7}\delta_i^l - \frac{1}{4}\delta_k^l\epsilon_{i_1\cdots i_7}\delta_i^j}{\sqrt{7!}} & 0 & 0 & 0
 \end{array}\right)\,,}} 
\\
 t_k{}^8 &\equiv \frac{1}{6!\,7!}\, \epsilon_{k_1\cdots k_7}\,\epsilon_{kl_1\cdots l_6}\,\eta^{k_1\cdots k_7,\,l_1\cdots l_6} 
\nn\\
 &= 
 \begin{pmatrix}
 0 & 0 & 0 & 0 \\
 0 & 0 & 0 & 0 \\
 0 & 0 & 0 & \frac{\epsilon_{j_1\cdots j_7}\,\epsilon_{k j i_1\cdots i_5}}{2 \sqrt{5!}} \\
 0 & 0 & \frac{\epsilon_{i_1\cdots i_7}\,\epsilon_{k i j_1\cdots j_5}}{2 \sqrt{5!}} & 0
 \end{pmatrix} .
\end{align}

If we use the conventional parameterization of the generalized coordinates,
\begin{align}
 (x^I) = \bigl( \tfrac{x^{\hat{i}_1\hat{i}_2}}{\sqrt{2!}},\, \tfrac{x_{\hat{i}_1\hat{i}_2}}{\sqrt{2!}}\bigr)\qquad 
 \bigl(\hat{i} =1,\dotsc,8\bigr) \,,
\end{align}
where
\begin{align}
 x^{i8} \equiv x^i\,,\quad x^{i_1i_2}\equiv -\frac{1}{5!}\,\epsilon^{i_1i_2j_1\cdots j_5}\,y_{j_1\cdots j_5}\,,\quad 
 x_{i8} \equiv \frac{1}{7!}\,\epsilon^{j_1\cdots j_7}\,y_{j_1\cdots j_7,\,i}\,,\quad x_{i_1i_2}\equiv y_{i_1i_2}\,, 
\end{align}
the above matrices take the following forms:
\begin{align}
 t_{k_1\cdots k_4}
 &= \begin{pmatrix}
 0 & 0 & 0 & 0 \\
 0 & \frac{4!}{\sqrt{2!\,2!}}\,\delta^{i_1i_2 j_1j_2}_{k_1\cdots k_4} & 0 & 0 \\
 0 & 0 & 0 & -\frac{\epsilon_{i j_1j_2 k_1\cdots k_4}}{\sqrt{2!}} \\
 0 & 0 & -\frac{\epsilon_{i_1i_2 j k_1\cdots k_4}}{\sqrt{2!}} & 0
 \end{pmatrix} , 
\\
 t_{k_1k_2k_38}
 &= \begin{pmatrix}
 0 & \frac{3!}{\sqrt{2!}}\,\delta^{i j_1j_2}_{k_1k_2k_3} & 0 & 0 \\
 \frac{3!}{\sqrt{2!}}\,\delta^{i_1i_2 j}_{k_1k_2k_3} & 0 & 0 & 0 \\
 0 & 0 & 0 & 0 \\
 0 & 0 & 0 & -\frac{\epsilon_{i_1i_2j_1j_2 k_1k_2k_3}}{\sqrt{2!\,2!}}
 \end{pmatrix} , 
\\
 t_8{}^k &= \begin{pmatrix}
 0 & 0 & 0 & -\frac{2!}{\sqrt{2!}}\,\delta^{k i}_{j_1j_2} \\
 0 & 0 & 0 & 0 \\
 0 & 0 & 0 & 0 \\
 -\frac{2!}{\sqrt{2!}}\,\delta^{k j}_{i_1i_2} & 0 & 0 & 0 
 \end{pmatrix} , 
\\
 t_k{}^l &{\arraycolsep=-0.8mm = \left(\begin{array}{cccc}
 0 & 0 & \delta_k^i\,\delta^l_j-\frac{1}{4}\,\delta_k^l\,\delta^i_j & 0 \\
 0 & 0 & 0 & 2\,\delta^{i_1i_2}_{km}\,\delta^{lm}_{j_1j_2}-\frac{1}{4}\,\delta_k^l\, \delta^{i_1i_2}_{j_1j_2} \\
 \delta_k^j\,\delta^l_i-\frac{1}{4}\,\delta_k^l\,\delta^j_i & 0 & 0 & 0 \\
 0 & 2\,\delta^{j_1j_2}_{km}\,\delta^{lm}_{i_1i_2}-\frac{1}{4}\,\delta_k^l\, \delta^{j_1j_2}_{i_1i_2} & 0 & 0
 \end{array}\right)\,,} 
\\
 t_k{}^8 &= 
 \begin{pmatrix}
 0 & 0 & 0 & 0 \\
 0 & 0 & -\frac{2!}{\sqrt{2!}}\,\delta^{i_1i_2}_{kj} & 0 \\
 0 & - \frac{2!}{\sqrt{2!}}\,\delta^{j_1j_2}_{ki} & 0 & 0 \\
 0 & 0 & 0 & 0
 \end{pmatrix} .
\end{align}
These can be summarized as the following familiar matrices (see, e.g., Appendix A.2 in \cite{LeDiffon:2011wt}):
\begin{align}
 t_{\hat{k}_1\cdots \hat{k}_4} &= \begin{pmatrix}
 \frac{4!}{\sqrt{2!\,2!}}\,\delta^{\hat{i}_1\hat{i}_2 \hat{j}_1\hat{j}_2}_{\hat{k}_1\cdots \hat{k}_4} & 0 \\
 0 & -\frac{\epsilon_{\hat{i}_1\hat{i}_2 \hat{j}_1\hat{j}_2 \hat{k}_1\cdots \hat{k}_4}}{\sqrt{2!\,2!}} 
 \end{pmatrix} , 
\\
 t_{\hat{k}}{}^{\hat{l}} &= \begin{pmatrix} 0 & 2\,\delta^{\hat{i}_1\hat{i}_2}_{\hat{k}\hat{m}}\,\delta^{\hat{l}\hat{m}}_{\hat{j}_1\hat{j}_2} -\frac{1}{4}\,\delta^{\hat{l}}_{\hat{k}}\,\delta^{\hat{i}_1\hat{i}_2}_{\hat{j}_1\hat{j}_2} \\
 2\,\delta^{\hat{j}_1\hat{j}_2}_{\hat{k}\hat{m}}\,\delta^{\hat{l}\hat{m}}_{\hat{i}_1\hat{i}_2} -\frac{1}{4}\,\delta^{\hat{l}}_{\hat{k}}\,\delta^{\hat{j}_1\hat{j}_2}_{\hat{i}_1\hat{i}_2} & 0
 \end{pmatrix} ,
\end{align}
where
\begin{align}
 t_{\hat{k}}{}^{\hat{k}} = 0 \,,\quad \epsilon_{\hat{1}\hat{2}\hat{3}\hat{4}\hat{5}\hat{6}\hat{7}\hat{8}} = \epsilon_{1234567} \,.
\end{align}
From these matrices, the $Y$-tensor and the generalized Lie derivative have been explicitly computed in \cite{Rosabal:2014rga}.


\begin{thebibliography}{99}
\bibitem{Siegel:1993xq} 
  W.~Siegel,
  ``Two vierbein formalism for string inspired axionic gravity,''
  Phys.\ Rev.\ D {\bf 47}, 5453 (1993)
  [hep-th/9302036].



\bibitem{Siegel:1993th} 
  W.~Siegel,
  ``Superspace duality in low-energy superstrings,''
  Phys.\ Rev.\ D {\bf 48}, 2826 (1993)
  [hep-th/9305073].



\bibitem{Siegel:1993bj} 
  W.~Siegel,
  ``Manifest duality in low-energy superstrings,''
  hep-th/9308133.



\bibitem{Hull:2009mi} 
  C.~Hull and B.~Zwiebach,
  ``Double Field Theory,''
  JHEP {\bf 0909}, 099 (2009)
  [arXiv:0904.4664 [hep-th]].



\bibitem{Hull:2009zb} 
  C.~Hull and B.~Zwiebach,
  ``The Gauge algebra of double field theory and Courant brackets,''
  JHEP {\bf 0909}, 090 (2009)
  [arXiv:0908.1792 [hep-th]].



\bibitem{Hohm:2010jy} 
  O.~Hohm, C.~Hull and B.~Zwiebach,
  ``Background independent action for double field theory,''
  JHEP {\bf 1007}, 016 (2010)
  [arXiv:1003.5027 [hep-th]].



\bibitem{Hohm:2010pp} 
  O.~Hohm, C.~Hull and B.~Zwiebach,
  ``Generalized metric formulation of double field theory,''
  JHEP {\bf 1008}, 008 (2010)
  [arXiv:1006.4823 [hep-th]].



\bibitem{Berman:2010is} 
  D.~S.~Berman and M.~J.~Perry,
  ``Generalized Geometry and M theory,''
  JHEP {\bf 1106}, 074 (2011)
  [arXiv:1008.1763 [hep-th]].



\bibitem{Berman:2011cg} 
  D.~S.~Berman, H.~Godazgar, M.~Godazgar and M.~J.~Perry,
  ``The Local symmetries of M-theory and their formulation in generalised geometry,''
  JHEP {\bf 1201}, 012 (2012)
  [arXiv:1110.3930 [hep-th]].



\bibitem{Berman:2011jh} 
  D.~S.~Berman, H.~Godazgar, M.~J.~Perry and P.~West,
  ``Duality Invariant Actions and Generalised Geometry,''
  JHEP {\bf 1202}, 108 (2012)
  [arXiv:1111.0459 [hep-th]].



\bibitem{Berman:2012vc} 
  D.~S.~Berman, M.~Cederwall, A.~Kleinschmidt and D.~C.~Thompson,
  ``The gauge structure of generalised diffeomorphisms,''
  JHEP {\bf 1301}, 064 (2013)
  [arXiv:1208.5884 [hep-th]].



\bibitem{Hohm:2013pua} 
  O.~Hohm and H.~Samtleben,
  ``Exceptional Form of D=11 Supergravity,''
  Phys.\ Rev.\ Lett.\  {\bf 111}, 231601 (2013)
  [arXiv:1308.1673 [hep-th]].



\bibitem{Hohm:2013vpa} 
  O.~Hohm and H.~Samtleben,
  ``Exceptional Field Theory I: $E_{6(6)}$ covariant Form of M-Theory and Type IIB,''
  Phys.\ Rev.\ D {\bf 89}, no. 6, 066016 (2014)
  [arXiv:1312.0614 [hep-th]].



\bibitem{Hohm:2013uia} 
  O.~Hohm and H.~Samtleben,
  ``Exceptional field theory. II. E$_{7(7)}$,''
  Phys.\ Rev.\ D {\bf 89}, 066017 (2014)
  [arXiv:1312.4542 [hep-th]].



\bibitem{Aldazabal:2013via} 
  G.~Aldazabal, M.~Gra\~na, D.~Marqu\'es and J.~A.~Rosabal,
  ``The gauge structure of Exceptional Field Theories and the tensor hierarchy,''
  JHEP {\bf 1404}, 049 (2014)
  [arXiv:1312.4549 [hep-th]].



\bibitem{Hohm:2014fxa} 
  O.~Hohm and H.~Samtleben,
  ``Exceptional field theory. III. E$_{8(8)}$,''
  Phys.\ Rev.\ D {\bf 90}, 066002 (2014)
  [arXiv:1406.3348 [hep-th]].



\bibitem{West:2001as} 
  P.~C.~West,
  ``E(11) and M theory,''
  Class.\ Quant.\ Grav.\  {\bf 18}, 4443 (2001)
  [hep-th/0104081].



\bibitem{West:2003fc} 
  P.~C.~West,
  ``E(11), SL(32) and central charges,''
  Phys.\ Lett.\ B {\bf 575}, 333 (2003)
  [hep-th/0307098].



\bibitem{West:2004st} 
  P.~C.~West,
  ``The IIA, IIB and eleven-dimensional theories and their common E(11) origin,''
  Nucl.\ Phys.\ B {\bf 693}, 76 (2004)
  [hep-th/0402140].



\bibitem{Hillmann:2009ci} 
  C.~Hillmann,
  ``Generalized E(7(7)) coset dynamics and D=11 supergravity,''
  JHEP {\bf 0903}, 135 (2009)
  [arXiv:0901.1581 [hep-th]].



\bibitem{Pacheco:2008ps} 
  P.~Pires Pacheco and D.~Waldram,
  ``M-theory, exceptional generalised geometry and superpotentials,''
  JHEP {\bf 0809}, 123 (2008)
  [arXiv:0804.1362 [hep-th]].



\bibitem{Coimbra:2011ky} 
  A.~Coimbra, C.~Strickland-Constable and D.~Waldram,
  ``$E_{d(d)} \times \mathbb{R}^+$ generalised geometry, connections and M theory,''
  JHEP {\bf 1402}, 054 (2014)
  [arXiv:1112.3989 [hep-th]].



\bibitem{Coimbra:2012af} 
  A.~Coimbra, C.~Strickland-Constable and D.~Waldram,
  ``Supergravity as Generalised Geometry II: $E_{d(d)} \times \mathbb{R}^+$ and M theory,''
  JHEP {\bf 1403}, 019 (2014)
  [arXiv:1212.1586 [hep-th]].



\bibitem{Bandos:2015rvs} 
  I.~Bandos,
  ``On section conditions of E$_{7(+7)}$ exceptional field theory and superparticle in $ \mathcal{N}=8 $ central charge superspace,''
  JHEP {\bf 1601}, 132 (2016)
  [arXiv:1512.02287 [hep-th]].



\bibitem{Bandos:2016ppv} 
  I.~Bandos,
  ``Exceptional field theories, superparticles in an enlarged 11D superspace and higher spin theories,''
  Nucl.\ Phys.\ B {\bf 925}, 28 (2017)
  [arXiv:1612.01321 [hep-th]].



\bibitem{Hatsuda:2012vm} 
  M.~Hatsuda and K.~Kamimura,
  ``SL(5) duality from canonical M2-brane,''
  JHEP {\bf 1211}, 001 (2012)
  [arXiv:1208.1232 [hep-th]].



\bibitem{Hatsuda:2013dya} 
  M.~Hatsuda and K.~Kamimura,
  ``M5 algebra and SO(5,5) duality,''
  JHEP {\bf 1306}, 095 (2013)
  [arXiv:1305.2258 [hep-th]].



\bibitem{Hatsuda:2012uk} 
  M.~Hatsuda and T.~Kimura,
  ``Canonical approach to Courant brackets for D-branes,''
  JHEP {\bf 1206}, 034 (2012)
  [arXiv:1203.5499 [hep-th]].



\bibitem{Obers:1998fb} 
  N.~A.~Obers and B.~Pioline,
  ``U duality and M theory,''
  Phys.\ Rept.\  {\bf 318}, 113 (1999)
  [hep-th/9809039].



\bibitem{Sakatani:2017nfr} 
  Y.~Sakatani and S.~Uehara,
  ``Connecting M-theory and type IIB parameterizations in Exceptional Field Theory,''
  PTEP {\bf 2017}, no. 4, 043B05 (2017)
  [arXiv:1701.07819 [hep-th]].



\bibitem{Linch:2015lwa} 
  W.~D.~Linch and W.~Siegel,
  ``F-theory Superspace,''
  arXiv:1501.02761 [hep-th].



\bibitem{Linch:2015fya} 
  W.~D.~Linch, III and W.~Siegel,
  ``F-theory from Fundamental Five-branes,''
  arXiv:1502.00510 [hep-th].



\bibitem{Linch:2015qva} 
  W.~D.~Linch and W.~Siegel,
  ``F-theory with Worldvolume Sectioning,''
  arXiv:1503.00940 [hep-th].



\bibitem{Linch:2015fca} 
  W.~D.~Linch and W.~Siegel,
  ``Critical Super F-theories,''
  arXiv:1507.01669 [hep-th].



\bibitem{Linch:2016ipx} 
  W.~D.~Linch and W.~Siegel,
  ``F-brane Dynamics,''
  arXiv:1610.01620 [hep-th].



\bibitem{Malek:2017njj} 
  E.~Malek,
  ``Half-Maximal Supersymmetry from Exceptional Field Theory,''
  Fortsch.\ Phys.\  {\bf 65}, no. 10-11, 1700061 (2017)
  [arXiv:1707.00714 [hep-th]].



\bibitem{Blair:2013gqa} 
  C.~D.~A.~Blair, E.~Malek and J.~H.~Park,
  ``M-theory and Type IIB from a Duality Manifest Action,''
  JHEP {\bf 1401}, 172 (2014)
  [arXiv:1311.5109 [hep-th]].



\bibitem{Rosabal:2014rga} 
  J.~A.~Rosabal,
  ``On the exceptional generalised Lie derivative for $d\geq7$,''
  JHEP {\bf 1509}, 153 (2015)
  [arXiv:1410.8148 [hep-th]].



\bibitem{Courant}
  T.~J.~Courant,
  ``Dirac manifolds,''
  Trans.\ Amer.\ Math.\ Soc.\ {\bf 319} 631--661 (1990).



\bibitem{Rey:2015mba} 
  S.~J.~Rey and Y.~Sakatani,
  ``Finite Transformations in Doubled and Exceptional Space,''
  arXiv:1510.06735 [hep-th].



\bibitem{Gualtieri:2003dx} 
  M.~Gualtieri,
  ``Generalized complex geometry,''
  math/0401221 [math-dg].



\bibitem{Cederwall:IIB}
  M.~Cederwall, ``(Brane) Charges for 1/2 BPS in Exceptional Geometry,''
  presented at the workshop ``Duality and Novel Geometry in M-theory'', 
  Asia Pacific Centre for Theoretical Physics, Postech, 2016.



\bibitem{Blair:2017gwn} 
  C.~D.~A.~Blair,
  ``Particle actions and brane tensions from double and exceptional geometry,''
  JHEP {\bf 1710}, 004 (2017)
  [arXiv:1707.07572 [hep-th]].



\bibitem{Duff:1989tf} 
  M.~J.~Duff,
  ``Duality Rotations in String Theory,''
  Nucl.\ Phys.\ B {\bf 335}, 610 (1990).



\bibitem{Sakatani:2016sko} 
  Y.~Sakatani and S.~Uehara,
  ``Branes in Extended Spacetime: Brane Worldvolume Theory Based on Duality Symmetry,''
  Phys.\ Rev.\ Lett.\  {\bf 117}, no. 19, 191601 (2016)
  [arXiv:1607.04265 [hep-th]].



\bibitem{Tumanov:2014pfa} 
  A.~G.~Tumanov and P.~West,
  ``Generalised vielbeins and non-linear realisations,''
  JHEP {\bf 1410}, 009 (2014)
  [arXiv:1405.7894 [hep-th]].



\bibitem{Lee:2016qwn} 
  K.~Lee, S.~J.~Rey and Y.~Sakatani,
  ``Effective action for non-geometric fluxes duality covariant actions,''
  JHEP {\bf 1707}, 075 (2017)
  [arXiv:1612.08738 [hep-th]].



\bibitem{LeDiffon:2011wt} 
  A.~Le Diffon, H.~Samtleben and M.~Trigiante,
  ``N=8 Supergravity with Local Scaling Symmetry,''
  JHEP {\bf 1104}, 079 (2011)
  [arXiv:1103.2785 [hep-th]].
\end{thebibliography}
\end{document}